\documentclass[12pt]{JHEP}
\usepackage{amsmath,epsfig}
\usepackage{amssymb,amsfonts}
\usepackage{latexsym}
\usepackage{graphicx}
\relax
\renewcommand{\theequation}{\arabic{section}.\arabic{equation}}
\def\hre#1#2{\href{http://arxiv.org/abs/#1/#2}{[ArXiv:#1/#2]}}

\def\be{\begin{equation}}
\def\ee{\end{equation}}
\def\bea{\begin{eqnarray}}
\def\eea{\end{eqnarray}}
\newcommand\fverb{\setbox\pippobox=\hbox\bgroup\verb}
\newcommand\fverbdo{\egroup\medskip\noindent%
                        \fbox{\unhbox\pippobox}\ }
\newcommand\fverbit{\egroup\item[\fbox{\unhbox\pippobox}]}

\newcommand{\la}{\lambda}

\newcommand{\bear}{\begin{eqnarray}}

\newcommand{\eear}{\end{eqnarray}}

\newbox\pippobox

\def\lab{\label}
\def\6{\partial}
\def\f{\Phi}
\def\a{\alpha}
\def\nn{\nonumber}
\def\of{\bar{f}}
\def\half{\frac12}

\def\le{\left}
\def\ri{\right}
\def\cO{{\cal O}}

\def\C0{{\bf C_0}}
\def\Y0{{\bf Y_0}}
\def\G0{{\bf G_0}}
\def\e{\epsilon}
\def\m{\mu}
\def\n{\nu}

\def\s{\sigma}

\def\sq
\def\a{\alpha}
\def\b{\beta}
\def\l{\lambda}

\def\tr{{\rm Tr}}

\def\eps{\epsilon}

\def\cL{{\cal L}}
\def\cH{{\cal H}}

\def\la{\langle}
\def\ra{\rangle}
\def\o{\omega}

\def\bz{\begin{itemize}}
\def\ez{\end{itemize}}
\def\bn{\begin{enumerate}}
\def\en{\end{enumerate}}
\def\parl{\parallel}
\def\lra{\leftrightarrow}
\def\ft{\tilde{f}}
\def\gt{\tilde{g}}
\def\of{\overline{\f}}
\def\dg{\dagger}

\title{Gravity/Spin-model correspondence and holographic superfluids}
\author{Umut G{\"u}rsoy\\
\href{http://www1.phys.uu.nl/wwwitf}{Institute for Theoretical
Physics, Utrecht University; Leuvenlaan 4, 3584 CE Utrecht, The
Netherlands.}}
\date{\today}
\preprint{
SPIN-10/17\\
ITP-UU-10/19}      % OR:
\abstract{We propose a general correspondence between gravity and
 spin models, inspired by the well-known
IR equivalence between lattice gauge theories and the spin models.
This suggests a connection between continuous type Hawking-phase
transitions in gravity and the continuous order-disorder
transitions in ferromagnets. The black-hole phase corresponds to
the ordered and the graviton gas corresponds to the disordered
phases respectively. A simple set-up based on Einstein-dilaton
gravity indicates that the vicinity of the phase transition is
governed by a linear-dilaton CFT. Employing this CFT we calculate
scaling of observables near $T_c$, and obtain mean-field scaling
in a semi-classical approximation. In case of the XY model the
Goldstone mode is identified with the zero mode of the NS-NS
two-form. We show that the second speed of sound vanishes at the
transition also with the mean field exponent.}

\begin{document}

\maketitle

\section{Introduction}

There has been great progress recently in applications of
holography \cite{malda,GKP,Witten1} to condensed matter systems
such as superconductors following the pioneering works of
\cite{Gubser} and \cite{Sean}. These authors managed to find a
simple gravitational background in Einstein-Maxwell gravity
coupled to a complex scalar field where a second order
normal-to-superfluid type transition occurs at finite
temperature. The basic interest behind application of holographic
ideas to condensed matter theory (CMT) lies in the hope that the
strongly correlated condensed matter systems may secretly possess
a gravitational description. Indeed, computations of certain
observables in the gravity picture, such as conductivity provides
supporting evidence, see
\cite{review1}\cite{review2}\cite{review3} for reviews. It is a
considerable possibility that \cite{SachdevBook} the underlying
dynamics behind the phase transition in high $T_c$ superconducting
materials is a strongly coupled quantum phase transition at zero
T. Then the hope is that, a dual gravity description of the
strongly coupled field theory around this critical point may also
shed light over the finite temperature transition in the quantum
critical region.

On the other hand, there are several issues of fundamental
importance in the proposed gravity-CMT models, such as the role of
the large N limit and the notion of weak-strong duality, that are
not entirely clarified. We have a much better understanding in the
holographic constructions of gauge theories, thanks to the basic
example \cite{malda,GKP,Witten1} of the ${\cal N}=4$ super
Yang-Mills theory where the D3 brane picture provide the link
between the gauge side and the gravity side. Such a ``top-bottom"
approach is missing in the gravity-CMT models.

In this work, we entertain the possibility that such a link may be
established under certain assumptions, at least for certain simple
condensed matter systems, i.e. spin models, by analogy with the
better understood gauge-gravity case.

The building blocks of such a connection are already present in
the well-known literature. First of all, we recall the famous
equivalence between lattice gauge theories (LGT) and spin-models
(SpM)\cite{Polyakov, Susskind}:
 Integrating out the gauge invariant degrees of freedom in the
 partition function of a LGT with gauge group $G$,
 one arrives at an effective action for the
lowest lying mode, namely the Polyakov loop $P$. This effective
action is invariant under the leftover center symmetry ${\cal C} =
Center(G)$ of the original gauge invariance. Identifying the
Polyakov loop $P$ with a spin field $\vec{s}$, one then obtains
the partition function of a spin-model with the global spin
invariance ${\cal C}$. Using this equivalence between lattice
gauge theories and spin-models Polyakov and Susskind were able to
show the existence of confinement-deconfinement phase transition
on the lattice, long time ago. Based on these works, than
Svetitsky and Yaffe \cite{Yaffe} further proposed that, if
continuous critical phenomena prevails in the continuum limit of a
certain lattice gauge theory, then it should fall in the same
universality class as the corresponding spin-model.

It is interesting to employ the same idea in the opposite
direction in order to study a spin-model that is strongly coupled
at criticality. In particular, one would like to compute the
critical exponents, the transition temperature $T_c$, certain
thermodynamic functions etc., by analytic methods. If one is lucky
enough to find a gauge-theory that corresponds to the spin-model
under the aforementioned equivalence, then one may be able  to
study the strongly coupled phenomena by the gauge-gravity
correspondence.

One purpose of this paper is to emphasize that this chain of
dualities may provide a well-defined setting in understanding
fundamental issues in the gravity-CMT correspondence. In
particular, if one can figure out the relevant D-brane
configuration that describes the gauge theory which arises in the
continuum limit of the LGT under question, then one may be able to
take the decoupling limit and obtain a gravity description of the
LGT---and of the equivalent spin model---around criticality.
Despite being abstract, in principle this provides a top-bottom
approach to the problem. In particular, such an approach would
hopefully provide a microscopic description that is long sought
for in holographic applications to CMT.

Another purpose of this paper is to provide a concrete realization
of these ideas in a simple setting. For this purpose we consider
$SU(N)$ gauge theory in $d$-dimensions (with possible adjoint
matter) in the strict $N\to\infty$ limit. In this limit the center
${\cal C}$ becomes $U(1)$\footnote{This idea in the AdS/CFT
context was considered before\cite{Witten}, see also
\cite{Minwalla2} and \cite{Yaffe} for earlier
discussions}. We imagine that the adjoint matter is arranged such
that the deconfinement transition of the gauge theory is of
continuous type. This transition is then in the same universality
class with the order-disorder transition in the corresponding
$U(1)$ rotor model in $d-1$ dimensions---that is sometimes called
the XY model. The XY-models---and their $O(n)$
generalizations---provide canonical examples of superfluidity that
arises as spontaneous breaking of the global $U(1)$ symmetry in a
continuous phase transition.

To realize this phenomenon in the dual gravity setting we consider
the NS-NS sector of non-critical string theory in $d+1$ dimensions
with Euclidean time direction $x^0$ compactified. It was shown in
\cite{exotic1} that this theory in the two-derivative sector
exhibits a continuous Hawking-Page transition at some finite
temperature $T_c$. The background is of the type $AdS^{d+1}$ near
the boundary and linear-dilaton in the deep-interior. Building
upon the ideas in \cite{Witten}, we argue that the $U(1)$ symmetry
(in the strict $N\to\infty$ limit) corresponds to the shift
symmetry $\int_M B \to \int_M B + const$ where $B$ is the NS-NS
two-form field and $M$ is the $(r,x^0)$ subspace of the background
geometry. The only objects that are charged under this symmetry
are string states winding the time circle. In the thermal gas
phase these states have infinite energy and cannot be excited,
hence the symmetry is unbroken and this phase corresponds to the
{\em normal phase} of the spin system. In the black-hole phase on
the other-hand they have finite energy (with an appropriate
regularization) and the black-hole corresponds to the {\em
superfluid phase}.

It was further observed in \cite{exotic1} that the geometry
becomes exactly linear-dilaton {\em in the transition region}.
Therefore, we argue that the {\em transition region of
the XY model is governed by the linear-dilaton CFT on the string
side.} Although in general the $\a'$ corrections can not be ignored 
in the type of backgrounds that we will consider in this paper\footnote{We recall that in 
the case of ${\cal N}=4$ sYM theory the $\a'$ corrections can be ignored 
both for the bulk and the string computations at strong 't Hooft coupling $\l$. 
In the theories we consider here we do not have a similar modulus that serves 
as a parameter to suppress the $\a'$ corrections. Generally, the string scale 
and the scale of the background geometry may be of the same order.}, 
one can still perform calculations in the critical regime, precisely because the 
linear-dilaton background is known to be an $\a'$-exact background in non-critical 
string theory\cite{PolchinskiBook1}.  In particular the calculations that involve 
probe strings  can be performed by employing the exact CFT description of the
linear-dilaton background, (in the limit $g_s\to 0$). 

The spin operator $\vec{s}(x)$ is related to a fundamental string that
wraps the time-circle and connected to the boundary at point $x$.
Consequently one can compute correlation functions of the operator
$\vec{s}$ by studying the string propagation in the linear-dilaton
CFT in the (single) winding sector. We perform such calculations
in a {\em semi-classical limit} where we only take into account
the contribution of the lowest-lying string states. It is shown
that in this approximation, one obtains mean-field scaling near
$T_c$. We find that the ``magnetization" behaves as
$$ M \sim (T-T_c)^{\half},\qquad as, \,\,\, T\to T_c. $$
A similar calculation with string propagation connecting the
points $x$ and $y$ on the (spatial) boundary corresponds to the
spin two-point function. We show that the expected behavior of the
spin-system arises in the large $|x-y|$ limit near $T_c$ indeed
arises from this calculation in a non-trivial manner. In
particular, in order to show that the correlation length $\xi$
diverges at $T_c$, one has to identify the transition with the
{\em Hagedorn temperature} where the lowest lying single-winding
mode becomes massless \cite{AtickWitten}. With such an
identification one indeed finds the expected behavior
$$\xi \sim (T-T_c)^{-\half},\qquad as, \,\,\, T\to T_c,$$
again in a semi-classical approximation.

One can also study scaling of the speed of second sound that is
associated with the Goldstone mode in the superfluid phase. This
mode is identified with fluctuations of the zero-mode of the
$NS$ two-form field $B$. We find that the speed of sound indeed vanishes at
$T_c$ precisely with the expected mean-field scaling,
$$c_s^2 \sim (T-T_c), \qquad as\,\,\, T\to T_c$$ in a second order
Hawking-Page transition. We also argue that this finding is not
altered by possible $\a'$ corrections.

The identification of spin operators with the F-strings suggest a
similar identification between the vortex configurations---that
play an important role in the 2D XY model---with D-strings in the
gravity dual. We study correlation functions of such D-string
configurations and find that they exhibit the expected behavior in
the spin-model.

The paper is organized as follows. In the next section, we review
basic ideas in the past literature which indicate a general
duality between spin-systems and gravity. We first focus on the
case of $SU(N)$ in the $N\to\infty$ limit and postpone the general
discussion to section 6. Section 3 reviews the Einstein-dilaton
system that was studied in \cite{exotic1}. In section 4 we argue
that the IR limit of the model is described by a linear-dilaton
CFT and review basic features of such CFTs. Section 5 contains
main technical results of this paper. We first review the basic
statistical mechanics results that are relevant in what follows.
Then we propose the precise identification between the F-string
configurations and the spin correlation functions. We calculate
the one-point and two-point functions near criticality in the
semi-classical approximation making use of the linear-dilaton CFT.
Finally we present calculations related to vortex configurations.
In section 6 we take a first step in formulating a gravity
spin-model duality in general. In the last section we discuss
various issues and possible future directions of research.

Several appendices detail our presentation.  In appendix A, we
review the simplest example od the equivalence between lattice
gauge theories and the spin-systems. In appendix B, we review the
connection between non-critical string theory and the
linear-dilaton background. Appendix C provides some basic
background material in statistical mechanics of the XY models for
the unfamiliar reader. Finally, Appendices D and E contain details
of our calculations in section 5.

\section{Gravity - spin model duality}
\lab{sec1}

Our goal in this section is to propose a particular approach to
the gravity-CMT correspondence that relates the spin-models in CMT
to gravity by a two step procedure: The first step is to employ
a well-known equivalence between spin-models and lattice gauge
theories \cite{Polyakov, Susskind} followed by a second step that
is to utilize the gauge-gravity duality to relate the (continuum
limit) of the lattice gauge theory to a dual gravitational
background.

\subsection{Correspondence between gauge theories and spin systems}
%\lab{sec1}

Existence of the confinement-deconfinement phase transition in
lattice gauge theories at strong coupling is rigorously proved
\cite{Polyakov, Susskind}
%\footnote{Of course the continuum limit
%as $g^2\to -1/log(a)$ ($a$ is the lattice spacing) poses the main
%problem in extending the proof to the continuum case. We shall
%assume that the desired continuum limit always exists in what
%follows.} 
long time ago. The proof is based on an equivalence
between lattice gauge theories (LGT) and spin systems with
nearest-neighbor ferromagnetic interactions, \cite{Polyakov,
Susskind, Yaffe}. In the original papers of Polyakov and Susskind,
this equivalence was established for the cases of $U(1)$ and
$SU(2)$ gauge theories. Subsequently it was generalized to general
Lie groups\footnote{See \cite{Pfeiffer} for a recent presentation
of how the map works in a general case.}. We shall refer to this
equivalence as the {\it LGT-spin model equivalence}. We review 
how the spin systems arise from the lattice gauge theories in the Hamiltonian formalism,  
and in the simplest case of $U(1)$ gauge group in Appendix A. 

This equivalence has profound implications in the continuum limit:
As argued and verified with various examples by Svetitsky and
Yaffe \cite{Yaffe}, the critical phenomena---if exists---in the
continuum limit of the LGT, should be in the same universality
class with the corresponding spin model. Therefore, {\em a
continuous order-disorder type transition in a $d-1$ dimensional
spin-model with global symmetry group ${\cal C}$ is directly
related to a continuous type confinement-deconfinement transition
of the gauge theory with gauge group $G$ where ${\cal C} =
Center(G)$}.\footnote{Of course, not all of the spin-models
exhibit continuous transitions. See \cite{Yaffe} for a list of
examples.}

\begin{figure}[h!]\begin{center}
\includegraphics[width=9cm]{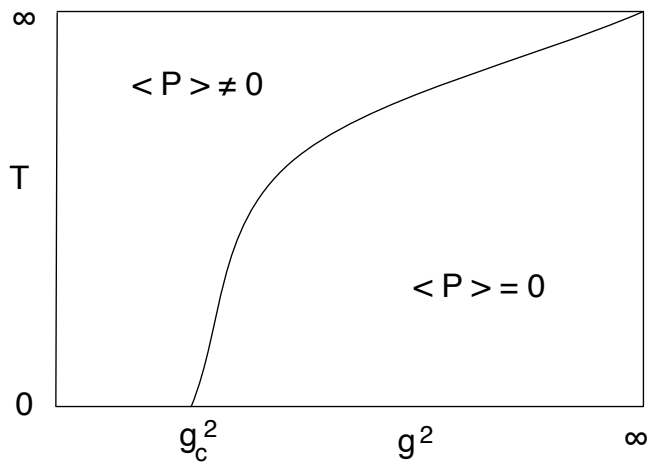}\end{center}
\caption{Typical phase diagram of a lattice gauge theory with non-trivial center. Low T phase is confining with vanishing expectation value for the Polyakov loop
$P$ and high T phase is de-confined. We assume that 
(at least a portion) of the phase boundary that separates these phases is of second or higher order. Then the critical phenomena around the phase boundary 
is determined by the corresponding spin-model.} \label{phase-diag}
\end{figure}

Let us briefly review the argument of \cite{Yaffe}. The basic observation is 
that the magnetic fluctuations are always gapped both in the high and the low T limit of the lattice gauge theory. 
Therefore, they are expected to be gapped for any T on a trajectory crossing the phase boundary in figure \ref{phase-diag}. 
This means that the magnetic fluctuations should not play an essential role at criticality in the vicinity of a continuous confinement-deconfinement 
transition. Integrating these short-range fluctuations, one indeed obtains an effective theory that only involves 
the Polyakov loops, which in turn can be mapped on a spin model. Therefore the critical phenomena, e.g. the critical exponents 
etc. of the lattice gauge theory around a continuous transition should be governed by the corresponding spin model. 

%%%%%%%%%%%%%%%%%%%%%%%%%%%
%%%%%%%%%%%%%%%%%%%%%%%%%%%

The magnetic sector is gapped at low T by assumption. We assume that the (bare) coupling constant is large enough (see figure \ref{phase-diag}) 
so that the low T theory is confined. The argument at high T is as follows. In the Lagrangian formulation of the LGT one can take the action to be, 
%%%%%%
\be\lab{latlang} 
{\cal A}_{lgt} = \sum_{\vec{r}} {\rm Re}\le\{ \b_t \sum_{i} \tr U_{\vec{r},0i} + \b_s \sum_{ij} \tr U_{\vec{r},ij}\ri\}; 
\qquad U_{\vec{r},\m\n} = U_{\vec{r},\m}U_{\vec{r}+\hat{\m},\n}U^\dg_{\vec{r}+\hat{\n},\m}U^\dg_{\vec{r},\n}
\ee
%%%%%%
where $\vec{r}$ labels sites on the square lattice, $U_{\vec{r},\m\n}$ are the product of link variables on a plaquette with corner $\vec{r}$. The first term 
in the action above corresponds to the electric contribution and the second to the magnetic. The electric and magnetic coupling constants are related to the bare coupling 
constant of the LGT and the temperature as follows:
%%%%%%
\be\lab{latlang2} 
\frac{2}{g^2} = a^{4-d}\sqrt{\b_t\b_s}, \qquad T = \sqrt{\frac{\b_t}{\b_s}} \frac{1}{N_t a}
\ee
%%%%%%
where $a$ is the lattice spacing and $N_t$ is the number of the lattice sites in the Euclidean time direction. 
One observes that, at fixed coupling, $\b_t \sim T$. Then, for sufficiently high T, only configurations with vanishing electric flux contributes in the 
partition function. This reduces the system to static configurations at high T, hence the theory can be thought of a $d-1$ dimensional 
LGT at zero T, with a coupling constant $g^2_{d-1} = g^2_d T$. Any such LGT with a non-trivial center is confined and 
exhibits magnetic screening at strong coupling \cite{Yaffe}. 
As mentioned before, the equivalence to the spin-models can be shown exactly at strong coupling, \cite{Polyakov, Susskind}.  

Svetitsky and Yaffe \cite{Yaffe} were able to make reliable
predictions concerning the critical phenomena of a wide range of
lattice gauge theories making use of this connection and the
well-known results on the critical phenomena of the corresponding
spin-models.

First of all, they correctly predicted that the 2nd order transition in the pure $SU(2)$ theory in 4D
is in the same  universality class with the 3D Ising model (see e.g. \cite{Ising} and references therein).   
As another check of these arguments \cite{Yaffe} presents the example of $SU(N)$ theory for $d-1=2$, $N>4$ 
where the dual spin model is again $Z_N$ symmetric and exhibit a BKT type continuous
transition. In this case, it was argued that for large N , the theory approximates that of a $U(1)$ LGT in 2+1 and 
the corresponding spin-model should be the XY-model in 2D.  It was explicitly checked in \cite{Yaffe} that, for the $U(1)$ LGT 
the critical phenomena is in the same universality class as that of the 2D XY model. 
More generally,  if the $d=2+1$ $SU(N)$ gauge theory---or a suitable deformation with additional adjoint matter---involves
continuous critical phenomena than it should be the BKT type. 

A particularly interesting case concerns $SU(N)$ gauge theory in
$d-1>2$ spatial dimensions with $N>4$ (that includes the large
$N$) where the dual spin model is $Z_N$ symmetric. We then
consider the large N limit that is most relevant for the
gauge-gravity duality. It is reasonable to believe that in the
strict $N\to\infty$ limit (with or without adjoint matter), {\em
the center $Z_N$ is promoted to $U(1)$}. See \cite{Witten} for an
argument in favor of this, in the case of ${\cal N}=4$ SYM at
strong-coupling\footnote{See however \cite{AharonyWitten} which
shows that the $U(1)$ symmetry is expected to arise only in the
strict $N\to\infty$ limit.}. Another indication that this happens
in $Z_N$ invariant LGT at $d=2$ is explained in \cite{Yaffe}.
Therefore, by the universality arguments above, if there exist a
continuous phase transition it should be governed by a $U(1)$
invariant spin model
%\footnote{It is also well-known that
%the large-N limit of $SU(N)$ gauge theory on a d-dimesional
%lattice can be reduced to theory of d unitary random matrix models
%\cite{EguchiKawai}\cite{Migdal}. It is interesting to see if this
%Eguchi-Kawai model can be mapped onto a $U(1)$ invariant spin
%model.}
\footnote{One can  ask whether there is any evidence, for or
against criticality at $N=\infty$. There are two independent
arguments that argue for a second order transition
\cite{Yaffe2}\cite{Pisarski} in the case of {\em pure} YM in 3+1
dimensions. On the other hand, there is the usual argument against
a continuous transition at large N that claims, since the number
of degrees of freedom in the system changes from $\cO(1)$ to
$\cO(N^2)$
 in a confinement-deconfinement transition, latent heat should be
 finite. In \cite{exotic1} we presented a counter-example to
this reasoning, albeit in a gravitational setting: although the
degrees of freedom change abruptly as the graviton gas deconfines
in the black-hole phase, the entropy difference may vanish at the
transition. See \cite{Minwalla2} for other examples of second
order transitions at large N. Finally, even if the transition is
first-order for pure YM, the situation may change when one adds
adjoint matter.}.

We review how the equivalence of LGTs and spin-systems work at strong coupling in 
Appendix \ref{sec2} for the unfamiliar reader. Here we shall mention two salient features.

\begin{itemize}

\item The temperature of the spin-system
is inversely related to temperature in the original gauge theory:
 \be\lab{inverse}
 T_s \sim T_l^{-1}.
 \ee
%With some more effort this computation can be generalized to the
%non-Abelian case \cite{Polyakov, Susskind}, and the general lesson
%remains the same: The temperatures of the LGT and the SM are
%inversely related.
Consequently, {\em the low temperature (confined) phase of the
gauge theory corresponds to the high temperature (disordered)
phase of the ferromagnet, whereas the high temperature
(de-confined) phase of the gauge theory corresponds to the low
temperature (ordered) phase of the ferromagnet.} \footnote{In
$d=2$, IR divergence of the spin waves prevent ordinary long range
order. Instead, a topological long-range order in terms of the
vortex-anti-vortex pairs arises \cite{BKT}\cite{Kosterlitz}. The
gauge theory partition function is capable of describing the
vortex configurations \cite{Yaffe}.}

%\item This example makes clear how the center
%of the original gauge theory arises as the global symmetry of the
%spin model\cite{Polyakov}: The Poisson resummation eats all the
%``particles'' $E(r,\hat{n})$ that transform under the gauge group
%but invariant under the center ${\cal G}$ and promotes the
%auxiliary fields $\a$ that impose the gauge invariance (and that
%do transform under ${\cal G}$) to particles of the new system.
%This feature immediately generalizes to non-Abeliean gauge groups where the symmetry
%group of the spin-model arises from the center symmetry of the gauge group.
%In the most general case, the duality transformation yields an
%effective theory of Polyakov loops in the continuum limit where
%the Polyakov loop become equivalent to the spin variables of the
%dual spin model.

\item Quite generally, the LGT-spin model equivalence can be
generalized to incorporate (adjoint) matter. This is mainly
because the basic ingredient in the calculation i.e. the center
symmetry of the LGT remains intact upon addition of adjoint
matter. See \cite{Minwalla2} for a related recent discussion.

\end{itemize}

%\section{Holographic superfluidity}

\subsection{Holographic superfluidity}
\lab{sec3}

Here and until section 6, we specify to the particular case of
$U(1)$ invariant spin-models. Continuous critical phenomena in
such models include the interesting case of {\em superfluidity},
that requires spontaneous breaking of the global $U(1)$ symmetry. As
reviewed above this transition is directly connected to the
confinement-deconfinement transition in the gauge theory. In the
original derivation of \cite{Polyakov}\cite{Susskind}, the
de-confined phase of the $U(1)$ invariant LGT was understood as an
ordered phase of the $U(1)$ spin model. This is clear from the
discussion of section \ref{sec1}, as the center of $U(1)$ is
$U(1)$ itself.

Instead, here we  shall adopt an alternative approach where the
$U(1)$ factor arises from the large N limit of an $SU(N)$ gauge
theory (pure or with adjoint matter). In this case the deconfinement
transition can be understood in the gravity dual as a {\em
Hawking-Page transition} by a generalization of the arguments in
\cite{Witten}. Assuming that the following assumptions hold,
\begin{itemize}
  %\item Center of $SU(N)$ can be thought of as $U(1)$ at large N,
  \item There exists a suitable $SU(N)$ lattice gauge theory with
  coupling to adjoint matter chosen such that, at large N  it
  flows to an IR fixed point with a continuous
  confinement-deconfinement transition,
  \item Gauge-gravity correspondence holds and maps this to a
  Hawking-Page type transition,
\end{itemize}
then one should be able to map the normal-to-superfluid transition
in the XY model to a {\em continuous} Hawking-Page type transition
on the gravity side.

The $U(1)$ symmetry of the spin-model follows on the GR side from the shift symmetry
$\psi\to \psi +const.$ where $\psi$ is the flux of the B-field 
%%%%%%%%%%%%%%%
\be\lab{psi} 
\psi = \int_M B = {\rm const.} 
\ee
%%%%%%%%%%%%%%%
on the subspace $M$ of the BH geometry that is spanned by the
coordinates r and $x_0$. 
In this paper we consider gravitational set-ups where the 
B-field is either constant or pure gauge $B = d \xi$ so that it does not back-react on the 
solution with $H=dB=0$.  Of course such a B-field has no visible effect on the gravitational 
solution and in the second case it can be removed by a gauge transformation. 
This ceases to be the case in presence of objects that are charged under this shift symmetry. 

In the classical approximation where one keeps
only the low-lying gravity fields, there are no bulk fields that
carry the extra $B$ charge. However,  strings that wind
around the time-circle couple to the B-field through the term $i \psi$, thus they are charged under the 
shift symmetry with the identification $\psi \sim \psi + 2\pi$. We shall denote this topological $U(1)$ symmetry 
as $U(1)_B$\footnote{This symmetry should be broken down to $Z_N$ for finite N by quantum
effects, see \cite{AharonyWitten}. However we only consider the $N\to\infty$ limit in this paper. } 
Therefore a non-vanishing string one-point function  signals a breakdown of the $U(1)_B$ symmetry. 
On the spin-model side this corresponds to an order-disorder transition upon identification of the 
$U(1)_B$ symmetry with the $U(1)$ spin symmetry of the spin-model. 
Below, we would like to review these ideas  in more detail.
%This is indeed what one expects from the arguments
%reviewed above, as the order parameter of the XY model i.e.
%magnetization $\la\vec{M}\ra$ corresponds to the Polyakov loop on
%the gauge theory side, that is dual to winding string solutions on
%the gravity side. In what follows we will treat the string
%solution in the probe approximation\footnote{This is entirely
%analogous to treating the complex bulk scalar $\Phi$ in
%\cite{Sean}---which plays the role of the order parameter in that
%case---in the probe approximation.} and review the arguments of
%\cite{Witten} in what follows.

\subsection{Spontaneous breaking of $U(1)_B$, the Goldstone mode
and the second speed of sound} \lab{sec4}

For simplicity, let us consider the (critical or non-critical)
bosonic string theory on a background with $U(1)\times E(d-1)$
isometry where the $U(1)$ corresponds to the temporal $S^1$, and
$E(d-1)$ to translations and rotations on the spatial part. The
general background with these symmetries is of the
form\footnote{In order to distinguish the dilaton and the scalar
field that appears in the Einstein-frame potential, which is
related to the dilaton by some rescaling, we denote the former
(dilaton itself) by $\of$ and the latter (rescaled dilaton) by
$\f$.} \be\lab{bg} ds^2 = A(r,\Omega) dx_0^2 + B(r,\Omega) dK^2 +
C(r,\Omega) dr^2 + D(r,\Omega) d\Omega; \quad \of =
\of(r,\Omega)\quad H=dB=0,\ee where $x_0\sim x_0+ 1/T$, $K$ is the
$d-1$ dimensional transverse part, and $\Omega$ is some internal
compact manifold. There can be additional bulk fields but we are
only interested in the NS-NS sector.

Most of the following traces the arguments in \cite{Witten}. The
order parameter for the transition is the vev of the Polyakov
loop, $\la P[C] \ra$, where $C$ is a loop isomorphic to the
time-circle. This maps to the expectation value of the F-string
path integral,
\begin{equation}\label{po1}
  \la P[C]\ra \propto \la {\cal W}_{F} \ra_{SG}
\end{equation}
where ${\cal W}_{F}$ denotes the F-string path integral over all
of the string configurations with the boundary ending on $C$, and
the final averaging is path integral over the super-gravity fields
that couple to the string. The string path integral is
\begin{equation}\label{OP}
  {\cal W}_{F} = \int {\cal D}X_\mu  {\cal D}h_{ab}~e^{-\int (G + iB + \of R^{(2)})},
\end{equation}
where $R^{(2)}$ is the Ricci scalar on the sub-manifold $M$ that
the F-string wraps and $X_\mu$ denotes the matter fields. We also
use the short-hand notation \be\lab{shorthand} G \equiv
\sqrt{det~h_{ab}}~h^{ab}\6_a X^{\m}\6_b X^{\n}  G_{\m\n},\qquad B
\equiv \sqrt{det~h_{ab}}~\epsilon^{ab}\6_a X^{\m}\6_b X^{\n}
B_{\m\n}. \ee One has to make sure that ${\cal W}$ is finite by an
appropriate regularization of infinite volume of the space
time\footnote{A cut-off in r that we call $\eps$ close to the
boundary would suffice for the sake of the discussion here. We
elaborate on this regularization in appendix \ref{AppP}.} and
factoring out diffeo-Weyl gauge volume a la Faddeev-Popov.

In the original discussion of \cite{Witten} ${\cal W}$ is
dominated by the classical saddles that minimize the action in
(\ref{OP}). The boundary condition for these classical strings is
such that at $\tau=0$, $X^{\m}(\sigma,\tau)$ ends on the temporal
circle $x_0$, some point x in $K$ and at the cut-off of the radial
coordinate $r=\eps$.

The string path integral is dominated by classical saddles when
$\ell/\ell_s\gg 1$ where $\ell$ is the typical curvature of the
target space and $\ell_s$ is the string length. In the original
AdS/CFT correspondence this ratio is proportional to the t 'Hooft
coupling  of the dual ${\cal N}=4$ SYM theory, $\ell/\ell_s
\propto \l^{\frac14}$ and indeed the classical strings dominate in
the limit of strong interactions. In the general case here one has
to consider the full path integral.

The vev of $P[C]$ is given by the path integral of ${\cal W_F}$ over
the super-gravity fields that couple the F-string, weighted by the SG action.
The non-trivial SG fields are the space-time metric $G_{\m\n}$,
the B-field $B_{\m\n}$ and the dilaton $\of$. Thus one has,
\begin{equation}\label{po2}
  \la P[C] \ra \propto \int {\cal D}G_{\m\n}~{\cal D}B_{\m\n}~{\cal D}
  \of~e^{-{\cal A}_{sg}} {\cal W}_{F},
\end{equation}
where ${\cal A}_{sg}$ is the gravity action. As we are interested
in the large N limit of the dual field theory, we can send the
string coupling $g_s\to 0$ and the SG path integral is dominated
by the classical saddles of ${\cal A}_{sg}$. For given asymptotic
boundary conditions of $G$, $\of$ and $B$, the saddles of interest
involve only two type of solutions, the thermal gas (TG) and the
black-hole (BH). At an arbitrary temperature T (that partially
determines the asymptotic boundary condition for $G$), only one of
these saddles will dominate the SG path integral as a result of
the classical limit $g_s\to 0$.
%\footnote{Exactly at $T_c$ the
%on-shell values of the SG action on TG and BH becomes exactly the
%same, thus one has to sum over both of the solutions.}

%We want to compute $P[C]$ on the two type of SG solutions of
%interest: the thermal gas (TG) and the black-hole (BH).
Let us assume that TG dominates at $T<T_c$ and BH dominates at
$T>T_c$. Let us also assume that the BH solution only exists above
a certain temperature $T_{min}$. Backgrounds that exhibit
confinement generically satisfy $T_c\ge T_{min}$ \cite{GKMN1,GKMN2}.
%This means that the as one heats up the background, first the BH
%solution forms at $T_{min}$, as a saddle sub-dominant to the TG
%solution, and than at $T_c$, the BH begins dominating the free
%energy giving rise to a first order hawking-Page transition.
As explained in detail in \cite{exotic1} and reviewed in the next
section, only in the case $T_{min} = T_c$ the transition is second
or higher order.

On the TG phase, the classical world-sheet $M$ has infinite area.
Therefore the string path-integral ${\cal W_F}$, hence $\la P[C]
\ra$ in (\ref{po2}) vanishes. One concludes that the TG solution
is $U(1)_B$ symmetric and the center in the dual gauge theory is
unbroken. This means that the dual spin-model is in the {\em
normal (disordered) phase}. This is precisely as one expects from
the behavior of the dual spin model in the high temperature phase,
recalling that the temperature of the spin model is inversely
proportional to the temperature on the gravity side $T_s \propto
T^{-1}$.

On the BH solution $T>T_{min}$ however, the classical string
saddle $M$ has finite area and one has to evaluate (\ref{po2})
carefully. One has to include all of the configurations over the
classical fields $G,B$ and $\of$ with the same on-shell value of
the SG action.

The path integral over $G$ and $\of$ in (\ref{po2}) is replaced by
the classical solution (\ref{bg}) that is a BH in this case. Sum
over these saddles  include the following important contribution
from the B-field. In the black-hole case the sub-manifold M has
finite area\footnote{The divergence near boundary is regularized
in the familiar way, cf. appendix \ref{AppP}.} and the B-field has
a flux $\psi =\int_{M} B$. $\psi$ in (\ref{OP}) has angular nature
because it appears with a factor of i and it can attain any value
in the range $\psi \sim \psi + 2\pi$. This identification yields
the $U(1)_B$ invariance\footnote{In the critical IIB theory this
identification arises as a result of discrete gauge
transformations that shift the value of $\psi$ by a multiple of
$2\pi$ \cite{Witten}.}. The sum over classical saddles then should
include various different values of $\psi$. As $dB=0$ all
different values of $\psi$ yield the same on-shell gravity action.
%This is the only non-trivial path integral over
%the SG fields that we have to consider.

%On the BH solution $T>T_{min}$ however, the saddle $M$ has finite
%area and one has to evaluate {\cal W} carefully. When $K$ is
%non-compact and $dim(K)>2$, then $\psi(K)$ viewed as a massless
%bosonic field on $K$ has long-range order, hence it should
%condense\footnote{The situation at $dim(K)=2$ exactly parallels
%the analogous situation in the 2D dual field theory, where IR
%divergence kill long-range order.} Thus, on the black-hole
%solution the $U(1)_B$ symmetry breaks down\footnote{As a technical
%aside, in the computation above, one should check that the dilaton
%term in the action does not spoil the arguments. In the particular
%case of the geometries considered in this paper, $\of$ diverges in
%the deep interior, hence this check especially becomes important.
%We check in App. E that this term indeed remains finite in our
%case.}. This happens exactly at the point where the black-hole
%forms, right above $T_{min}$. {\em As a result, the fluctuation
%$\delta\psi$ in (\ref{OP2}) becomes a Goldstone mode on the
%transverse space $K$.}

%Fluctuations of $K$ however are important.
  We can thus write,
\begin{equation}\label{OP2}
  P[C] \propto \int {\cal D}\psi e^{-S_{sg}[\psi]} \int
   {\cal D}X_\mu~{\cal D}h_{ab}~
  e^{i\psi}~e^{-\int_{M} (G  + \of R^{(2)})},
\end{equation}
where $S_{sg}$ now is evaluated on the saddle solution and is
only a functional of $\psi$. On the other hand the $\psi$ path
integral includes the classical saddle $\psi = const$ and the
fluctuations
 $\delta \psi(K)$ around it.

When $K$ is non-compact and $dim(K)>2$, then the fluctuations
$\delta \psi(K)$ viewed as a massless bosonic field on $K$ has
long-range order, hence $\psi$ should condense\footnote{The
situation at $dim(K)=2$ exactly parallels the analogous situation
in the 2D dual field theory, where IR divergences kill long-range
order.}. Thus, on the black-hole solution the $U(1)_B$ symmetry
breaks down\footnote{As a technical aside, in the computation
above, one should check that the dilaton term in the action does
not spoil the arguments. In the particular case of the geometries
considered in this paper, $\of$ diverges in the deep interior,
hence this check especially becomes important. We check in
appendix \ref{AppP} that this term indeed remains finite in our
case.}. This happens exactly at the point where the black-hole
forms, right above $T_{min}$. {\em As a result, the fluctuation
$\delta\psi$ in (\ref{OP2}) becomes a Goldstone mode on the
transverse space $K$.}
%
%
%
%One has,
%
%One also has to integrate over the fluctuations of $X$ in $K$ and
%$\Omega$ around these saddles. Fluctuations on $\Omega$, being
%compact, can be factored out and do not give interesting physics.
%where $\delta\psi$ denote fluctuations of $\psi$ in the transverse
%space $K$.
%\footnote{Faddeev-Popov determinant can be omitted in the
%following}
%
%{\bf\Large rewrite this and refer to the gravity calculation for
%the vanishing of $c_\psi$. }

Considering the wave
equation for $\delta\psi$
%\footnote{There is possible mixing with
%the scalar $\psi$, fluctuations the dilaton $\of$ and the isotropic
%fluctuations of the metric. We do not attempt to derive the full
%set of coupled equations here. It is a subject of future work.
%However the Goldstone
%mode is not spoiled by this mixing.}
one expects to find,
\begin{equation}\label{cpsi}
  \o^2 = c_{\psi}^2(T)~{\bf q}^2 + \cO({\bf q}^4),
\end{equation}
where $c_{\psi}$ is the speed of sound of $\psi$ and there is no
mass term for $\psi$ for $T>T_{c}$. It is well-known that (see
appendix \ref{statmech} for a review, and section \ref{grss} for a
holographic derivation in gravity), the speed of sound $c_{\psi}$
of the Goldstone mode vanishes continuously as one approaches the
transition temperature $T_c$ from above, {\em only if the
transition is of  continuous type}. This is exactly what happens
in super-fluidity, where the ``{\em second speed of sound}", i.e.
the speed of sound associated with the entropy waves vanish as one
approaches $T_c$ of the XY model from below (recall that
temperature in gravity and in the XY model are inversely related).
In order to mimic this property of the spin model, we should
require that the Hawking-Page transition in gravity is of
continuous type, hence $T_c = T_{min}$ \cite{exotic1}. In section
\ref{grss} we show by an explicit gravity calculation that indeed
the second sound vanishes with the expected mean-field exponents.

{\bf Our conclusion is:} whenever a second order (or higher order)
Hawking phase transition occurs in the gravitational background,
it is natural to associate it with super-fluidity. Here the
thermal gas phase is dual to the normal phase of the system, and
the black-hole phase is dual to the super-fluid. The ``first speed
of sound" i.e. the sound of the density waves is associated with
the graviton fluctuations (that we are considered in
\cite{exotic1}), and the ``second speed of sound" is associated
with the fluctuations of the B-field that we consider in section
\ref{grss}. \vspace{.3cm}
\newline
\begin{center}
\begin{picture}(5,20)
%\lab{table2}
\put(-10,-10){\vector(0,1){40}} \put(-15,-22){{\small
$T$}} \put(-5,-7){$T_c$}
\end{picture}
\begin{tabular}{|c|c|c|c|}
\hline
  {\bf Lattice gauge theory} & {\bf Gravity} & \hspace{.01cm} & {\bf Spin model} \\
  \hline
  Deconfined, $U\!\!(1)_{\cal C}\!\!\!\!\!\!\!\!\!\!\!\!\!/$ & Black-hole, $U\!\!(1)_B\!\!\!\!\!\!\!\!\!\!\!\!\!/$ & \hspace{.01cm} & Superfluid ,
  $U\!\!(1)_S\!\!\!\!\!\!\!\!\!\!\!\!\!/$ \\
  \hline
  Confined, $U\!\!(1)_{\cal C}$ & Thermal gas, $U\!\!(1)_B$ & \hspace{.01cm} & Normal phase, $U\!\!(1)_S$ \\
  \hline
\end{tabular}
\begin{picture}(5,20)
\put(18,13){\vector(0,-1){40}} \put(14,18){$T$}
\put(-3,-7){{\small $T_c^{-1}$}}
\end{picture}
\end{center}
\vspace{.3cm}

One can summarize the various phases of the theories by the table above.
The various $U(1)$ factors in this table are as follows: The
$U(1)_B$ is the dual symmetry that arises from compactifying the B
field on the temporal circle. The $U(1)_{\cal C}$ is the center symmetry
of the corresponding lattice gauge theory that is proposed to
arise in the large N limit of $SU(N)$ (with or without) adjoint
matter. Finally the $U(1)_S$ is the spin symmetry of the
corresponding XY model. The arrow of increasing T is the same
for the LGT and gravity picture and opposite in the spin model
picture.

\section{A model based on Einstein-scalar
gravity}
\lab{model}

The arguments put forward in favor of a gravity-spin model
correspondence above are general. In this section we would like to
introduce a simple set-up which allows for computations of
quantities such as the scaling of magnetization and spin-spin
correlation function on the gravity side. The model is inspired by
non-critical string theory and it becomes precisely non-critical
string theory in the interesting regime near the continuous phase
transition.

\subsection{The model}

The action in the Einstein frame reads,
\begin{equation}\label{action}
 {\cal A} = \frac{1}{16\pi G_N} \int d^{d+1}x \sqrt{-g}\le( R - \frac{4}{d-1} (\6\f)^2 + V(\f)
   -\frac{1}{12}e^{-\frac{8}{d-1}\f} H^2 + \cdots\ri)+\,\,\, G.H.
\end{equation}
where the kinetic terms of the dilaton\footnote{The scalar field
$\f$ here is related to the original dilaton of the non-critical
string $\of$ by some rescaling that is defined in section
\ref{largeN}. By $\f$ we will always mean the ``rescaled dilaton"
throughout the paper.}
 and the B-field $H = d B$
are inspired by non-critical string theory in $d+1$ dimensions.
The ellipsis denote higher derivative corrections. The last term
in (\ref{action}), that we shall not need to specify here, is the
Gibbons-Hawking term on the boundary.

We allow for a non-trivial dilaton potential $V(\f)$ that should
be specified by matching the thermodynamics of the dual field
theory. In the case of non-critical string theory in $d+1$
dimensions the potential is given by,
\begin{equation}\label{vncst}
  V_{nc}(\f) = \frac{\delta c}{\ell_s^2} e^{\frac{4}{d-1}\f},
\end{equation}
where $\ell_s$ is the string length and $\delta c $ is the central
deficit, see section \ref{largeN} for more detail. $G_N$ in
(\ref{action}) is the Newton's constant in $D=d+1$ dimensions. It
is related to N of the dual field theory\footnote{As explained
above, $N$ may either be the number of colors in $SU(N)$ gauge
theory or the number of spin states at each site in a $Z_N$
spin-model.} by,
\begin{equation}\label{Mp}
\frac{1}{16\pi G_N} = M_p^{d-1} N^2,
\end{equation}
where $M_p$ is a ``normalized" Planck scale, that is generally of
the same order as the typical curvature of the background $\ell$.
The limit of large $N$ corresponds to classical gravity as usual.
One should be careful in attaining this classical limit: The
correct way of achieving this is described in section
\ref{largeN}. On the gravity side the parameter N arises from the
RR-sector, where it is the integration constant of a space-filling
$F_{(d+1)}$ form, $F_{(d+1)}\propto N$. Then the large N limit is
defined as sending this value to infinity and sending the boundary
value of the dilaton $\f_0$ to $-\infty$ such that $N \exp(\f_0)$
remains constant and yields $M_p$ in (\ref{Mp}). We refer to
section \ref{largeN} for details.

%Before going further, we would like to digress and comment on the
%issue of higher derivative corrections in (\ref{action}). Ignoring
%these corrections in holographic applications can be justified in
%two cases: 1) The gravity sector is decoupled from the string
%sector by presence of an adjustable parameter such as the 't Hooft
%coupling of ${\cal N}=4$ sYM. 2) There is no such a parameter but
%the size of higher derivative corrections that are measured by
%$\ell_s/\ell$ happens to be small. In a phenomenological dual of a
%field theory with no tunable parameters one can only hope for the
%second possibility. For example in holographic models of pure
%Yang-Mills theories\cite{GKN} the second possibility is realized
%where one finds $\ell_s/\ell \sim \cO(10^{-1})$. In the case of
%spin-models, both options may be possible. There is no a priori
%reason that the entire tower of higher spin states in string
%theory is necessary to describe a dual spin-model. It may also be
%that, depending on the particular spin-model the parameter
%$\ell_s/\ell$ may indeed be small. Even if this parameter is
%$\cO(1)$, one can still hope the arrive at a qualitatively (but
%not quantitatively)  viable description in gravity. We shall
%discuss this issue further in section \ref{discuss}.

In what follows we shall only consider solutions with either constant or pure-gauge
$B$-field whose legs are taken to lie along r and $x_0$
directions: 
%%%%%
\be\lab{constB} B_{\m\n} = B_{r0}, \ee
%%%%%
In this case $H=0$ in (\ref{action}) and the B-field contributes to
neither the equations of motion nor the on-shell value of the
action. However, it contributes the F-string and D-string
solutions as we study in section 5.

There are only  two types of backgrounds at finite T (with
Euclidean time compactified), with  Poincar\'e symmetries in $d-1$
spatial dimensions, and an additional $U(1)$ symmetry in the
Euclidean time direction. These are the {\it thermal graviton
gas},
\begin{equation}\label{TG}
  ds^2 = e^{2A_0(r)}\le( dr^2 + dx_{d-1}^2 + dx_0^2\ri), \qquad \f= \f_0(r),
\end{equation}
and the {\it black-hole},
\begin{equation}\label{BH}
  ds^2 = e^{2A(r)}\le( f^{-1}(r)dr^2 + dx_{d-1}^2 + dx_0^2 f(r) \ri),
  \qquad \f= \f(r).
\end{equation}
We define the coordinate system such that the boundary is located
at $r=0$. For the potentials $V$ that we consider in this paper,
there is a curvature singularity in the deep interior, at $r=r_s$.
In (\ref{TG}), r runs up to singularity $r_s$. In (\ref{BH}) there
is a horizon that cloaks this singularity at $r_h<r_s$ where
$f(r_h)=0$. $x_0$ is the Euclidean time that is identified as
$x_0\sim x_0+1/T$. This defines the temperature T of the
associated thermodynamics. In the black-hole solution, the
relation between the temperature and $r_h$ is obtained in the
standard way, by demanding absence of a conical singularity at the
horizon:
\begin{equation}\label{Trh}
4\pi T  = -f'(r_h).
\end{equation}
This identifies T and the surface gravity in the BH solution.

In the r-frame defined by (\ref{TG}) and (\ref{BH}) one derives
the following Einstein and scalar equations of motion from
(\ref{action}):
%%%%%%%%%%%%%%%%%%%%%%%%%%%%%%%%%%%%%%%%%
\bea \label{E1}
  A'' - A'^2 + \frac{\xi}{d-1} \f^{'2} &=& 0, \\
\lab{E3}
f'' + (d-1)A'f' &=& 0, \\
\label{E2}
   (d-1)A'^2 + A'f'+A''f -\frac{V}{d-1} e^{2A} &=& 0 .
\eea
%%%%%%%%%%%%%%%%%%%%%%%%%%%%%%%%%%%%%%%%%
One easily solves (\ref{E3}) to obtain the ``blackness function"
$f(r)$ in terms of the scale factor as,
\begin{equation}\label{f}
  f(r) = 1- \frac{\int_0^r e^{-(d-1)A}}{\int_0^{r_h} e^{-(d-1)A}}.
\end{equation}
Then the temperature of the BH is given by eq. (\ref{Trh}):
\begin{equation}\label{T}
  T^{-1} = 4\pi e^{(d-1)A(r_h)}\int_0^{r_h} e^{-(d-1)A(r)}dr.
\end{equation}
The difference between the entropy densities of the BH and the TG
solutions is given by the BH entropy density up to $1/N^2$
corrections\footnote{We choose to normalize the thermodynamic
quantities by an extra factor of $1/N^2$ so that the entropy on
the BH becomes $\cO(1)$ and on the TG it becomes $\cO(1/N^2)$.}
that we ignore from now on\cite{exotic1}:
\begin{equation}\label{S}
\Delta S = \frac{1}{4G_N N^2}e^{(d-1)A(r_h)}.
\end{equation}
 The difference in the free energy densities
 can be evaluated by integrating  the first law of thermodynamics,
 \cite{GKMN2}:
\begin{equation}\label{F}
  \Delta F(r_h) = -\frac{1}{4G_N N^2} \int_{r_c}^{r_h}
  e^{(d-1)A(\tilde{r}_h)}\frac{dT}{d\tilde{r}_h} d\tilde{r}_h,
\end{equation}
where $r_c$ is the value of the horizon size that corresponds to
the phase transition temperature $T(r_c) = T_c$, at which the
difference in free energies should vanish.

\subsection{Scaling of the free energy}

In \cite{exotic1} we showed that there exists a continuous type
Hawking-Page transition between the TG and the BH solutions when
the black-hole horizon marginally traps a curvature singularity:
$r_h = r_c \to \infty$. This happens only when the IR asymptotics
of the dilaton potential is chosen such that,
\begin{equation}\label{Vs}
  V(\f) \to V_\infty~e^{\frac{4}{d-1}\f}\le(1 + V_{sub}(\f)\ri), \qquad \f\to\infty
\end{equation}
where $V_\infty$ is a constant and $V_{sub}$ denote subleading
corrections that vanish as $\f \to \infty$. It is also shown in
\cite{exotic1} that the transition temperature $T_c$ that follows
from (\ref{T}) with $r_h\to\infty$ stays finite.

Given the asymptotics in (\ref{Vs}) one solves the equations of
motion (\ref{E1}) and (\ref{E1}) to obtain the IR behavior, as
$r\to \infty$,
%%%%%%%%%%%%%%%%%%%%%
\bea\lab{fs} \f(r) &\to& \frac{\sqrt{V_\infty}}{2} r + \cdots \\
\lab{As} A(r) &\to& -\frac{\sqrt{V_\infty}}{d-1} r + \cdots
\eea
%%%%%%%%%%%%%%%%%%%%%
where the subleading terms vanish in the limit.

Depending on $V_{sub}$ there are various different possibilities
for types of transitions. We consider only two classes of
potentials with:
%%%%%%%%%%%%%%%%%%%%%%%%%%%%%%%%%%%%%%%%%%%%%%%%%%%%%%%%%%%%%%%%%
\bea
  \mathrm{Case\,\, i:}\qquad V_{sub} &=& C~e^{-\kappa \f}, \quad \kappa>0, \qquad
  \f\to\infty\lab{case1}\\
  \mathrm{Case\,\, ii:}\qquad V_{sub} &=& C~\f^{-\a}, \quad \a>0, \qquad
  \f\to\infty\lab{case2}
\eea
%%%%%%%%%%%%%%%%%%%%%%%%%%%%%%%%%%%%%%%%%%%%%%%%%%%%%%%%%%%%%%%%%
Defining the normalized temperature,
\begin{equation}\label{t}
  t=\frac{T-T_c}{T_c},
\end{equation}
the scaling of thermodynamic functions with $t$ can be found from
the following set of formulae: The reduced temperature directly
follows from the subleading term in the potential,
\begin{equation}\label{ast}
  t = V_{sub}(\f_h),
\end{equation}
where $\f_h$ is the value of the dilaton at the horizon. Then the
free-energy as a function of $t$ follows from by (\ref{F}) as,
\begin{equation}\label{asF}
  \Delta F(t) \propto \int_0^t d\tilde{t}~e^{(d-1)A(\tilde{t})}.
\end{equation}
Here, the dependence of the scale factor on $t$ should be found by
inverting (\ref{ast}), and comparing the (leading term)
asymptotics of the scale factor $A(r)$ with the dilaton $\f(r)$
\cite{exotic1}. In the cases (\ref{case1}) and (\ref{case2}) one
finds that,
%%%%%%%%%%%%%%%%%%%%%%%%%%%%%%%%%%%%%%%%%%%%%%%%%%%%%%%%%%%%%%%%%
\bea
  \mathrm{Case\,\, i:}\qquad A(t) &=& \frac{2}{\kappa (d-1)}\log(t/C) + \cdots, \qquad
  t\to 0^+ \lab{Acase1}\\
  \mathrm{Case\,\, ii:}\qquad A(t) &=& -\frac{2}{\kappa (d-1)} \le(t/C\ri)^{-\frac{1}{\a}} + \cdots, \qquad
  t\to 0^+ \lab{Acase2}.
\eea
%%%%%%%%%%%%%%%%%%%%%%%%%%%%%%%%%%%%%%%%%%%%%%%%%%%%%%%%%%%%%%%%%
The free energy then follows from (\ref{asF}) as :
%%%%%%%%%%%%%%%%%%%%%%%%%%%%%%%%%%%%%%%%%%%%%%%%%%%%%%%%%%%%%%%%%
\bea
  \mathrm{Case\,\, i:}\qquad \Delta F(t) &\propto & t^{\frac{2}{\kappa}+1}, \qquad
  t\to 0^+ \lab{Fcase1}\\
  \mathrm{Case\,\, ii:}\qquad \Delta F(t) &\propto & e^{C' t^{-\frac{1}{\a}}} t^{1+\frac{1}{\a}}, \qquad
  t\to 0^+ \lab{Fcase2},
\eea
%%%%%%%%%%%%%%%%%%%%%%%%%%%%%%%%%%%%%%%%%%%%%%%%%%%%%%%%%%%%%%%%%
where $C'= 2C^{\frac{1}{\a}}$ in the second equation. We see that
$F$ vanishes, as it should, for arbitrary but positive constants
$\xi$, $\kappa$ and $\a$. Other thermodynamic quantities such as
the entropy, specific heat, speed of sound etc, all follow from
the free energy above \cite{exotic1}.

In the special case of
%%%%%%%%%%%%%%%%%%%%%
\be\lab{kappaNC} \kappa =\frac{2}{n-1}, \ee
%%%%%%%%%%%%%%%%%%%%%
in (\ref{Fcase1}) one finds an nth order phase transition. On the
other hand, the special case of $\a=2$ in (\ref{case2})
corresponds to the BKT type scaling\footnote{Very recently
holographic realizations of (quantum) BKT scalings were obtained
in \cite{SonBKT} and \cite{Liu2}.}.

One can also obtain the value of the transition temperature $T_c$
in terms of the coefficient of the dilaton potential in the IR as
\cite{exotic1}:
%%%%%%%%%%%%%%%%%%%%
\be\lab{Tc} T_c = \frac{\sqrt{V_\infty}}{4\pi}.\ee
%%%%%%%%%%%%%%%%%%%%

Finally, we should note the following issue. As mentioned above,
the transition region $t\approx 0$ generically coincides with the
singular region $\f_h \gg 1$ in this setting. We do not need to
worry about the $\a'$ corrections because they vanish in the
interesting region $r\gg 1$ in the interesting limit $r_h\gg 1$
\cite{exotic1}. However, one should worry about the string loops.
In a generic situation the higher string loops cannot be ignored
near the transition region. We are however interested in the
situation with $g_s\to 0$ ($N\to\infty$) that corresponds to the
$U(1)$ invariant spin-model. This can be achieved by sending the
boundary value of the dilaton to $-\infty$. We will now dwell on
this point in more detail.
% In the phenomenological approach, there is a crude way to avoid
%this problem: one can fine-tune the constant $C$ in the subleading
%potential (\ref{case1}) and (\ref{case2}) to very small values
%such that the transition region corresponds to milder values of
%$\f_h$, see equation (\ref{ast}). We refer the reader to
%\cite{exotic1} for a detailed discussion on this point.

\subsection{The large N limit and string perturbation theory}
\lab{largeN}

The effective Einstein frame action in (\ref{action}) is supposed to arise from a (fermionic)
non-critical string theory which also involves an RR-sector. The string frame action is,
\begin{equation}\label{saction}
 {\cal A}_s =\frac{1}{g_s^2 \ell_s^{d-1}} \int d^{d+1}x \sqrt{-g_s} e^{-2\of} \le( R_s + 4
  (\6\of)^2 + \frac{\delta c}{\ell_s^2} - \frac{1}{12} H_{(3)}^2\ri) - \frac{1}{2(d+1)!} F^2_{(d+1)} + \cdots
\end{equation}
The ellipsis denote higher derivative ($\a'$) corrections,
subscript s denote string-frame objects and $\delta c$ is the
central deficit that---depending on the fermionic or the bosonic
string theory---reads\footnote{The constants $c_f$, $c_b$ depend
on the particular CFT on the world-sheet as there are various
possibilities for the boundary conditions and GSO projections on
the world-sheet fermions possible twisted or shifted boundary
conditions for the scalar matter $X^{\m}$ \cite{Chemseddine}. In
the case of bosonic world-sheet with periodic scalars, one has
$c_b=2/3$ which is indeed what one obtains from solving (\ref{E2})
with the asymptotics (\ref{As}) and (\ref{fs}). See the next
section for details of the IR CFT.}, \be\lab{cdef} \delta c=
c_f~(9-d), \qquad fermionic; \quad \delta c = c_b~(25-d), \qquad
bosonic. \ee $F_{(d+1)}$ is a space filling RR-form whose presence
is motivated by holography: it should couple to the $D_{d-1}$
branes that are responsible for producing the $SU(N)$ gauge group.
As it is space-filling, its effect in the theory can be obtained
by replacing it in the action by its on-shell solution \cite{GK}.
This solution in general will be very complicated as the higher
derivative corrections will also depend on $F_{(d+1)}$. Let us
ignore these higher derivative solutions for the moment in order
to be definite---the following discussion will not qualitatively
depend on the higher derivative corrections.

The equation of motion for $F_{(d+1)}$ is $d*F_{(d+1)} = 0.$
The solution is
\be\lab{solF}
F_{(d+1)}  = \frac{c_F N}{\ell_s^2} \frac{\epsilon_{(d+1)}}{\sqrt{-g_s}},
\ee
where $\epsilon_{(d+1)}$ is the Levi-Civita symbol in $d+1$ dimensions and $c_F$ is some $\cO(1)$ constant.
We chose the integration constant to be proportional to N motivated by the fact that F should couple to N $D_{d-1}$ branes
before the decoupling limit.
 Inserting the solution in the action gives (we ignore the NS-NS two-form in the following discussion),
\begin{equation}\label{saction1}
 {\cal A}_s =\frac{1}{g_s^2 \ell_s^{d-1}} \int d^{d+1}x \sqrt{-g_s} e^{-2\of} \le( R_s + 4
  (\6\of)^2 + \frac{\delta c}{\ell_s^2}\ri) + \frac{c_F^2}{2\ell_s^2} N^2  + \cdots
\end{equation}
Now we define a shifted dilaton field
\be\lab{shdil}
\f = \of + \log N,
\ee
and go to the Einstein frame by
\be\lab{weylE}
g_{s,\m\n}= e^{\frac{4}{d-1}\f} g_{\m\n}.
\ee
We obtain,
\begin{equation}\label{action1}
 {\cal A} = \frac{N^2}{g_s^2\ell_s^{d-1}} \int d^{d+1}x \sqrt{-g}\le( R - \frac{4}{d-1} (\6\f)^2 + V(\f) \ri) +\cdots
\end{equation}
where the dilaton potential becomes,
\be\lab{dilpot1}
V(\f) =  \frac{1}{\ell_s^2} \le( \delta c~e^{\frac{4}{d-1}\f} + \frac{c_F^2}{2}  e^{\frac{2(d+1)}{d-1} \f} + \cdots \ri)
\ee
We denote the corrections coming from the higher-derivative  terms by the ellipsis.
This is what one would obtain by ignoring the higher derivative terms\footnote{In the phenomenological approach
that we adopted in the previous section, one {\em assumes} that there exist a string theory that would
produce a potential of the form (\ref{Vs}) instead of (\ref{dilpot1}). In particular the leading term with exponent
$2(d+1)/(d-1)$ should either be absent or renormalized to $4/(d-1)$.}.

On the other hand the solution of the dilaton equation of motion follows from (\ref{saction})
generically involves an integration constant that we shall denote as $\of_0$. For example in the kink solutions
of \cite{exotic1} this corresponds to the boundary value of the dilaton on the AdS boundary.
One can write
\be\lab{defdf}
\of = \of_ 0 + \delta \of(r)
\ee
to make explicit the integration constant. Now, we are ready to define the large-N limit. We send $N\to \infty$,
$\of_0\to -\infty$ such that the shifted dilaton in (\ref{shdil}) remains constant
\be\lab{limN}
e^{\of_0} N \to \l  \Longrightarrow e^{\f} = \lambda e^{\delta \of}
\ee
where $\l$ is some $\cO(1)$ constant.

The shifted dilaton $\f$ is the one that we used in the previous section to discuss thermodynamics and it is what we
will refer in the next sections to study the observables of the spin-system from the gravity point of view.
Whether $\f$  is large  or small does not matter neither for the loop-counting of strings nor  for the strength of gravitational
interactions: The latter is determined by the coefficient in the action (\ref{action1}). Identification with
(\ref{action}) yields the Newton's constant
\be\lab{gravst}
G_N = \frac{g_s^2\ell_s^{d-1}}{16\pi N^2},
\ee
which shows that the gravitational interactions among the bulk fields can be safely ignored in the large N limit. This equation
also defines the ``rescaled"  Planck energy that was introduced in (\ref{Mp}) in terms of $g_s$ and $\ell_s$ as,
\be\lab{Mp1}
M_p = \ell_s^{-1} g_s^{-\frac{2}{d-1}}.
\ee
The string loops on the other hand are counted by the coupling of the original dilaton $\of$ to a world-sheet M with genus
g as,
\be\lab{cpws}
e^{-\frac{1}{4\pi} \int_M \sqrt{h} R^{(2)} \of} = e^{-\of_0\chi(M)} e^{-\frac{1}{4\pi} \int_M \sqrt{h} R^{(2)} \delta \of}
= N^{\chi(M)} e^{-\frac{1}{4\pi} \int_M \sqrt{h} R^{(2)} \f},
\ee
where $\chi(M) = 2(1-g)$ is the Euler characteristic. We observe that the above definition of the large N limit
does the job and suppresses the strings with higher genus.

One might still worry about the viability of the string
perturbation expansion if the additional term proportional to
$\int_M \sqrt{h} R^{(2)} \f$ in (\ref{cpws}) becomes very large in
some limit. Indeed, as we argued above the interesting physics
concerns the region $\f\gg 1$ which corresponds to the vicinity of
the phase transition. We check in appendix \ref{AppP} and
\ref{AppPP} that for all of the string paths that we consider in
this paper the world-sheet Ricci scalar suppresses the linear
divergence in $\f$. For example in case of (\ref{case1}) one finds
that in the transition region $\sqrt{h} R^{(2)}\sim \exp(-\kappa
\f)$.

All of the discussion we presented above can be understood in the following equivalent way. To be definite
let us consider the simplest  effective (rescaled) dilaton potential that corresponds to case \ref{case1}:
\be\lab{efdil}
  V(\f) = V_\infty~e^{\frac{4}{d-1}\f}\le(1 + Ce^{-\kappa \f} \ri).
\end{equation}
It was shown in \cite{exotic1} that this potential has a kink solution that flows from the AdS extremum at
\be\lab{extr}
 e^{\f_0} = C^{\frac{1}{\kappa}} a
\ee
---where $a$ is some number independent of $C$---to the linear dilaton geometry in the IR $\f\to\infty$.
Then the subleading term in the potential can be written as,
\be\lab{subterm} V_{sub}(\f) = a^{-\kappa} e^{\kappa \f_0 - \kappa
\f} = a^{-\kappa} e^{\kappa \of_0} e^{-\kappa \of} \ee where we
used (\ref{shdil}). The statement that ``the transition region
corresponds to large dilaton" now can be quantified. What we
really mean by this is that the reduced temperature $t$  (\ref{t})
is small enough, so that the scaling behavior of observables set
in. Now, from (\ref{ast}) we see that this is given as,
\be\lab{subterm2} t = a^{-\kappa} e^{\kappa \of_0} e^{-\kappa
\of}. \ee On the other hand the large N limit (\ref{limN})
involves $\of_0\to -\infty$, therefore we see that in order for
$t$ to be small, one needs not the actual dilaton $\of$ but the
difference $\delta \f = \of - \of_0$ to be large. The same
reasoning can be generalized to general potentials that involve an
AdS extremum.

To conclude, we can safely ignore higher string loops in the computations below.

\subsection{Parameters of the model}
\lab{parameters}

In the model constructed above there are various parameters.
%The
%model is of the type discussed in \cite{GKMN2} and the reader is
%recommended to consult this paper for a detailed discussion of the
%parameters.
Here we shall list the parameters without derivation
and refer to \cite{GKMN2} for a detailed discussion.
\begin{itemize}
\item{Parameters of the action}: In the weak gravity limit,
 $G_N\to 0$, $N\to\infty$ and $M_p^{1-d} = 16\pi G_N N^2 = fixed.$,
there are two parameters in the action: $M_p$ and the overall size
of the potential $\ell$. The latter fixes the units in the theory.
One can construct a single dimensionless parameter from the two:
$M_p\ell$ which determines the overall size of thermodynamics
functions in the dual field theory and it can be fixed e.g. by
comparison with the value of the free energy at high temperatures,
see \cite{GKMN2}. In the present paper we are only interested in
scaling of functions near $T_c$, thus this parameter will play no
role in what follows.
\item{Parameters of the potential:} We have not specified the
potential apart from its IR asymptotics. The IR piece will be
enough to determine the scaling behaviors and also the transition
temperature through equation (\ref{Tc}). Therefore we have only
three (dimensionless) parameters: $V_\infty\ell^2$, $C$ and
$\kappa$ or $\alpha$ that appear in  (\ref{Vs}), (\ref{case1}) and
(\ref{case2}). The first determines the (dimensionless) transition
temperature $T_c\ell$ through (\ref{Tc}), the second one is
related to the boundary value of the dilaton (cf. the discussion
above), and the third one determines the type of the transition.
For example $\kappa = 2$ for a second order transition, equation
(\ref{kappaNC}).
\item{Integration constants}: In \cite{exotic1} we solve the
Einstein-dilaton system and work out the thermodynamics in the
reduced system of ``scalar variables" that is a coupled system of
two first order differential equations. One boundary condition can
be interpreted as the value of $T$, and the other is just
regularity of the solution at the horizon. Therefore the only
dimensionless parameter that arise among the integration constants
is $T\ell$.\footnote{In the fifth order system of
(\ref{E1}-\ref{E3}) it is a little harder to work out the
non-trivial integration constants. There it works as follows
\cite{GKMN2}: In (\ref{E3}), one requires $f\to 1$ as $r\to 0$.
This fix one constant, and the other is gives T. In the rest, one
is fixed by requirement of regularity at the horizon, one is just
a reparametrization in $r$, and the last is fixed either by the
asymptotic value of the dilaton in the case $\f(r)= \f_0$ is
constant at the boundary, or the integration constant $\Lambda$
that determines the running of the dilaton near the boundary
$\f\sim \log(-\log(\Lambda r))$ near $r\to 0$. In either case the
thermodynamic functions can be shown to be independent of this
constant \cite{GKMN2}.}

\end{itemize}

\section{Non-critical string theory and the IR CFT}
\lab{ncst}
%In the phenomenological model above we had a list of parameters.
%In the case of pure non-critical string theory the situation is
%much more constrained.

\subsection{Linear-dilaton in the deep interior}

The leading asymptotics (\ref{Vs}) of the dilaton potential which
follows from the requirement of a continuous Hawking-Page
transition is precisely the same as the potential that follows from
 $d+1$ dimensional non-critical string theory. This is easily
seen by transforming (\ref{action}) with the potential (\ref{Vs}) to string frame with $g_{s,\m\n} =
\exp(2\f/(d-1)) g_{\m\n}$. Not only that but we also have the
asymptotics (\ref{fs}), which imply  that the asymptotic solution
in the IR corresponds to a linear-dilaton background that
is---very conveniently---an $\a'$ exact solution to
(\ref{saction}) and corresponds to an exact world-sheet CFT.
Indeed, in \cite{exotic1} it is shown that, with the subleading
terms of the form (\ref{Vs}), the string-frame curvature
invariants both on the TG and the BH backgrounds vanish in the
deep interior region near criticality i.e. for $r_h\to\infty$,
($T\to T_c$). Hence the higher derivative terms denoted by
ellipsis in (\ref{saction}) become unimportant in the IR theory.

This implies that {\em the dynamics in the  transition region
should be governed by the linear-dilaton CFT. More precisely, we
expect that quantities that receive dominant contributions from
the deep interior region near criticality should be determined by
the linear-dilaton CFT.}

In the next section we shall make use of this observation to argue
that the various observables in the corresponding spin-model scale
precisely with the expected critical exponents near $T_c$.

Another implication of this is that an asymptotically linear
dilaton geometry (with corrections governed by the subleading
terms in (\ref{Vs}) develops an instability at a finite
temperature $T_c$ into formation of black-holes. It is quite
reasonable to expect that in the limit of weak $g_s$ this point
coincides with the Hagedorn temperature of strings on the
linear-dilaton background \cite{AtickWitten}. We have more to say
on this in section \ref{tpfquant}.

Finally, we note that  in the case when the model is embedded in non-critical string theory, all of the
parameters in the model are entirely fixed. To illustrate this let us assume
that the entire potential is given by the leading term, ignoring the sub-leading terms
etc. Then the
coefficient $V_\infty$ in (\ref{Vs}) and the transition
temperature would be given as,
%%%%%%%%%%%%%
\be\lab{Vsnc}  V_{\infty,nc} = \frac{c_b(25-d)}{\ell_s^2}, \qquad
T_{c,nc} = \frac{1}{4\pi\ell_s} \sqrt{c_b(25-d)}, \ee
%%%%%%%%%%%%
in the case of bosonic world-sheet CFT and
%%%%%%%%%%%%%
\be\lab{Vsncf}  V_{\infty,nc} = \frac{c_f(9-d)}{\ell_s^2}, \qquad
T_{c,nc} = \frac{1}{4\pi\ell_s} \sqrt{c_f(9-d)}, \ee
%%%%%%%%%%%%
in the case of  fermionic world-sheet CFT.  These results follow
from  (\ref{cdef}) and (\ref{Tc}). Of course, in reality these
numbers should be renormalized because the theory is not just
given by the leading piece: a potential with only the leading
exponential behavior do not possess any phase transition. The
corrections will depend on the UV physics where the $\a'$
corrections kick in and renormalize these coefficients. We shall
argue for another way to fix these numbers in section  \ref{sec8}.
We will also show in that section that the scaling exponents are
also determined completely, once the CFT is fixed.

\subsection{The CFT in the IR}
\lab{IRCFT}

The arguments presented above point towards the conclusion that,
on the string side the criticality of the dual spin-system should
be governed by a linear-dilaton CFT. Here we want to
spell out some of the salient features of this IR CFT. We start
with the bosonic case and then mention generalization to fermionic
CFT in the end.

We reviewed the intimate connection between non-critical string
theory and the linear-dilaton background in appendix
\ref{lindilNC}. Utilizing this relation one can obtain the
stress-tensor of the (bosonic) linear-dilaton CFT as
\cite{Chemseddine}, \be\lab{wsemt} T(z) = -\frac{1}{\a'} :\6
X^{\m} \6 X_{\m}: + v_{\m} \6^2 X^{\m} \ee for the left-movers,
with an analogous expression for the right movers. $v_\mu$ are the
proportionality constants in the dilaton solution $\of = v_\mu
X^\mu$. The indices are raised and lowered by the flat metric. The
total central charge of the theory (including the ghost sector)
vanishes for $v_\mu$ satisfying (\ref{condv}).
%The zero-point
%energies are given by $a=\tilde{a} = (1-d)/24$ for the bosonic
%theory, $a= (1-d)/8$ for the NS sector of the fermionic theory and
%$a=0$ for the Ramond sector of the fermionic theory\footnote{We
%list these values for completeness, although we shall assume a
%general value for the zero-point energies in what follows.}.
In our case we have, \be\lab{ldcftus} v_\mu =
\frac{\sqrt{V_\infty}}{2}~\delta_{\mu,r}\equiv m_0~\delta_{\mu,r}.
\ee The reason for denoting this constant $m_0$ will be clear when
we analyze the spectrum of fluctuations in this geometry, see
appendix \ref{Appflucs}. The total central charge of the theory
(including ghosts) vanishes only for, \be\lab{m0sq} m_0^2 =
\left\{
\begin{array}{ll}
\frac{25-d}{6\ell_s^2} & \textrm{bosonic}\\
\frac{8-d}{4\ell_s^2} & \textrm{fermionic}
\end{array} \right.
\ee for the bosonic and fermionic CFT's, \cite{Myers}.

Now we discuss the spectrum in the case we are interested in: The
{\em Euclidean} $d+1$ dimensional world sheet with (\ref{ldcftus})
and the Euclidean $X^0$ dimension compactified on a radius $R =
1/2\pi T$. There are various ways to obtain the spectrum. Both the
light-cone and the covariant quantization is discussed in
\cite{Myers,Chemseddine}. Here we trivially extend these results
in our case.

The Virasoro generators are now \be\lab{Virasoro} L_m = \half
\sum_{n=-\infty}^{\infty} : \a_{m-n}^{\m} \a_{n,\m}: + i
\frac{\sqrt{\a'}}{\sqrt{2}} (m+1) m_0 \a_m^r. \ee The
center-of-mass momenta are related to the zero mode oscillators as
usual, $p^{\m}_L = \sqrt{\frac{2}{\a'}} \a_0^{\m}$ and $p^{\m}_R =
\sqrt{\frac{2}{\a'}} \tilde{\a}_0^{\m}$. Decomposing into
components one has, \bea\lab{zero}
p_{0,L} &=& 2\pi T k + \frac{w}{2\pi T \a'}, \qquad p_{0,R} = 2\pi T k - \frac{w}{2\pi T \a'},\\
p_{i,L} &=& p_{i,R} = p_i,\qquad\qquad\,\,\,\,\,\, p_{r,L} =
p_{r,R} = p_r.\lab{spar} \eea In the first line the integer $k$
denotes the {\em Matsubara frequency}  and the  integer $w$
denotes the {\em winding number} on the time-circle. As a result
of the linear piece in the zeroth level Virasoro generator
(\ref{Virasoro}) one obtains the following mass-shell conditions
(we adopt the definition of mass in \cite{Chemseddine}) in the
light-cone gauge: \bea\lab{mashel1} -m^2_{d+1} &=& p_\perp^2  +
p_r^2 +2im_0 p_r + (2\pi k T)^2 +
 \le(\frac{w}{2\pi T \a'}\ri)^2  = -\frac{2}{\a'}(N + \tilde{N} - 2),\\
0 &=& kw + N -\tilde{N}\lab{mashel2}, \eea where $p_\perp$, $N$
and $\tilde{N}$ denote the center-of-mass momentum, the left
(right) number of oscillations in the space transverse to motion,
respectively, \be\lab{oscil} N = \sum_{n=1}^\infty \a_{\perp,-n}
\cdot \a_{\perp,n}, \qquad \tilde{N} = \sum_{n=1}^\infty
\tilde{\a}_{\perp,-n} \cdot \tilde{\a}_{\perp,n}. \ee

In (\ref{mashel1}) $m^2_{d+1}$ denote the $d+1$ dimensional mass.
One important difference between the linear-dilaton and the flat
case is that the definition of the mass of the string excitations
in terms of their momentum gets modified\cite{Chemseddine} due to
the linear oscillator piece in (\ref{wsemt}). The flat case
follows by setting $m_0=0$, hence sending dilaton to constant.

Once the modified definition of mass is attained, the physical
spectrum of the linear dilaton is exactly the same as the critical
string: the lowest level $N=\tilde{N}=0$ is a tachyon with mass
$-4/\a'$, the next level is massless and corresponds to the
fluctuations of the metric, the B-field and the dilaton, etc.
\cite{Chemseddine}.

All of these results are readily extended to the fermionic case
with $N=1$ world-sheet supersymmetry \cite{Chemseddine}. In the
light-cone gauge, one obtains the following spectra for the NS and
Ramond sectors, \bea\lab{NS} m^2_{d+1} &=& N + \sum_{q>0} q
b^{\perp}_{-q}
b^{\perp}_q  - \half, \qquad (NS)\\
\lab{R} m^2_{d+1} &=& N + \sum_{q>0} q b^{\perp}_{-q} b^{\perp}_q,
\qquad (R), \eea where $N$ denotes the number of bosonic
oscillations in the transverse space (\ref{oscil}) and $q\in
{\mathbb Z}$ for the R-fermions and $q\in {\mathbb Z}+\half$ for
the NS-fermions. This is again the same spectra that one finds in
the critical super-string.

However one finds crucial differences at the one-loop level:
Modular invariance does not allow for NS-R fermions except in
particular dimensions given by multiples of 8. This is quite
convenient for our holographic purposes, because we do not want
any fermionic operators in the dual spin-model. Thus in a generic
dimension $d+1<8$ one has only two sectors R-R and NS-NS.
Furthermore, in the generic case, there is no analog of the GSO
projection of the superstring. {\em Therefore the tachyon in the
NS-NS sector survives}.

Existence of tachyon in the physical spectrum is a very generic
feature of the linear-dilaton CFT in any dimensions. The mass of
the tachyon changes depending on which particular CFT chosen. With
the definition of mass adopted above it is given by $m^2_T =
-4/\a'$ for the bosonic case, $m^2_T=-2/\a'$ for the NS-NS
fermions, $m^2=-15/4\a'$ for an orbifold in the r-direction, etc.,
but we stress that the ground state for $k=w=0$ in linear-dilaton
CFT in arbitrary dimensions is always a tachyon.\footnote{With a
more conventional definition of mass \cite{Myers}, one finds a
tachyon only for $d>1$ in a $d+1$ dimensional theory. In our case,
the equivalent statement is that if we consider propagation of the
tachyonic mode, we find a smooth propagation for $d\leq 1$ but
oscillatory behavior for $d>1$.}

This fact renders the linear-dilaton theory unattractive from many
perspectives. In our case however, it is a desired feature of the
IR CFT. We recall that the background geometry becomes
asymptotically linear-dilaton {\em only} in the transition region
$T\sim T_c$ and {\em only} in the large $r$ region. We do not
expect that the complete sigma-model which corresponds to the
black-hole for an arbitrary T have tachyon as a ground state. This
would imply that the black-hole geometry is unstable at any
temperature. Instead the linear dilaton CFT describes the physics
near the transition and we do expect instability in this region.
In fact, as we show in sections (\ref{opfquant}) and
(\ref{tpfquant}), {\em it is the presence of the tachyon} which
guarantees vanishing of magnetization as $M\sim (T-T_c)^{\beta}$
and divergence of the correlation length as $\xi \sim
|T-T_c|^{-\nu}$ at the transition!

\section{Spin-model observables from strings}
\lab{sec8}

F-strings and D-branes are important probes in the standard
examples of the gauge-gravity correspondence. In case of the
holographic models for QCD-like theories, the phase of the field
theory at finite temperature, the quark-anti-quark potential, the
force between the magnetic quarks etc, can all be read off from
classical F-string and D-string solutions in the dual
gravitational background. In this section we argue that the probe
strings constitute indispensable tools also in the
spin-model-gravity correspondence. In particular, the Landau
potential, the correlation length, the various critical exponents,
the scaling of order parameters near the transition, the phase of
the system,  spin-spin correlation function, etc. can all be
computed from the probe strings in the dual background. In this
section we discuss how to obtain the various observables of the
spin-model from the probe string solutions.
%This will serve as a
%consistency check on the duality proposed in the previous section.

\subsection{What can we learn from the Gravity-Spin model duality?}
\lab{what}

In order to answer this question, one has to identify the Landau
and the mean-field approximations on the gravity side. The Landau
approach is based on integrating out the ``fast" degrees of
freedom in the spin-model in order to obtain a free-energy
functional for the ``slow" degrees of freedom, i.e. the order
parameter $\vec{M}$. We refer to appendix \ref{statmech} for a
review of the statistical mechanics background and in particular a
description of the Landau approach. This is exactly analogous to
integrating out the gauge invariant states to obtain an effective
action for the Polyakov loop on the LGT side, as illustrated in
appendix \ref{sec2}. In the context of the gauge-gravity
correspondence, this is, in essence, very similar to keeping only
the lowest lying degrees of freedom in string theory, i.e. the
supergravity multiplet. It is tempting to think that the
complicated step of integrating over the spin configurations in
(\ref{SpinZ}) to obtain (\ref{SpinL}) can be side-stepped by use
of the gravity-spin model duality\footnote{We note a very
interesting paper \cite{Headrick} that dwells on these issues. In
this paper Headrick argues that one can generate the Landau
functional at strong coupling in terms of classical string
solutions.}.

In the most general case, the correspondence between the spin
model and gravity should relate the Landau functional
(\ref{SpinL}) with the string path integral\footnote{We shall be
schematic in what follows.}:
%%%%%%%%%%%%%%%
\be\lab{grlan1} Z_L = Z_{st}. \ee
%%%%%%%%%%%%%%%
As in the original gauge-gravity duality we expect that there is a
simple corner of the correspondence where both sides of
(\ref{grlan1}) become classical and one approximates the path
integrals by the classical saddles.

This is the large N limit: On the LHS this is given by the {\em
Landau approximation} (\ref{Landau}). On the RHS, it is given by
the saddle-point approximation to the string theory where one can
ignore string interactions $g_s\to 0$. Then (\ref{grlan1}) reduces
to,
%%%%%%%%%%%%%%%
\be\lab{grlan2} e^{-\beta F_L} = e^{-{\cal A}_{st}}, \ee
%%%%%%%%%%%%%%%
where the action on the RHS is the full target-space action
including all the $\a'$ corrections, evaluated on the classical
saddle. It is the effective action for all excitations of a single
string\footnote{It is important to note that this is not a string
field theory action, the excitations governed by ${\cal A}_{st}$
are particles, rather than strings.} and in principle it can be
obtained from the sigma model on the world-sheet.

At this point, it is clear that scaling of any quantity on both
side of (\ref{grlan2}) near $T_c$ should be characterized by the
{\em mean-field scaling}. This is just a consequence of the saddle
point approximation. Therefore, any operator in the spin-model
that is given by a fluctuation of ${\cal A}_{st}$ should obey the
standard mean-field scaling. We shall refer to these operators as
{\em local operators}. The only possible exceptions to
this---within the classical approximation of (\ref{grlan2})---are
operators that {\em can not} be obtained as fluctuations of
${\cal A}_{st}$. These correspond to {\em non-local operators} on the
gauge theory, they are governed by probe F-strings or D-branes on
the string side. Yet, as we will show in the next section, they
can correspond to quite ordinary quantities such as the {\em
magnetization} on the spin-model side. Thus, magnetization is an
example of a {\em non-local operator}.
%In section \ref{magna} we
%indeed confirm the expectation that the scaling of magnetization
%is not necessarily mean-field in the gravity description. This is
%easy to understand on the string side: stringy fluctuations are
%characterized by a length scale $\ell_s$ different than the
%ordinary bulk fluctuations that are typically characterized by
%$\ell$. Therefore they may exhibit non-mean-field scaling
%behavior.
Even for the ``non-local observables" though the mean-field
scaling is expected to hold in a semi-classical approximation,
where one only keeps the lowest-lying string excitations in string
path integrals. These excitations correspond to bulk gravity modes (levels $N=0$
and $N=1$ of the string spectrum). We confirm this expectation in the sections
(\ref{opfquant}) and (\ref{tpfquant}) below.

In practice, it is usually very hard to reckon with
(\ref{grlan2}), and one further considers the weak-curvature limit
where one can replace the RHS with the (super)gravity action:
%%%%%%%%%%%%%%%
\be\lab{grlan3}  \frac{1}{T} F_L \approx T V_{d-1} {\cal L}_{gr}.
\ee
%%%%%%%%%%%%%%%
Here, ${\cal L}_{gr}$ is the (super)gravity action evaluated
on-shell, on the classical saddle. We also assumed a trivial
dependence on the spatial volume and made use of the fact that the
temperatures on the spin-model and the gravity sides are inversely
related, cf. appendix \ref{sec2}.

Influenced by the standard lore of the gauge-gravity
correspondence, we expect that the weak curvature limit
corresponds to strong correlations on the spin-model side. On the
other hand,  one quantifies ``strong correlations" by the Ginzburg
criterion in the spin-model, as reviewed in section
\ref{statmech}. Quite generally, the system will be in a regime
of strong correlations around the phase transition where the
mean-field approximation usually breaks down. Therefore, one may hope that
the  gravity side provides a better description in (\ref{grlan3}) precisely within this interesting region.
This can be checked explicitly by computing curvature invariants in the string frame. 
Even though one show that the Ricci scalar (and the various contractions of Ricci two-form and the Riemann tensor) 
vanish in the limit (see \cite{exotic1}) there exists invariants such as $d\f^2$ that asymptote to a constant that is 
generically the same order as the string length scale $\ell_s^{-2}$. Therefore, in a generic case one is forced to 
include the higher derivative corrections. Luckily this can be done precisely in the interesting critical regime, because 
the background asymptotes to a linear-dilaton theory.   

What observables can we actually calculate on the gravity side?
Because going beyond the large N limit is very hard, one can (at
present) only hope to obtain results in the Landau approximation.
The main observables then include the Landau
coefficients\footnote{We refer to appendix \ref{statmech} for a
definition of these coefficients.} $\a_0(T)$, $\a_1(T)$,
$\a_2(T)$, the basic scaling exponents $\beta$, $\nu$, $\eta$,
$\gamma$ etc., and the spin correlation functions. Moreover, the
scaling exponents of operators that are dual to fluctuations of
the bulk fields in ${\cal L}_{gr}$ in (\ref{grlan3}) are
necessarily given by the mean-field scaling. Therefore one can
only hope to obtain results beyond mean-field in the scaling exponents of
operators dual to stringy objects, such as magnetization or the spin-spin correlator. 

Once again, we would like to emphasize the distinction between
``mean-field scaling" and the ``mean-field approximation". The
former is unavoidable for {\em local operators} in the Landau
approximation (large N). On the other hand, gravity description is
expected to go beyond the latter. Therefore for quantities such as
$T_c$, the Landau coefficients at $T_c$, etc., and correlation
functions of the non-local observables we expect gravity to
provide better answers than the mean-field approximation.

One may still ask the question, what is the use gravity-spin-model
duality if one can compute all of these quantities by employing
Monte-Carlo simulations, or RG techniques? First of all, the RG
techniques are limited in the case of strong correlations.
Secondly, the calculations on the gravity side are much easier to
perform, much easier than the Monte-Carlo simulations, and one can
usually obtain analytic results. However a more fundamental reason
is that, there are situations where applicability of the
Monte-Carlo simulations are limited. The well-known examples are
the computation of real-time correlators or spin-models with
fermionic degrees of freedom. By the gravity-spin-model
correspondence, one expects to overcome such fundamental
difficulties.

\subsection{Identification of observables}
\lab{probes}

%Along the discussion of ``gravity-spin-model'' duality of section
%2, this picture is also very suggestive.
%We want to determine the Landau functional near $T_c$ by gravity
%methods. A crucial part of the calculation is to determine how the
%magnetization $\vec{M}$ scales near the transition.
The duality between the lattice gauge theories and spin-models
\cite{Polyakov}, \cite{Susskind} relate the magnetization directly
to the Polyakov loop. On the other hand, the Polyakov loop is
related to the classical F-string solution as discussed in section
\ref{sec4}. Therefore we propose the following chain of relations:
%On the spin-model side the Polyakov loops are nothing but the
%spin fields $\vec{s}$. Therefore we propose to study the spin
%fields and their correlators by the fundamental strings in the
%gravity dual:
\be\lab{Pspin} \la P(x) \ra \leftrightarrow \la \vec{m}(x)\ra
\leftrightarrow e^{-S_{NG}[C_x]}. \ee Here the boundary condition
$C_x$ for the string is just a point x in the spatial part and a
loop on the temporal circle.

The spin field is valued under $U(1)_S$. Similarly the Polyakov
loop is valued under the center ${\cal C}= Z_N$ that becomes a
$U(1)$ in the large N limit. One should think of this as the
exponents becoming angles in the transformation,
$$P\to e^{2\pi i\frac{k}{N}} P, \qquad k=1,\,2 \cdots  N$$
at large N. We shall denote this $U(1)$ as $U(1)_{\cal C}$.
Similarly, as discussed in section \ref{sec3} at length, the F-string that
winds the time-circle is charged under the
$U(1)_B$\footnote{The charge is determined by the winding number.
Here we are only interested in strings that wind the time
circle once.}, because it couples to the B-field. Thus one should
identify
%%%%%%%%%%%%%%%%%%
\be\lab{id} U(1)_S = U(1)_{\cal C}=U(1)_B. \ee
%%%%%%%%%%%%%%%%%%
as in table in section \ref{sec4}.

One should work out the identification in (\ref{Pspin}) carefully.
In particular the first entry is a complex number and the second
entry is a vector in 2D spin space. The precise identification of
the two is provided with the standard isomorphism between $U(1)$
and $O(2)$ representations. We imagine the vector $\vec{m}$ on
the XY plane represented by the magnitude $|\vec{m}|$ and the
phase $\psi$. Then the simplest option is to set $m_x = Re(P)$ and
$m_y = Im(P)$. There is a little complication though, because in
fact the identification should depend on the value of $\psi$. This
is because the physically preferred reference frame is set by the
direction of the magnetization vector $v_i$ in (\ref{magun}). All
of the correlation functions should be decomposed into components
parallel and perpendicular to $v_i$. Represented by the phase, the direction of magnetization
reads
%%%%%%%%%%%%
\be\lab{vpsi} \vec{v} = (\cos(\psi),\sin(\psi)). \ee
%%%%%%%%%%%%
Thus, the naive identification mentioned above is correct only for $\psi=0$.
For a different value of $\psi$ one should obtain the correct
identification by a $U(1)$ rotation: $P = \exp(i\psi) (m_x + i
m_y)$. Thus, in general we have,
\begin{equation}\label{idt}
Re(P) = m_{\parl} = \cos(\psi) m_x - \sin(\psi) m_y, \quad Im(P)=
m_{\perp} = \sin(\psi) m_x + \cos(\psi) m_y,
\end{equation}
where
\begin{equation}\label{parperl}
  m_{\parl,i} = \vec{v}\cdot\vec{m}~v_i, \qquad m_{\perp,i} =
  (\delta_{ij} - v_i v_j) m_j.
\end{equation}

The identification of the second and the third entries in
(\ref{Pspin}) is straightforward. One can schematically write,
\be\lab{WF} \la P[C] \ra = \la e^{ -\int G + i \int B }\ra, \ee using
(\ref{OP}) and (\ref{po2}), where we dropped the dilaton
coupling\footnote{We check in appendix \ref{AppP} that this contribution is
sub-leading and do not contribute to the scaling near $T_c$.}.
Thus the magnitude of $P$ is determined by the space-time metric
and the phase is determined by the B-field as explained in section
\ref{sec4} and one should identify the $\psi$ angle defined in
(\ref{psi}) and the angle defined in (\ref{vpsi}).

In this picture the spin correlation function should be given by a
fundamental string solution that ends on two separate points $x$
and $y$: \be\lab{Pspin2} \la m_i(x) m_j(y) \ra \leftrightarrow
e^{-S_{NG}[C_{xy}]}. \ee The boundary condition is such that the
string ends on the points x and y on the spatial parts and wraps
the temporal circle.

Again, one has to be careful in the identification (\ref{Pspin2})
and has to split the correlator into the parts perpendicular and
parallel to the magnetization vector $v_i$ as in (\ref{sscorg}):
\begin{equation}\label{2pf}
  \la m_i(x)~m_j(0)\ra = \la \vec{m}_{\parl}(x)\cdot \vec{m}_{\parl}(0) \ra v_iv_j +
   \la \vec{m}_{\perp}(x)\cdot \vec{m}_{\perp}(0) \ra(\delta_{ij} -
  v_i v_j).
\end{equation}
On the other hand one has the identification,
\begin{equation}\label{idcor}
\la \vec{m}(x)\cdot \vec{m}(0)\ra = \la P^*(x) P(0)\ra.
\end{equation}
Given the identification (\ref{idt}) one obtains,
%%%%%%%%%%%%
\bea\lab{idt21} \la \vec{m}_{\parl}(x)\cdot \vec{m}_{\parl}(0)\ra
&=& |\vec{M}|^2 +  \la \vec{s}_{\parl}(x)\cdot
\vec{s}_{\parl}(0)\ra = \la Re~P(x)Re~P(0)\ra, \\
\lab{idt22} \la \vec{m}_{\perp}(x)\cdot \vec{m}_{\perp}(0) \ra &=&
\la \vec{s}_{\perp}(x)\cdot \vec{s}_{\perp}(0) \ra= \la
Im~P(x)Im~P(0)\ra\qquad, \eea
%%%%%%%%%%%%
where we also decomposed the magnetization $\vec{M}$ and the
fluctuations $\vec{s}$ according to (\ref{flus}), assuming that
$\vec{M}$ is isotropic. This is of course in the black-hole phase.
In the thermal-gas phase $|\vec{M}|$ vanishes and any direction
$v_i$ is identical.

\subsection{One-point function}
\lab{magna}

Having identified the observables on the spin model side with the
observables on the gravity side, we are ready to determine the
magnetization $\vec{M}$ of the spin model on the gravity side by a
one-point function calculation. As we argued in the previous
section the magnetization should be given by the  real part of the
F-string solution that wraps the time circle. As a warp-up
exercise, we shall first assume that the string path integral is
dominated by the classical saddle and obtain the resulting scaling
law for the magnetization. After this, we will loosen the
assumption and perform the same calculation in a semi-classical
regime in section ({\ref{opfquant}).

\subsubsection{Warm-up: classical computation}
\lab{opfclas}

The definition of the Polyakov action and the boundary conditions
are  given in detail in section \ref{sec3}. In Appendix D we prove
that in all of the cases we consider in this paper, the dilaton
coupling term in $S_{NG}$, that is given by $\f R^{(2)}$ gives
finite contributions, hence do not alter the scaling. Also, the
effect of the B-field is discussed in detail in section 3.2. Thus
we shall only restrict our attention to the area term (see eq.
(\ref{WF})):
\begin{equation}\label{areaterm}
  |\vec{M}| = |\la P[C]\ra | \propto \la e^{-\int G} \ra
\end{equation}
and replace the fundamental string action with the Nambu-Goto
action.

To compute the energy of the string, we fix the gauge $\s = x^0$,
$\tau = r$ where $(\tau,\sigma)$ are the coordinates on the
world-sheet, $x^0$ is the Euclidean time that is identified as $
x^0\sim x^0+1/T$ and $r$ is the radial variable in the coordinate
system given by (\ref{BH}). Then,
\begin{equation}\label{NG}
  S_{NG}  = \frac{T(r_h)^{-1}}{2\pi\ell_s^2}
  \int_{\epsilon}^{r_h} dr \sqrt{det\,\, h_{ab}},
  \qquad h_{ab} = \6_a x^{\m} \6_b x^{\n} g^s_{\m\n},
\end{equation}
where $\ell_s$ is the string length, $g^s$ is the BH metric in the
string frame:
\begin{equation}\label{sBH}
  ds^2_s = e^{2A_s(r)}\le( f^{-1}(r)dr^2 + dx_{d-1}^2 + dt^2 f(r) \ri), \qquad  A_s(r) = A(r) + \frac{2}{d-1}\f(r),
\end{equation}
 and $\epsilon$ is a point near the boundary\footnote{The target space metric typically diverges on the boundary and this cut-off
guarantees finiteness of (\ref{NG}). One can remove the dependence
on $\epsilon$ by some renormalization procedure, however we do not need
this as we are only interested in the dependence of $S_{NG}$ on
$r_h$ in the limit $r_h\to \infty$ that corresponds to $T\to
T_c$. We provide an appropriate renormalization scheme in appendix \ref{AppP}.}.

In passing we review the discussion in section \ref{largeN}. There
we introduced the rescaled field $\f$ in relation to the ``real"
dilaton as in equation (\ref{shdil}) as $\f = \of + \log N = \of_0
+ \delta \of + \log N  =  \log\l + \delta \of$ the last two lines
follow from (\ref{defdf}) and (\ref{limN}). The constant $\l$ is
$\cO(1)$. Thus, when large, $\f$ just corresponds to the
difference of the real dilaton and its  boundary value $\of_0$.
Large $\f$ does not mean large $\of$ when $\of_0$ is chosen very
small. This choice indeed corresponds to the large N limit in the
gravity language. Therefore we can safely ignore the loop
corrections here, and in the next sections.

On the TG solution one replaces $f\to1$, $A(r)\to A_0(r)$ and
$\f(r)\to \f_0(r)$ in the above formulae. As described in section
appendix \ref{AppP}  the exponential of $-S_{NG}$ vanishes on the TG solution. Thus,
\begin{equation}\label{MTG}
  \vec{M}_{TG} = 0.
\end{equation}
This confirms our discussion in section \ref{sec4}, that indeed
{\em the TG phase of the gravity   corresponds to the disordered
phase of the spin model.}

Let's turn to the BH phase. It is shown in (\ref{AppP}) that
(\ref{NG}) is finite, unless $d=2$. In the latter case the
fluctuations of the zero mode of the B-field in the 2D transverse
space makes it vanish---see the discussion in section \ref{sec4}.
Thus one finds,
\begin{equation}\label{MBH}
  \vec{M}_{BH} \ne 0 \quad (d>2),\qquad \vec{M}_{BH} = 0 \quad
  (d=2),
\end{equation}
and  {\em the black-hole solution indeed corresponds to the
ordered phase of the spin-model for $d>2$. One also confirms on the gravity side, that no
long-range order in the spin field is possible in $d=2$}.

We would also like to see how $\vec{M}$ scales near the phase
transition (as one approaches from below in the spin model). Using
eqs. (\ref{As}) and (\ref{fs}),  one finds that the scale factor
vanishes  $A_s(r)\to 0$ as $r\to\infty$, see (\ref{delAs}). Thus
the integrand in (\ref{NG}) becomes constant in the limit
$r_h\to\infty$. Using also the fact that $T\to T_c$ in this limit
one finds, (see App. \ref{AppP} for details),
\begin{equation}\label{NG2}
  S_{NG}  \to  \frac{T_c^{-1}}{2\pi\ell_s^2}  r_h, \qquad T\to T_c.
\end{equation}
In order to determine the scaling of $S_{NG}$ with the reduced
temperature $t$, one needs to find the dependence of $r_h$
on $t$ . This is done in App. \ref{AppP}.
%easily determined in the cases (\ref{case1})
%and (\ref{case2}) using (\ref{As}) and the asymptotics of the
%Einstein-frame scale factor (\ref{Acase1}) and (\ref{Acase2}).
% We present the details of
%this calculation in App. \ref{vinftc}.
Define the dimensionless constant:
\begin{equation}\label{vs}
V_s =V_\infty\ell_s^2 = (4\pi T_c\ell_s)^2,
\end{equation}
where $V_\infty$ is defined in (\ref{Vs}) and (\ref{Tc}) is used
to relate it to $T_c$. Then the result is,
%%%%%%%%%%%%%%%%%%%%%%%%%%%%%%%%%%%%%%%%%%%%%%%%%%%%%%%%%%%%%%%%%
\bea
  \mathrm{Case\,\, i:} \qquad \vec{M}_{BH} = e^{-S_{NG}} &\propto & t^{\frac{4}{\kappa~V_s}}\qquad
  t\to 0 \lab{mcase1}\\
  \mathrm{Case\,\, ii:}\qquad \vec{M}_{BH} = e^{-S_{NG}} &\propto & e^{-\frac{4}{V_s} \le(\frac{t}{C}\ri)^{-\frac{1}{\a}}}, \qquad
  t\to 0 \lab{mcase2},
\eea
%%%%%%%%%%%%%%%%%%%%%%%%%%%%%%%%%%%%%%%%%%%%%%%%%%%%%%%%%%%%%%%%%
where the constants $\a,\kappa$ and $C$ are defined in (\ref{case1}) and (\ref{case2}).\\
We note that (\ref{mcase2}) is valid strictly for $d>2$. As
mentioned before, for $d=2$ we obtain $\vec{M}=0$ below and above
the transition.

Let us now specify to the case of {\em second-order transitions}.
Then the coefficient $\kappa$ is given by (\ref{kappaNC}) with
$n=2$:
%%%%%%%%%%%%%%%%%%%
\be\lab{sopt} \kappa =2, \qquad {\rm second-order\,\,\,
transition.} \ee
%%%%%%%%%%%%%%%%%%%
Then, comparison of (\ref{mcase1}) with (\ref{beta}) yields the critical exponent of the magnetization as,
\begin{equation}\label{nudet}
  \beta = 2 V_s^{-1},
\end{equation}
where $V_s$ is given by (\ref{vs}).

%It is essentially determined
%by the phenomenological parameter $\ell_s$ as follows. The value
%of $V_\infty$ in (\ref{vs}) can be fixed by the observed $T_c$.
%Then, from (\ref{nudet}) one obtains $\nu \sim (\ell/\ell_s)$. In
%a top-down approach to the gravity-spin model duality---see
%comment 1 below eq. (\ref{map3}) of section \ref{sec2}---one
%should be able to obtain the precise relation between this
%quantity and the interaction strength of the spin model $J$. This
%would provide a non-trivial prediction of the holographic
%approach. In our phenomenological approach here, $\ell_s$ is a phenomenological parameter,
%that is so far unspecified. We shall, see below that all of the scaling exponents will be related to this
%parameter, or equivalently $V_s$.

Finally, we note that the mean-field scaling $|\vec{M}| \sim t^\half$ corresponds to a particular value of the parameter $V_s$:
\begin{equation}\label{VsMF}
  V_s^{MF} = 4.
\end{equation}
It seems like a contradiction that one does not automatically
obtain the mean-field scaling for $|\vec{M}|$ directly from the
gravity action. However, it is not. As explained in section
\ref{what}, the magnetization is a ``non-local operator" which
maps onto a non-local object in the string theory side, i.e. the
expectation value of the F-string that wraps the time-circle.
Therefore the classical string computation is not bound to produce
the mean-field result.

On the other hand, we remind the reader that this section  is just
meant to be a warm-up exercise. The classical computation is not
at all guaranteed to be self-consistent. In particular it assumes
that the string path integral is dominated by classical saddles,
which only holds when $\ell/\ell_s$ is parametrically large. This
is not guaranteed in the backgrounds that we discuss in this
paper. Below, we consider the semi-classical computation and argue
that the classical result is altered non-trivially due to large
quantum fluctuations.  We shall observe that the mean-field
scaling arises as a result of the semi-classical computation.

%On the other hand, there are {\em local operators} on the spin-model, such as the Goldstone mode of the phase fluctuations
%in the super-fluid phase, which maps to bulk-fields. These phase fluctuations are mapped onto the fluctuations of the NS-NS
%B-field in (\ref{action}) and we shall see in section {\ref{grss}) that indeed they produce the expected mean-field scaling
%automatically.

\subsubsection{Semi-classical computation}
\lab{opfquant}

In principle the classical saddle  dominates the path integral
only in a regime where the typical curvature radius of the
geometry---that is determined by the asymptotic AdS radius
$\ell$---is much larger than the string length $\ell/\ell_s\gg 1$.
This is indeed the case for a pure AdS black-hole geometry when
the dual ${\cal N}=4$ theory is at strong coupling, because the
AdS/CFT prescription relates the ratio $\ell/\ell_s$ to the 't
Hooft coupling $\l_t$ of the dual gauge theory as $\ell^2/\ell_s^2
\propto \sqrt{\lambda_t}$. In the theories we are interested in,
when there is no tunable moduli like $\lambda_t$, this assumption
will generically fail---unless there is some physical reason for
$\ell/\ell_s$ to be large. In this section we consider a full path
integral computation.

What kind of a string propagator do we want to compute? In the
classical approximation of the previous section, the recipe
\cite{Malda2, Rey,Witten} to compute $\la P[C]\ra$ can be
described as follows. Consider a string that stretches between the
boundary at $r=0$ and a probe D-brane just outside the horizon at
$r= r_h-\epsilon$. The boundary operator wraps the time-circle,
hence the string that couples to it also should. In the Euclidean
BH the length of the time circle measured by an observer sitting
at $r$ goes to zero as $r\to r_h$ thus the string world-sheet
wraps a 2D ball, and yields a finite answer when the UV divergence
regularized properly. This string world-sheet is the classical
saddle of the Nambu-Goto action and it provides the correct answer
for $\ell\gg \ell_s$. We can generalize this picture to the case
$\ell\sim \ell_s$ simply by considering a string that stretches
between the boundary and the horizon, wrapping the time circle,
but this time computing the full path integral including all
quantum fluctuations on the string. This is similar to an
open-string annulus diagram. It is very hard to compute this
however because the string stretches over the entire range between
the boundary and the horizon and one needs the full CFT that
governs the physics everywhere on the target-space. Instead one
can think of this diagram as propagation of a closed string that
is created at the boundary, travels the distance from $r=0$ to
$r=r_h$ on the BH and absorbed at the horizon. This is easier to
handle because, at least we know the CFT close to $r_h$ in the
limit $r_h\to\infty$, which corresponds to the phase transition
regime. This is the linear-dilaton CFT described in section
\ref{ncst}. Indeed, this CFT will prove important in determining
the critical exponents of the corresponding spin system.

%We would like to perform the full path integral of the  closed
%string with the boundary conditions specified above, in a $d+1$
%dimensional background with AdS asymptotics near the boundary and
%linear dilaton asymptotics in the deep interior. We shall confine
%our interest in the case i \ref{case1} with arbitrary $\kappa$.
%Such an analytic background with these asymptotics at zero T was
%constructed in \cite{exotic1}. We are interested in a black-hole
%solution turned on this zero T background. Performing the full
%path integral of the string in such a curved background is beyond
%our present technical ability. We will only be able to present a
%heuristic argument to obtain the scaling of the one-point function
%near $T_c$ by using the general features of the IR CFT outlined in
%section \ref{IRCFT}.

Then the idea is to divide the closed string paths into two parts:
from the boundary to a point $r_m$ and from $r_m$ to the horizon
$r_h$. The point $r_m$ should be chosen such that the string
propagation from $r_m$ to $r_h$ be governed by the IR CFT, see
figure  \ref{fig1}. What is meant by ``semi-classical
approximation" will be to consider  the contribution of the lowest
mass string states at levels $N=0$ and $N=1$ in the IR CFT.

\paragraph{Field theory analogy:}

It is helpful to introduce the idea first in a similar situation
in quantum-field theory. Generalization to the string will then be
clear. First consider the correlator of a free massive scalar
field with mass $m^2$ in flat $d+1$ dimensions $\la \phi(0)
\phi(y) \ra$. This can easily be given a point-particle
interpretation\footnote{We consider Euclidean case for
simplicity.}, \be\lab{fifi} \la \phi(x) \phi(y) \ra =
\int_0^\infty d\tau \la 0,0| \tau,y \ra_{op} \ee where the
integrand is just the propagator of a point-particle in proper
time $\tau$ with Hamiltonian $p^2 + m^2$.

Now let us consider the
more general situation of computing the propagator of a field
$\phi$ in curved space time. The field should  be specified  with
some  quantum numbers such as momenta, charge etc.
determined by the isometries of the background. We assume that
there are no self-interactions, hence no Feynman loops. We also
assume that the back-reaction on gravity can be ignored. Finally
we consider a background geometry of the ``domain-wall type"
$\ref{BH}$ where one coordinate $r$ is singled out. We denote the
$d+1$ coordinates as $(r,\vec{x})$.  Now, $\la \phi(0,\vec{0})
\phi(r,\vec{y}) \ra$ can still be formulated in proper-time as in
(\ref{fifi}) but this time the one-particle Hamiltonian will be
much more complicated.

However, let us consider a situation when the background geometry
simplifies in some asymptotic region, when $r \gg 1$, where we
know how to write down the one-particle Hamiltonian. Then the idea
is to divide one-particle paths in (\ref{fifi}) from 0 to $r_m$
and from $r_m$ to $r_h$, where $r_h>r_m\gg 1$. For this purpose we
decompose the correlator as, \bea\lab{fifi1} \la \phi(0,\vec{0})
\phi(r_h,\vec{y})\ra
 &=& \int_0^\infty d\tau \int d r d^{d}x_m \la 0;0,\vec{0}|\tau_m; r,
 \vec{x}_m \ra
\la \tau_m;r,\vec{x}_m|\tau; r_h, \vec{y}_m \ra \\
{}&\approx& \int_0^\infty d\tau \int d\tau_m~J_m~d^d x_m \la
0;0,\vec{0}|\tau_m; r_m,
 \vec{x}_m \ra
\la \tau_m;r_m,\vec{x}_m|\tau; r_h, \vec{y}_m \ra_{IR}, \nn\eea
where in the second line we exchanged the integral over the
intermediate point $r$ with an integral over $\tau_m$ producing a
Jacobian $J_m$ \footnote{This is achieved by making use of the
freedom  to choose  $\tau_m$ anywhere in between $0$ and $\tau$
and inserting inside the integral $1 = \int d\tau_m
\frac{\delta(r(\tau_m) - r_m)}{\frac{dr}{d\tau}\big|_{r(\tau)=
r_m}}$ a la Faddeev-Popov. This is not to be confused with the
usual re-parametrization invariance of the relativistic
point-particle. Here we describe propagation of a quantum field in
the Schwinger's proper-time formulation, not a relativistic
particle. In particular the Lagrangian that generates the
propagation is not re-parametrization invariant.}.

In the second line of (\ref{fifi1}) the propagator in the region
$0\leq r \leq r_m$ is governed by an unknown one-particle
Hamiltonian $H_{UV}$ that is the full Hamiltonian valid everywhere
on the target-space. The second propagator is governed by $H_{IR}$
for which we assume the knowledge of the spectrum. The two should
be continuously connected at $r_m$. The approximation in the
second line is to replace the full Hamiltonian in the second
propagator with this IR Hamiltonian. The entire procedure will in
general depend on the matching point $r_m$. But, if  we are
interested in how the object scales as a function of the end-point
$r_h$, this dependence will be irrelevant.

A technical but crucial point is that the division of the paths as
in the second line of (\ref{fifi}) makes sense only for the paths
with $\dot{r}(\tau)>0$. This will certainly be  satisfied for
``slight deformations" from the classical path, if the classical
path itself satisfies $\dot{r}(\tau)_{cl}>0$. It is reasonable to
assume that these are the paths that dominate because they
minimize the kinetic energy in the one-particle lagrangian $L_{op}
\propto \dot{r}^2 + \cdots.$ The procedure can be extended to
non-monotonic paths in an interesting way. Let us consider
one-dimensional case for simplicity (the generalization to
arbitrary dimensions is trivial). Suppose that we want to divide
the path integrals in two different regions in space, for $r\leq
r_m$  and $r > r_m$. Then, one has to classify all of the paths
according to their  ``crossing number" $c_m$ that is defined as
the number of solutions to $r(\tau) = r_m$.  The monotonic paths
have crossing number $c_m=1$. This is obviously an odd number and
the next case have $c_m=3$, see figure \ref{figC}. All of the
paths from $r=0$ to $r=r_h>r_m$ are classified by $c_m$. For
non-trivial paths, with $c_m>1$ one can apply the same procedure
by defining $\tau_m$ to be the greatest solution to $r(\tau_m)=
r_m$. Then, the procedure applies smoothly. For sake of the
argument  here, we will restrict only to the paths with $c_m=1$.
This can be achieved by choosing $r_m$ to be close enough to
$r_h$. This is indeed the case  in the physical situation we are
interested in, because the region where the CFT on the string is
governed by the IR CFT corresponds to $r_m\lesssim r_h$ for $r_h
\gg 1$.

\begin{figure}[h!]\begin{center}
\includegraphics[width=9cm]{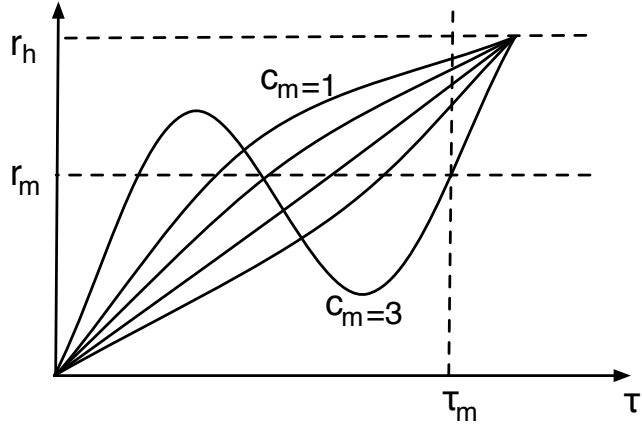}\end{center}
\caption{Division of the one-particle paths that contribute to the
QFT correlation function into regions with $r<r_m$ and $r>r_m$.
The paths are classified according to their crossing number $c_m$.
For crossing $c_m>1$, $\tau_m$ should be chosen as the greatest
node.} \label{figC}
\end{figure}

At this point it is helpful to switch to canonical formulation and
express the propagators in terms of the eigenstates of $H_{UV}$
and $H_{IR}$ that we denote as $\xi'$ and $\xi$ respectively:
\bea\lab{fifi3} \la \phi(0,0) \phi(r_h,\vec{y})\ra
 &=& \int_0^\infty d\tau  \int d\tau_m J_m d^d x_m \sum_{\xi'\in {\cal H}_{UV}}
 \tilde{\Psi}_{\xi'}(0,\vec{0})\tilde{\Psi}_{\xi'}^*(r_m,\vec{x}_m) e^{-\tau_m
 H_{UV}(\xi')}\nn\\
{}&& \,\,\,\,\,\,\,\,\,\,\,\,\,\,\,\ \times \sum_{\xi\in {\cal
H}_{IR}}
 \Psi_{\xi}(r_m,\vec{x}_m)\Psi_{\xi}^*(r_h,\vec{y})
 e^{-(\tau-\tau_m) H_{IR}(\xi)},
 \eea
where ${\cal H}$ denote the Hilbert spaces of the respective
Hamiltonians. $\Psi_\xi$ and $\tilde{\Psi}_{\xi'}$ denote the
wave-functions of the eigenstates $\xi$ and $\xi'$ of $H_{IR}$ and
$H_{UV}$ respectively.

One can now carry out the integration at the matching subspace
$r=r_m$ over $\vec{x}_m$. This integration will produce the
overlap of the wave function $\tilde{\Psi}$ of the UV Hamiltonian
with the wave function $\Psi$ of the IR Hamiltonian \footnote{We
do not assume that the  Hilbert spaces of the UV and the IR
Hamiltonians have same dimensionality, the overlap matrix may be
rectangular.} : \be\lab{overlap} C_{\xi \xi'}  = \int d^d x_m
\Psi^*_\xi (x_m) \tilde{\Psi}_{\xi'} (x_m). \ee One can further
sum over the UV HIlbert space, by defining the overlap function
\be\lab{ofunc} A_\xi(0,\tau_m) = \sum_{\xi'\in {\cal H}_{UV}}
C_{\xi\xi'} \tilde{\Psi}_{\xi'}(0) e^{-H_{UV}(\xi')\tau_m}. \ee
This is the amplitude for production of a state $\xi$ of {\em the
IR Hamiltonian} at $\tau_m$. Thus one has \be\lab{fifi4} \la
\phi(0,0) \phi(r_h,\vec{y})\ra
 = \int_0^\infty d\tau  \int d\tau_m J_m \sum_{\xi\in {\cal
H}_{IR}} A_\xi(0,\tau_m)
 \Psi_{\xi}^*(r_h,\vec{y})
 e^{-(\tau-\tau_m) H_{IR}(\xi)}.
 \ee
Now one can carry out the $\tau$ integral; this will produce $H_{IR}(\xi)$ in the denominator
and then the sum over the IR states $\xi$ will only get contribution from the {\em on-shell state} $\xi_*$
with  $H_{UV}(\xi^*) = 0$.
This is the analog of the on-shell state $p^2 + m^2 = 0$ in the free-field case.
Therefore our final expression is,
\be\lab{fifi5} \la \phi(0,0) \phi(r_h,\vec{y})\ra
 = \le (\int d\tau_m J_m  A_{\xi_*}(0,\tau_m)
 e^{\tau_m H_{IR}(\xi_*)}\ri) \Psi_{\xi_*}^*(r_h,\vec{y})
 \ee
Dependence on the matching point $r_m$ of the UV and IR  regions
is hidden in the Jacobian $J_m$. Our ignorance  about the UV
region of the target space is summarized by the function
$A_{\xi_*}(0,\tau_m)$. More generally this may be replaced by a
sum over the on-shell states $\xi_*$ as will be the case for the
string propagation below.

\paragraph{Closed string case:}

Having outlined the procedure for the simpler case of quantum
field theory, let us now consider the closed string propagation.
As we argued in the beginning of this section, the one-point
function $\la P[0]\ra$ is given by the propagator $\la \Psi_i,
0,\vec{0}| \Psi_f, r_h,\vec{x}_f\ra$ with some initial state
$\Psi_i$ on the boundary that corresponds to the Polyakov loop at
the transverse point $\vec{0}$, and some final state $\Psi_f$ at
the horizon at a transverse point $\vec{x}_f$. In the end of the
computation everything that is not determined by the boundary
condition at $r=0$ should be summed over. In particular we should
integrate over $\vec{x}_f$. When comes to $\Psi_f$ the situation
is as follows: In the CFT language, the path integral we want to
compute is a sphere diagram with two insertions of vertex
operators $V_{\Psi_i}(\sigma_1,\tau_1)$ and
$V_{\Psi_f}(\sigma_2,\tau_2)$. These operators should be defined
in the full CFT. Quite generally, the $PSL(2,C)$ invariance of the
sphere 1) allows to fix locations of the insertion points
$(\sigma_1,\tau_1)$ and $(\sigma_2,\tau_2)$;  2) it restricts the
conformal weights $(h_f,\tilde{h}_f)$ of the operator $V_{\Psi_f}$
in terms of the ones of the initial state $(h_i,\tilde{h}_i)$.
Therefore the final state will be fixed automatically. However, we
will still have a sum because the matching procedure described
above will effectively yield a decomposition of $\Psi_f$ in terms
of the spectrum of the IR CFT.

What do we know  about the initial string state $\Psi_i$? It
should correspond to the Polyakov loop on the boundary gauge
theory. Clearly it should be a winding $w=1$ state in the full CFT
with zero transverse momentum $\vec{p}_\perp=0$ and zero Matsubara
frequency $k=0$\footnote{It is also reasonable to assume that it
corresponds to a state with no string excitation numbers but we
will not assume this.}. Apart from these we cannot say much. In
particular we do not know the precise form of the vertex operator
that corresponds to this state, as we do not know the details of
the full CFT. However, this ignorance will not affect our final
result.

\paragraph{Gauge fixing:}
One important  complication in comparison to the QFT case above is
that now we have two re-parametrization + one Weyl invariance on
the world-sheet that should be gauge fixed. String paths are
parametrized by the world-sheet coordinates $(\sigma,\tau)$ with
$\sigma \sim \sigma + 2\pi$. In the path integral we can fix the
two re-parametrizations by  fixing the world-sheet metric to be of
the form $h_{ab} = \hat{h}_{ab} e^{\sigma_L}$ with some reference
metric $\hat{h}$. The remaining freedom $\sigma_L$ is the
Liouville mode, which can be left as unfixed. It is
well-known that under quantum effects $\sigma_L$ becomes a
space-like dimension and the target space becomes flat with one
additional dimension plus  a linear dilaton. It  was further shown
in \cite{Chemseddine}---see appendix \ref{lindilNC} for a
review---that for non-critical strings on flat $d-1$
dimensional target-space to make sense at higher genera, one is
forced to introduce an extra {\em world-sheet} field $\phi$ that
couples to the world-sheet Ricci scalar. Then, after fixing the
re-paratmetrization invariance, the Liouville field $\sigma_L$
combines with $\phi$ to produce two extra dimensions on the target
space plus a linear dilaton field. Thus the end-result is
non-critical string theory in $d+1$ dimensional flat target-space
with a linear dilaton $\f$, which is exactly what we have in the
IR in the model of section \ref{model}  for $r_h\gg 1$.  Another option is
to make sure that the CFT has vanishing total central charge. In this case
$\sigma_L$ decouples and one has fixed the entire gauge symmetry on the world-sheet.
The background becomes linear-dilaton in the range $r\gg1$ for $r_h\to\infty$.
The two options are totally equivalent and for definiteness let us adopt the latter
option. Then we start with $d+1$ dimensions and we fix both re-parametrization and
Weyl.

At this point there are two options that one can choose to work
with. One can either keep the ghosts that arise from the
re-parametrization fixing or one can ignore them and include only
the transverse string fluctuations in the canonical formalism and
treat the propagation in the {\em light-cone gauge}. In the asymptotic linear-dilaton regime
it is known that the ghosts exactly cancel the excitations of the string along
the $r$ and the $x^0$ directions in the linear dilaton background,
just like the flat case \cite{Chemseddine}. We will assume that this is true also in
the more general case when we have the correction terms in
(\ref{Vs}).

The calculation becomes more transparent in the light-cone gauge
which can easily be generalized to the linear-dilaton background
\cite{Myers, Chemseddine}. Here, one ignores the contribution of
the re-parametrization ghosts and fixes the metric $h_{ab} =
\hat{h}_{ab} e^{\sigma_L}$ by hand. There is a residual freedom
from the combination of diffeo-Weyl that leaves $\hat{h}$
invariant which can be fixed by, \be\lab{lcg} X^+  =  p^+ \tau +
x^+, \qquad X^{\pm} = \frac{1}{\sqrt{2}}(X^0 \pm r). \ee $X^-$ is
also fixed through the Virasoro constraint and one is left with
only the transverse oscillators along $X^i$.

\paragraph{Calculation:}

For the purpose of identifying the contribution from the IR region
we divide the propagation into two parts. The procedure we
outlined for the field-theory case to separate the paths in the
two different regions and to sew the propagators at the matching
region $r=r_m$ has a direct generalization to the closed string
propagation in our background: One only has to \begin{enumerate}
\item replace the Hamiltonians in (\ref{fifi3})with the Virasoro
generators $H = L_0 + \tilde{L}_0$; \item extend the integration
over $\tau$ to a complex parameter $w = \tau + i\sigma$ whose
imaginary part couples to $L_0 - \tilde{L}_0$, hence the integral
over it produces the left-right matching of the on-shell states;
\item replace the wave-functions $\Psi_\xi$ and
$\tilde{\Psi}_{\xi'}$ of the IR and UV Hamiltonian with vertex
operators in the CFTs.
%\item generalize the Jacobian in
%(\ref{fifi1}) to a two-dimensional one $J_m \to \det(\6(r,
%x^0)/\6(\sigma, \tau))$ where $x^0$ is the compact time circle.
\end{enumerate}
%As we are interested only in the far IR region, and in particular
%how the result scales with $r_h$, the details of the matching
%procedure will be irrelevant.

%The $\sigma$ dependence on the world-sheet is trivial, therefore
%it is better to define the propagation of the string only in
%reference to the world-sheet time $\tau$. This is managed by
%creating states by the ``averaged vertex operator" \cite{GSW1}
%\be\lab{avo} |\chi(\tau)\ra = V_\chi(\tau)|0\ra \equiv
%\int_0^{2\pi} \frac{d\sigma}{2\pi}
%V_{\chi,L}(\tau-\sigma)V_{\chi,R}(\tau+\sigma) |0\ra, \ee where L
%and R refer to holomorphic and anti-holomorphic sectors of the
%CFT. It is also useful to factor out the center-of-mass part,
%\be\lab{factor} V_{\chi}(\tau) = e^{i

\begin{figure}
 \begin{center}
\includegraphics[height=6cm,width=9cm]{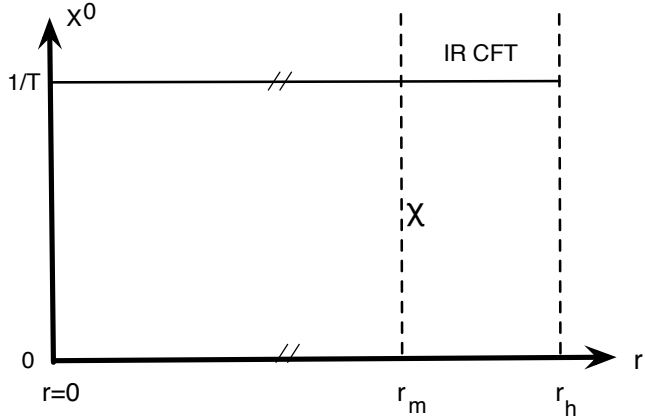}
 \end{center}
 \caption[]{The string world-sheet that wraps the time-circle
 and hangs from the boundary to the horizon as the dual of the
 one-point function of the Polyakov loop.
 The Euclidean time is identified as indicated by the dashes.
 We insert a complete set of string states $\chi$ at an
 intermediate point $r_m$ chosen close enough to $r_h$, for large $r_h$,
  such that the IR CFT description holds for $r>r_m$.} \label{fig1}
 \end{figure}

Formally we first decompose the path integral from $\tau=0$ to
$\tau_m$ and $\tau_m$ to $\infty$: \bea\lab{sisi1} \la
Psi_i,0|\Psi_f,\infty\ra\equiv \la \Psi_i, 0,\vec{0}| \Psi_f,
r_h,\vec{x}_f\ra &=& \int_{bc_i, bc_f}\!\!\!\!\!\! {\cal D} X^\mu
e^{-{\cal A}[X] - {\cal A}[\sigma_L]}\\ {}&=& \int  d
X^\mu_m(\sigma) \int_{bc_i, bc_m} \!\!\!\!\!\! {\cal D} X^\mu
e^{-{\cal A}[X]} \int_{bc_m, bc_f} \!\!\!\!\!\! {\cal D} X^\mu
e^{-{\cal A}[X]}\nn, \eea where the boundary conditions are
defined by the sets, \bea\lab{bcs} bc_i &=& \{ X^\mu(0,\s) =
X_i^\mu(\s): X_i^0(\s+2\pi) = X_i^0(\s) +
\frac{1}{T}; \vec{X}_{i,0}= r_0 = 0\}\\
bc_m &=& \{ X^\mu(\tau_m,\s) = X_m^\mu(\s)\}\\
bc_f &=& \{ X^\mu(\infty,\s) = X_f^\mu(\s): \vec{X}_{f,0}=
\vec{x}_f, r_0 = r_h \}.\eea
Quantities with subscript 0 refer to the center-of-mass positions
in the first and the last lines.

The r-component of the intermediate string $r_m(\s)$ has a
center-of-mass piece \be\lab{compiece} r(\tau_m,\s) =
\int_0^{2\pi} \frac{d\s}{2\pi} r(\tau_m ,\s) + \cdots \equiv r_0 +
\cdots \ee Just as in the field theory computation above, we can
use the freedom to choose $\tau_m$ to replace the integration over
$r_0$ with an integration over $\tau_m$ by inserting \be\lab{FPs}
1 = \int d\tau_m \delta(r_0(\tau_m) - r_m) J_m,\qquad J_m^{-1} =
\frac{dr_0}{d\tau}\big|_{r_0(\tau)= r_m}\ee inside the integral
over $X_m^{\m}(\s)$. Now, we assume that only the paths with
crossing number $c_m=1$ dominate the path integral. This is
certainly the case for $r_m$ chosen close enough\footnote{We
remind that the ``crossing number" was defined in the field theory
analogy above.  This assumption can be lifted as explained in the
field theory analogy above by modifying the procedure of dividing
paths. This is an unimportant detail however, which has no effect
on the final result. Here we just restrict the analysis to $c_m=1$
paths for simplicity.} to $r_h$ for $r_h\gg 1$. This means that,
\be\lab{cm1} r(\tau') > r(\tau), \qquad \tau'>\tau, \qquad for\,\,
\tau\geq \tau_m, \ee where \be\lab{rtau} r(\tau) \equiv
\int_0^{2\pi} \frac{d\s}{2\pi} r(\tau ,\s).\ee Therefore we
achieved the division of path integrals from $r_0=0$ to $r_0 =
r_m$ and $r_0= r_m$ to $r_0=r_h$. We will approximate the second
class of paths by replacing the action with that of the
linear-dilaton CFT: \be\lab{sisi2} \la \Psi_i, 0| \Psi_f, \infty
\ra \approx \int_0^{\infty} d\tau_m J_m \int dX^{'\mu}_m(\sigma)
PI(0,\tau_m) PI_{IR}(\tau_m,\infty), \ee where the first path
integral is \be\lab{1PI} PI(0,\tau_m) = \int_{bc_i,
bc_m}\!\!\!\!\!\! {\cal D} \sigma_L {\cal D} X^\mu e^{-{\cal A}[X]
},\ee with ${\cal A}$ the full world-sheet action and the second
one is \be\lab{2PI} PI_{IR}(\tau_m,\infty) = \int_{bc'_m,
bc_f}\!\!\!\!\!\! {\cal D}X^\mu e^{-{\cal A}_{IR}}.\ee The primes
in (\ref{sisi2}) and (\ref{2PI}) denote omission of the
center-of-mass piece in $r(\tau_m)$ as it is fixed to $r_m$ by
(\ref{FPs}).

The approximation in (\ref{sisi2}) is two-fold: First of all we
approximate the action in the region $r(\tau)>r_m$ by the IR CFT:
\be\lab{IRac} {\cal A}_{IR} = \frac{1}{4\pi \a'} \int_0^{2\pi}
d\s\int_{\tau_m}^{\infty} d\tau \sqrt{\hat{h}} \le[ \hat{h}^{ab}
\6_a X^\mu \6_b X^{\nu}\eta_{\m\n} + 4\a' v_\m X^\m \hat{R} + 2\a'
b_{ab} \nabla^a c^b\ri], \ee where we idsplay explicitly the
reparametrization ghosts. The proportionality factor in the
dilaton in our case is given by (\ref{ldcftus}). We consider the
bosonic linear-dilaton theory for definiteness; the final result
(in the semi-classical approximation that is to be defined below)
is independent of the particular linear-dilaton CFT chosen.
%Here we used the non-trivial
%relation between the flat non-critical string in $d$ dimensions
%and linear-dilaton in $d+1$ dimensions, appendix \ref{lindilNC}
%\footnote{The appendix discusses the more general case of
%including also a  scalar coupling $\phi$ on the world-sheet. In
%this case one should start from a $d-1$ dimensional flat NCST
%instead of $d$-dimensional case as in here. These two theories are
%equivalent for the tree diagrams we discuss in this section.}.

The second approximation is that we restrict the analysis to the
$c_m=1$ paths. Both of these approximations become better as the
point $r_m$ is chosen closer to $r_h$. In fact, the second
assumption is automatically satisfied in the light-cone gauge
(\ref{lcg}) where the role of $r(\tau)$ is played by $X^+(\tau)$.

Now, we focus on the second path integral $PI_{IR}$. In the
canonical formalism this can be written as, \be\lab{sisi3}
PI_{IR}(\tau_m,\infty) = \sum_{\chi\in {\cal H}_\perp} \la
V_\chi(X_m, \tau_m) V^*_\chi(X_f, \infty)\ra \Delta_{IR} (\chi),
\ee where $\chi$ runs in the transverse Fock space of the
linear-dilaton CFT and $V_\chi(X,\tau)$ denotes the
vertex-operator for creating a closed string $X$ at world-sheet
time $\tau$ in the $\chi$ eigenstate of the Hamiltonian $L_0+
\tilde{L}_0$ \cite{GSW1}: \be\lab{vos} V_\chi(X,\tau) =
\int_0^{2\pi} \frac{d\s}{2\pi} V_L(\chi,
X_L(\tau-\sigma))V_R(\chi, X_R(\tau-\sigma)). \ee The eigenstates
$\chi$ of our linear-dilaton CFT are labelled by the
center-of-mass momenta in r and transverse directions $p_r$,
$\vec{p}_\perp$, the Matsubara frequency $k$ and the left (right)
oscillator numbers $N$ ($\tilde{N}$).

%\footnote{We note that winding is not conserved in
%the Euclidean black-hole phase, hence all of the physical states
%have $w=0$.}
The propagator of a state $\chi(p_r,
p_\perp,k,w, N,\tilde{N})$ is then given by \be\lab{opf}
 \Delta_{IR} (\chi)  = \int_{|z|<1}
 \frac{d^2z}{|z|^2} z^{L_0(\chi)-1} \bar{z}^{\tilde{L}_0(\chi)-1},
\ee
where the Virasoro generators are given by (\ref{Virasoro}).
%To be definite we perform the calculation in bosonic CFT with
%periodic boundary conditions on all $X^{\mu}$. The end result does
%not depend on the particular linear-dilaton CFT chosen.
%Although we do not know much about the IR CFT, it is reasonable to assume that it has similar properties to the linear-dilaton CFT in bosonic (or fermionic) non-critical string theory, because the background in the IR asymptotes to a linear-dilaton background  in the limit $r_h\gg 1$.
%Let us then first review some facts about the linear-dilaton CFT\cite{Myers}.
In the propagator (\ref{opf}), the integral over $z$ projects the
states on the mass-shell (\ref{mashel1}), (\ref{mashel2}).
% onto
%the states with $L_0=\tilde{L}_0$ i.e. the states that satisfy
%(\ref{mashel2}). The modulus integral then yields $(L_0+
%\tilde{L}_0 - 2a)^{-1} = (m_r^2 + p_r^2)^{-1}$ by (\ref{mashel2}).
%For notational simplicity we defined, \be\lab{mr} m_r = \sqrt{ p_i
%p_i + \frac{2}{\a'}(N + \tilde{N} - 2a) + (2\pi k T)^2 +
%\le(\frac{w}{2\pi T \a'}\ri)^2 }. \ee

The IR path integral (\ref{sisi3}) contains the $r_h$ dependence
that we seek for, inside the vertex operator for the final state.
It is contained in the center-of-mass position term in the
r-direction, see (\ref{bcs}). Let us make it explicit by factoring
out
%It is useful to factor out the center of mass part in (\ref{vos}) for the final-state as,
\be\lab{vosf} V^*_\chi(X_f,\infty) = e^{-ip_r r_h}
\overline{V}_\chi^*(X,\infty), \ee where $\overline{V}$ contains
no dependence on $r_h$. On the other hand, the sum over $\chi$ in
(\ref{sisi3}) contain integrals over $p_r$ and $p_\perp$ and sums
over $k$, $w$ $N$ and $\tilde{N}$. Noting also that the integral
over $z$ in (\ref{opf}) projects onto the mass-shell states
(\ref{mashel1}) and (\ref{mashel2}) one can directly perform the
integral over $p_r$ in (\ref{sisi3}) and find, \be\lab{sisi4}
PI_{IR}(\tau_m,\infty)  = \sum_{\chi} C_\chi
 e^{-ip_r^*(\chi)(r_h-r_m)},
\ee
%where the  proportionality factor depends on $(r_m-r_i)$ but
%this dependence will be subleading with respect to the exponent in
%the large $r_h$ limit.
%The subscript instructs us to include only
%the states that conserve world-sheet momenta and
where the constant $C_\chi$ does not depend on $r_h$ and  $p_r^*(\chi)$
denotes the solution of the mass-shell condition (\ref{mashel1}):
\bea\lab{pstar} p_r^* &=& -i m_0 \le(1 +
\sqrt{1+\frac{m_*^2(\chi)}{m_0^2}}\ri),\\
m_*^2 &\equiv&
\frac{2}{\a'}\le(N+\tilde{N} - 2\ri) + p_\perp^2 + (2\pi kT)^2+\le(\frac{w}{2\pi T \a'}\ri)^2.
\lab{mstar}
\eea
%It is reasonable to assume that the overlap ${\cal
%F}(\chi,\Psi_f,r_h)$ is finite for all $\chi$ and $r_h$. This just
%means that any state $\chi$ inserted at $r_m$ has a finite
%probability to fall on the horizon. The integral over $x_f$
%restricts to the states with zero transverse momentum $p_\perp
%=0$. This is just momentum conservation in the transverse
%directions, as the initial state $\Psi_i$ had zero transverse
%momentum. One obtains a generic form,
Substituting (\ref{sisi4}) in (\ref{sisi2}) we find that the
entire $r_h$ dependence of the Polyakov loop becomes,
\be\lab{sisi5} \la P[0]\ra \propto \sum_{\chi} {\cal
 C}(\chi) e^{-i p_r^*(\chi)r_h},
 \ee
 where the states $\chi$ have zero transverse momenta and ${\cal
 C}$ is some c-number whose value is independent of $r_h$. This
 result can be thought of a direct generalization of the field
 theory analog in (\ref{fifi5}). The exponential term above is the
 analog of the wave-function $\Psi_{\chi^*}$ and the coefficient
 $C_\chi$ above is the analog of the expression inside the
 brackets in (\ref{fifi5}). The important difference is that here
 we have an infinite sum over all possible on-shell states of the
 string.

The expression inside the square-root in (\ref{pstar}) has the
following behavior for the various physical states. For any state
other than the tachyon ($N=\tilde{N}=k=w=0$) it is larger than 1.
For the tachyon it equals, \be\lab{sqrtt} 1+\frac{m_*^2}{m_0^2} =
\frac{1-d}{25-d}. \ee Thus it is negative for any non-zero spatial
dimension $d-1$. For a ``winding tachyon" $N=\tilde{N} = 0$,
$k>0$, its sign is determined by the value of the temperature $T$.
This latter case will prove important in the evaluation of the
correlation length in section \ref{tpfquant}. Here, the important
point is to realize that (\ref{sisi5}) is always dominated by the
tachyonic ground state in the limit $r\to r_h$ because all of the
higher states result in bigger suppression in the exponential.

For the tachyon on the other hand the square bracket in
(\ref{pstar}) is oscillatory, thus it gives an imaginary
contribution to the exponent in (\ref{sisi5}) and the modulus of
$\la P[0] \ra$ is determined by the first term in (\ref{pstar}),
\be\lab{opf4} \lim_{r_h\to\infty} |\la P[0]\ra| \propto e^{-m_0
r_h}. \ee We can now translate the variable from $r_h$ to the
reduced temperature $t$ using (\ref{D18}) near the transition
region, \be\lab{trns} e^{-m_0~r_h} = t^{\frac{1}{\kappa}}. \ee
Consequently, we obtain the scaling of magnetization as,
\be\lab{magnacarta} |\vec{M}| \propto |P[C]| \propto
t^{\frac{1}{\kappa}}. \ee In particular, for the second order
transition $\kappa = 2$ one obtains the mean-field scaling,
\be\lab{magnacartaMF} |\vec{M}| \propto t^{\frac{1}{2}}. \ee This
provides a non-trivial check on the proposed duality. Although
this is a stringy computation in principle, we observe that the
scaling exponent is determined by the lowest lying fluctuation of
the string that always correspond to the tachyon in a
linear-dilaton CFT in arbitrary dimension. As this is a gravity
mode, it is quite reasonable to expect mean-field scaling by the
arguments in section \ref{what}. The computation above confirms
this expectation non-trivially.

It can also be shown that all of the arguments that we made
throughout the derivation readily extends to the other
linear-dilaton CFTs, including the fermionic ones. This is because
there always exists a tachyon in the spectrum for which the
square-root in (\ref{pstar}) becomes imaginary, whereas it becomes
real for all of the other states in the physical spectrum. This
result can easily be understood intuitively: in the transition
region $T\approx T_c$ where the linear dilaton CFT governs the
scaling behavior, the existence of the tachyon signals
instability, hence phase transition, and it is this tachyonic
state which dominates and determines the scaling law of
observables. In section (\ref{tpfquant}) we provide another
example of this phenomenon.

There are various possible modifications of the mean-field result.
First of all, it would be interesting to see whether going beyond
the semi-classical approximation would modify the critical
exponent. For this one has to sum over all of the string states
instead of focusing only on the dominant tachyonic contribution.
It will be very interesting to obtain corrections to mean-field
scaling as a result of this computation. We hope to investigate
this issue in the future.

Second type of possible modification involves the $1/N$
corrections. In the calculation above we assumed that the boundary
value of the dilaton can be tuned strictly to zero, so that we can
ignore gravitational interactions.
%However, even for a tiny value
%of the dilaton VeV, the loop corrections to the string propagator
%would become important in the propagator and they are entitled to
One expects that the $1/N$ corrections modify the mean-field
scaling as, \be\lab{mfN} |\vec{M}| \propto t^{\frac{1}{2} +
\cO(N^{-2})}. \ee

%\subsection{The correlation length}\lab{ximf}

%As explained  in appendix \ref{statmech}, in the {\em mean-field
%approximation}, the correlation length and the magnetization are
%inversely related near $T_c$ (\ref{Msca}). Thus within this
%approximation, one can directly obtain the scaling of the
%correlation length from (\ref{mcase1}) and (\ref{mcase2}). Setting
%$V_s=4$ (\ref{VsMF}) as required by the mean-field approximation,
%we obtain,
%%%%%%%%%%%%%%%%%%%%%%%%%%%%%%%%%%%%%%%%%%%%%%%%%%%%%%%%%%%%%%%%%
%\bea
%  \mathrm{Case\,\, i:} \qquad \vec{\xi}_{MF} &\propto & t^{-\frac{1}{\kappa}}\qquad
%  t\to 0 \lab{xicase1}\\
%  \mathrm{Case\,\, ii:}\qquad \vec{\xi}_{MF} &\propto &  e^{\le(\frac{t}{C}\ri)^{-\frac{1}{\a}}}, \qquad
%  t\to 0 \lab{xicase2},
%\eea
%%%%%%%%%%%%%%%%%%%%%%%%%%%%%%%%%%%%%%%%%%%%%%%%%%%%%%%%%%%%%%%%%
%The second eq. is in principle valid for $d>2$, because the
%mean-field computation here relates $\xi$ to inverse of the
%magnetization which strictly vanishes for $d=2$. {\bf \Large
%Compare with the scaling in KT?}
%
%There is another method to compute the correlation length. The
%correlation length appears in the spin-spin correlator as in
%(\ref{sscorg}). This computation can be carried out on the gravity
%side, see section \ref{ssCor}, and it is in principle valid beyond
%the mean-field approximation. See section  \ref{ssCor} for further
%details.

\subsection{The two-point function}
\lab{ssCor}

As explained in section \ref{probes}, the spin-spin correlation
function is represented by an F-string solution on the gravity
side, (\ref{Pspin2}) that wraps the Euclidean-time circle and is
connected to two separate points on the boundary that we take as
$x$ and $0$. As in (\ref{opfclas}), we first compute this quantity
classically as a warm-up exercise. We then generalize to the
realistic case where one has to consider the full path integral.
The classical computation was first carried out in the case of AdS
black-holes in \cite{PP1}\cite{PP2} (see also \cite{YK}).

\subsubsection{Warm-up: classical computation}
\lab{tpfclass}

We will perform the computation in the black-hole phase (that
corresponds to the super-fluid phase in the XY-model). The
computation in the TG phase is very similar and as we are
essentially interested in how the correlator scales near $T_c$,
the two results will yield very similar results. Although all of
the NS-NS fields $G_{\mu\nu}$, $B_{\mu\nu}$ and $\f$ couple to the
F-string, we show that the G-coupling yield the dominant term in
App. \ref{AppPP}. Thus one can replace the F-string action with
the Nambu-Goto action in this section, unless specified otherwise.

\begin{figure}[h!]\begin{center}\begin{tabular}{ccc}
\includegraphics[width=5cm]{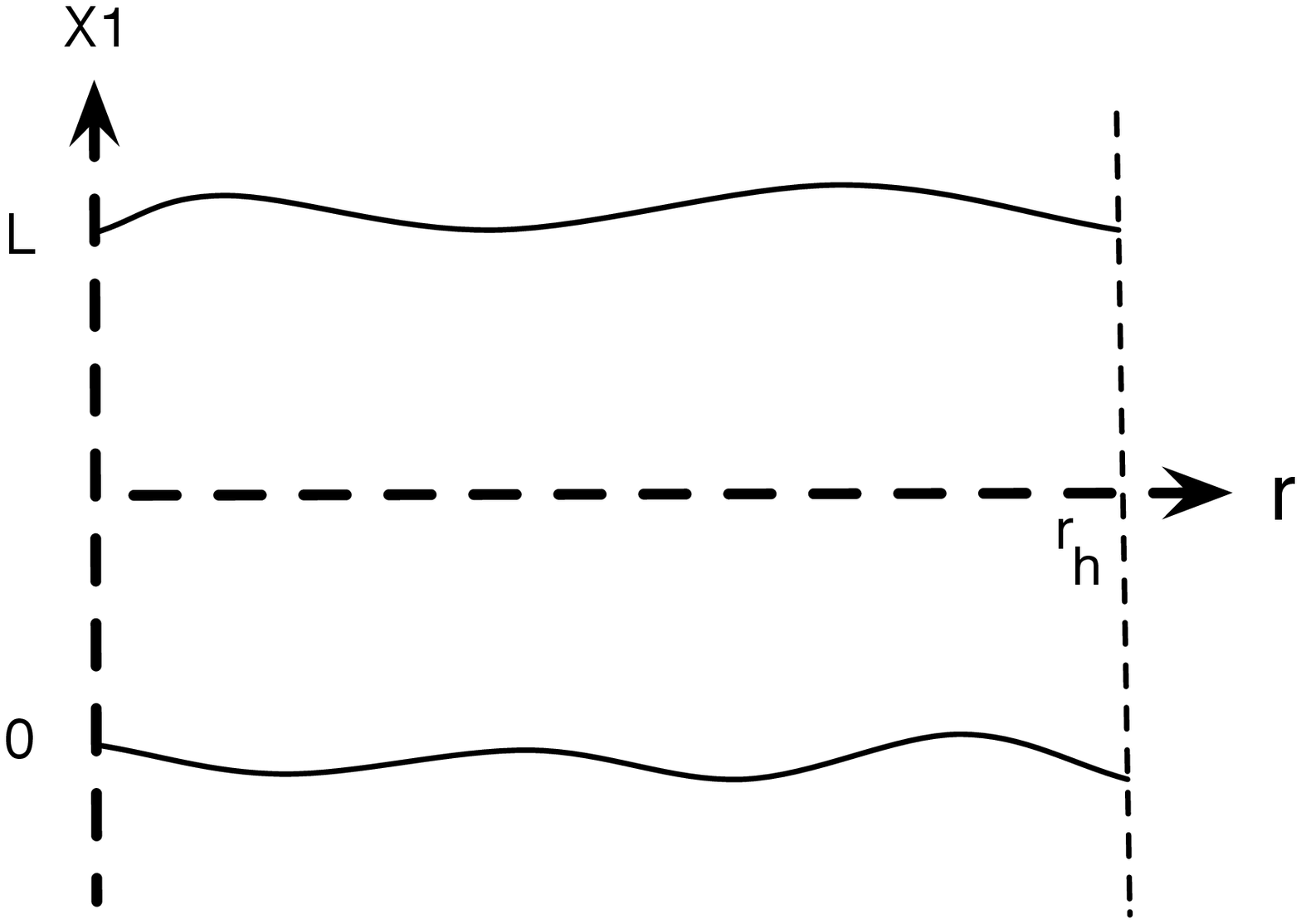} &
\includegraphics[width=5cm]{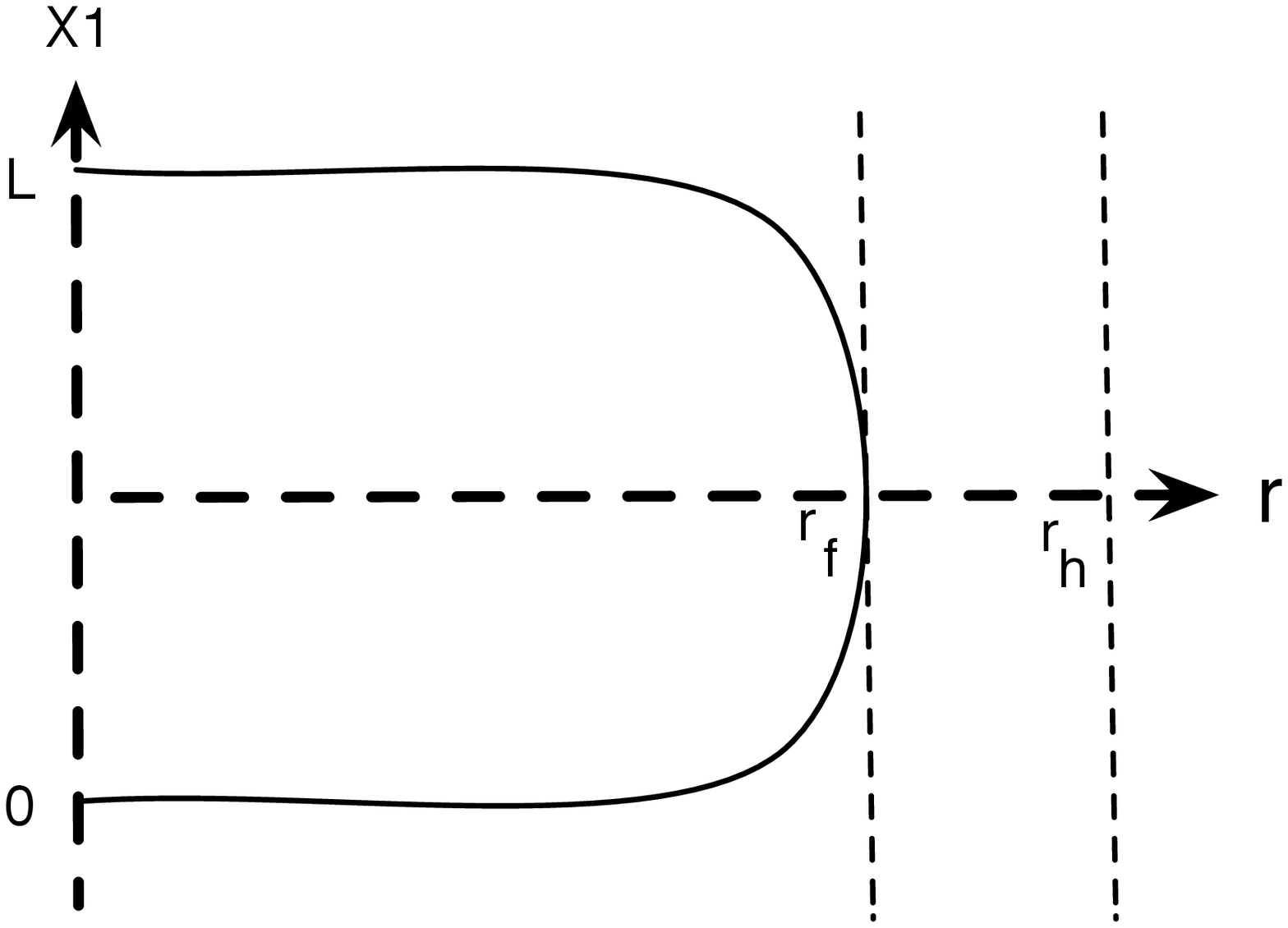} &
\includegraphics[width=5cm]{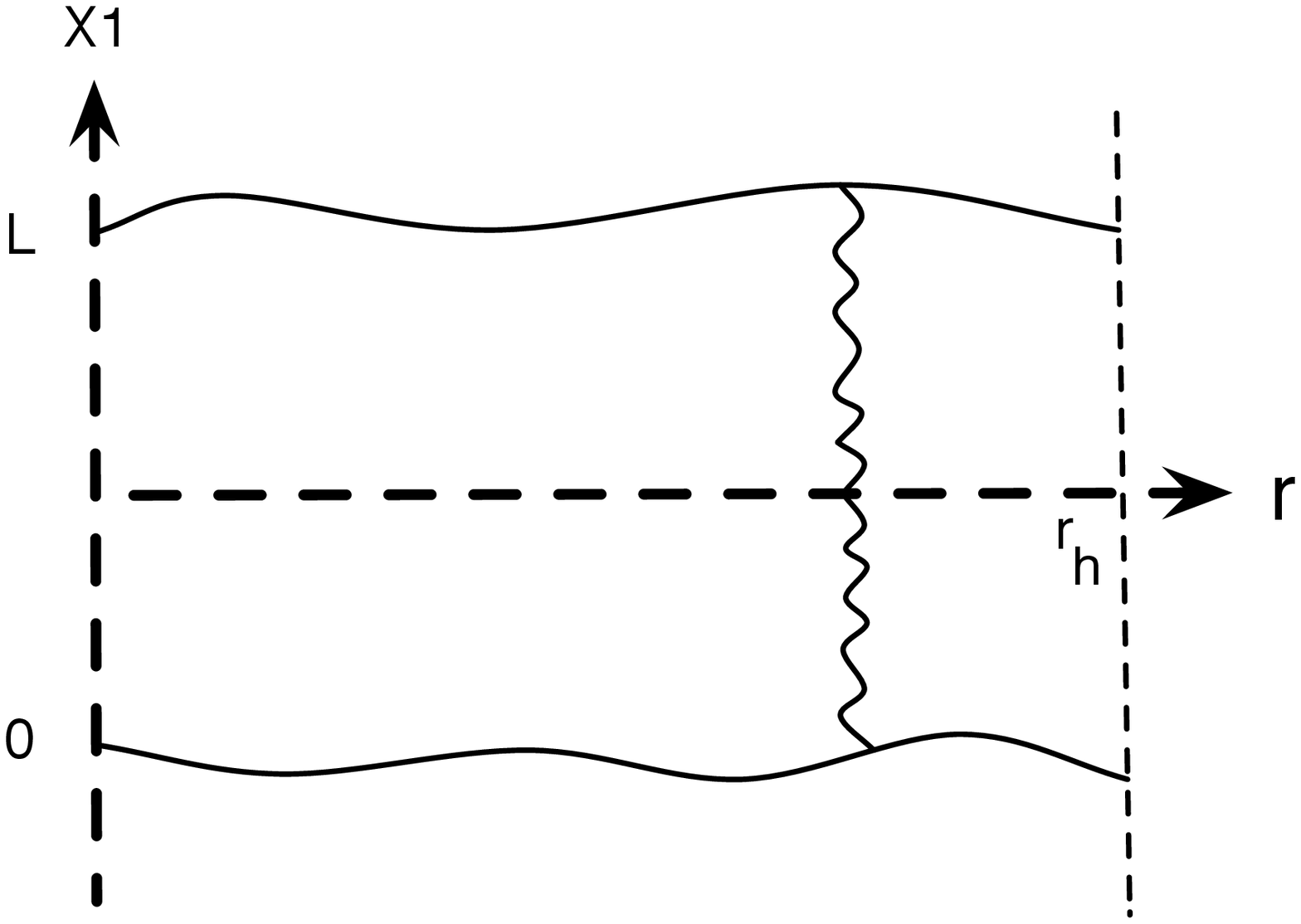}\\
(a) & (b) & (c)
\end{tabular}\end{center}
\caption{The classical string saddles that contribute the
spin-spin two point function in the ordered phase of the
ferromagnet. (a) The disconnected diagram. Strings are falling on
the horizon.  Its value is proportional to the square of the
magnetization expectation value $|\vec{M}|^2$. (b) Connected
string diagram. (c) A bulk-mode exchanged between two disconnected
strings.}\label{fig2}
\end{figure}

There are three string embeddings that contribute \cite{YK}, see
figure (\ref{fig2}):
\begin{enumerate}
\item {\bf Disconnected diagram}: This consists of two straight
strings that connect to the boundary at points $x_1= L = |x|$ and
$x_1=0$ and hang from the boundary to the horizon. The result is
twice the NG action in (\ref{NG}), thus it is finite as long as
$T>T_c$. As $T\to T_c$ it diverges as in (\ref{NG2}), that
corresponds to the fact that the magnetization vanishes smoothly
as $T\to T_c$. Thus the disconnected contribution yields,
\begin{equation}\label{dis}
  \la \vec{m}(x)\cdot \vec{m}(0)\ra_{dis} = \la |P|\ra^2  =
  |\vec{M}|^2 = finite, \qquad T>T_c.
\end{equation}
We are not interested in the actual value of the disconnected
piece, it depends on the normalization of the Polyakov loop. As
mentioned before this piece is absent in the TG phase. As it is
disconnected, this piece is actually $\cO(g_s^{-2})$ enhanced with
respected to the interesting connected contributions \cite{YK}
that should be handled separately.

\item {\bf Connected diagram}: This is a differentiable world-sheet
whose end-points are connected to the boundary at points $x$ and
0. To compute its on-shell action we fix the gauge $\tau = x_1$,
$\sigma = x_0$ where $x_1$ is the coordinate on the boundary on
which the end-points of the string $L$ and $0$ lie. The result is
(see App. \ref{AppPP}):
\begin{equation}\label{con}
  S_{NG}^{con} = \frac{1}{2\pi \ell_s^2 T} \int_{\epsilon}^{r_f} dr
  \frac{e^{2A_s(r)}}{\sqrt{1-\frac{e^{4A_s(r_f)}f(r_f)}{e^{4A_s(r)}f(r)}}},
\end{equation}
where $r_f<r_h$ is the turning point of the string that
corresponds to $dr/dx_1=0$ and $\epsilon$ is a cut-off near the
boundary\footnote{For the sake of the discussion here, we do not
need to renormalize the action by subtracting counter-terms. This
can be done in a standard way, if desired.}. $A_s$ denotes the
scale factor of the metric in the string-frame, see (\ref{sBH}).

The length between the end-points of the string $L$ is given by,
\begin{equation}\label{Lrf}
  L = 2\int_{\epsilon}^{r_f} dx_1  = 2\int_{\epsilon}^{r_f} dr
  \frac{1}{\sqrt{f(r)}
  \sqrt{\frac{e^{4A_s(r)}f(r)}{e^{4A_s(r_f)}f(r_f)}-1}}.
\end{equation}
Eqs. (\ref{con}) and (\ref{Lrf}) parametrically define the
function $S^{con}_{NG}(L)$. One can easily show that $L$ is
monotonically increasing in $r_f$. When the distance $L$ reaches a
certain value that corresponds to $r_f= r_h$, the connected string
solution falls into the horizon, thus ceases to exist. Beyond this
point this diagram gets replaced by the ``exchange diagram" that
we discuss below.

We are eventually interested in the scaling of the spin-spin
correlation function near $T_c$. Therefore let us focus on the
limit $r_h\gg r_f\gg 1$. In this limit $T$ is very close to $T_c$
and at the same time $L$ is large. One can easily compute the
function $S_{NG}^{con}$ in this limit (see App. \ref{AppPP}) and
finds,
\begin{equation}\label{scon}
  S_{NG}^{con} \to m_T~L + \cdots, \qquad m_T \equiv \frac{1}{2\pi
  \ell_s^2 T_c}, \quad\quad (T\to T_c)
\end{equation}
where $m_T(t)$ defines an effective mass term, that stays finite
at all $T$ and whose value in the limit $t\to 0$ is shown above.
The ellipsis denote contributions that are sub-leading in $L$.
Thus, we can identify the first contribution to the spin-spin
correlator:
\begin{equation}\label{sscon}
  \la \vec{m}(x)\cdot\vec{m}(0)\ra_{con} \sim e^{-m_T~L + \cdots}
\end{equation}
Comparing with (\ref{sscor}), one indeed finds qualitative
agreement where $m_T^{-1}$ gives a {\em finite} contribution to
the spin-spin correlation length $\xi$, that stays finite {\em
even at} $T_c$. On the other hand, we expect $\xi$ to diverge at
$T_c$. To see how this divergence arises one has to perform the
full path integral computation that we turn in section
\ref{tpfquant}.
%This missing contribution that is {\em dominant} near $T_c$
%is identified below.

\item {\bf Exchange diagram}: When the curvature on the string
world-sheet becomes strong, one should also take into account the
fluctuations of the string, that are no-longer negligible. It was
first observed in \cite{YK} that this contribution yields a
crucial correction to the Polyakov correlator, that actually
resolved a puzzle that was encountered in \cite{PP1}\cite{PP2}: As
the connected contribution above ceases to exist beyond $r_f=
r_h$, one may naively think that the connected part of the
correlator $\la P^*(L) P(0)\ra$ vanishes beyond a certain value of
L. This is in contradiction with a generic QFT as the connected
piece of a generic correlator should be a convex function which
smoothly decreases with increasing $L$.

The missing contribution, in fact, comes from the world-sheet
fluctuations that become crucial in the regions where the
world-sheet curvature $R^{(2)}$ becomes large. In App. \ref{AppPP}
we show that  $R^{(2)}$ indeed becomes large near the horizon,
hence another type of connected diagram that arise from
world-sheet fluctuations become more dominant at $r_f=r_h$. This
diagram can be calculated in the limit $L$ becomes large. In this
limit, the contribution is given by the diagram that consists of
two disconnected world-sheets connected by the exchange of gravity
modes. In the large $L$ limit, the exchange is dominated by the
gravity mode with lowest mass. Thus, in this limit $\exp{-S_{NG}}$
is given by the propagator of the lowest mass gravity mode in
$d-1$ dimensions\footnote{The exchange mode propagates in $d-1$
dimensions because the propagator is fixed at a certain value of
$r$ and it's compactified on the time-circle.}:
\begin{equation}\label{prog}
  e^{-S_{NG}} \sim \frac{e^{- m_{min}~L}}{L^{d-3}},
\end{equation}
where $m_{min}$ is the lowest mass bulk mode.

Now, it is crucial to figure out which gravity modes contribute to
which parts of the correlator. As explained in \cite{YK}, in the
gauge theory, only the $C{\cal T}$-even modes couple to the real part
of $P[C]$, and $C{\cal T}$-odd modes to the imaginary part. Here
$\tau$ denotes reflections in Euclidean time. The analogous
statement in gravity is that only the $C{\cal T}^+$ bulk modes are
exchanged in the part of $S_{NG}^{exc}$ that corresponds to $\la
Re~P Re~P\ra$ and the $C{\cal T}^-$ bulk modes are exchanged in the part
that corresponds to $\la Im~P Im~P\ra$. Using the identification
(\ref{idt21}) and (\ref{idt22}), we find that,
%%%%%%%%%%%%
\bea\lab{exc1} \la \vec{m}_{\parl}(x)\cdot
\vec{m}_{\parl}(0)\ra_{exc} & \sim & e^{-m_+~L}/L^{d-3}, \\
\lab{exc2} \la \vec{m}_{\perp}(x)\cdot \vec{m}_{\perp}(0)
\ra_{exc} & \sim & e^{-m_-~L}/L^{d-3},\eea
%%%%%%%%%%%%
where $L=|x|$, the $m_{\pm}$ are the lowest masses of the bulk
modes in the $C{\cal T}^{\pm}$ channels and the result is valid in
the limit $L\gg 1$.

The NS-NS modes in the $C{\cal T}^+$ channel with their $J^{C{\cal T}}$
designations are, $G_{00}$ $(0^{++})$, $G_{ij}^{TT}$ $(2^{++})$,
$\f$  $(0^{++})$, $B_{ij}$ $1^{--}$, $G_{ii}$  $(0^{++})$. One
should solve the fluctuation eqs. on the BH background in order to
figure out the lowest mass one.

This is done in Appendix \ref{Appflucs}, where we showed that
all of the masses that correspond to gravitational fluctuations are bounded from below and
non-zero, i.e. there is a mass-gap in the $C{\cal T}^+$ channel
that is given by $m_0 = \sqrt{V_\infty}/2$, equation (\ref{spec}).
Hence, in the large distance limit $L\gg 1$ the exchange diagram gives the
contribution,
%\be\lab{excres}
%\vec{m}_{\parl}(0)\ra_{exc} & \sim & e^{-m_0~L}/L^{d-3}.
%\ee
\begin{equation}\label{ssexcparl}
  \la \vec{m}_{\parl}(x)\cdot\vec{m}_{\parl}(0)\ra_{exc}
  \sim \frac{e^{-m_o~L + \cdots}}{L^{d-3}}, \qquad L\gg 1.
\end{equation}
%where $m_+(t)$ is expected to behave as in (\ref{mplus}).

On the other hand, the NS-NS modes in the $C{\cal T}^-$ channel with
their $J^{C{\cal T}}$ designations read, $G_{i0}$ $(1^{+-})$, $B_{i0}$
$(1^{-+})$ and $B_{r0}$ $0^{-+}$. Here, the crucial observation is
that, as explained in section \ref{sec3}, the zero-mode of the
latter is nothing other than the field $\psi$ in (\ref{psi}) that
was identified with the {\em Goldstone mode}! Thus the lowest
lying mass in the $C{\cal T}^-$ channel is the zero mode of the
$B_{r0}$ field and it is zero: $m_- = 0$. Thus we find no
attenuation term in the corresponding part of the spin correlation
function:
\begin{equation}\label{ssexcperp}
  \la \vec{m}_{\perp}(x)\cdot\vec{m}_{\perp}(0)\ra_{exc}
  \sim \frac{e^{-m_-(t)~L}}{L^{d-3}}= \frac{1}{L^{d-3}},
\end{equation}
where the result is valid for large $L=|x|$, and at any
temperature $T>T_c$ which corresponds to the ordered phase in the
corresponding spin-system (we recall that the temperature on the
gravity side and the spin-model side are inversely related to each
other).
\end{enumerate}

Combining (\ref{dis}), (\ref{sscon}), (\ref{ssexcparl}) and
(\ref{ssexcperp}) we arrive at the total result for the spin
correlation function (in the large $L$ limit):
\begin{equation}\label{sscongr}
  \la \vec{m}(x)\cdot\vec{m}(0)\ra \sim \vec{M}^2 + c_1~e^{-m_T~L +
  \cdots} + c_2~\frac{e^{-m_0~L}}{L^{d-3}} + \frac{c_3}{L^{d-3}},
\end{equation}
where $c_i$ are some constants.
%As explained above $m_+$ is
%supposed to vanish in the limit $t\to 0$, (\ref{mplus}) whereas $m_T$ is always finite.

Comparison with the mean-field spin-model result (\ref{sscor})
shows {\em perfect qualitative agreement for temperatures $T >
T_c$} (which corresponds to the low T regime of the super-fluid).

However, we also observe that {\em the classical computation fails
to reproduce a very crucial feature of the spin-model at $T_c$}.
Namely, the longitudinal component of the two-point function
should in fact have a vanishing exponent as $T\to T_c$:
\begin{equation}\label{should}
  \la \vec{m}_{\parl}(x)\cdot\vec{m}_{\parl}(0)\ra
  \sim \frac{e^{-m_\parl(t)~L + \cdots}}{L^{d-3}}, \qquad with \,\,\, m_\parl(t) \sim t^\nu, \,\,\, as\,\,\, t\to 0.
\end{equation}
This corresponds to the fact that the longitudinal correlation
length also diverges at $T_c$. On the other hand, in our classical
string computation we found an exponent $m_\parl$ bound from below
as $m_\parl \to min(m_o, m_T)$ where $m_0$ and $m_T$ are given by
(\ref{spec}) and (\ref{scon}). We will argue below that the full
path integral computation of the two-point function yields the
desired result.

\subsubsection{Semi-classical computation}
\lab{tpfquant}

Now, we look  at the more generic situation when the assumption
$\ell/\ell_s\gg 1$ fails---as expected in non-critical string
theory---and the string path integral that corresponds to the
two-point function $ \la P^*(x) P(0)\ra =   \la \vec{m}(x)\cdot
\vec{m}(0)\ra$  is given by the full string path integral. As in
section ({\ref{opfquant}) we shall calculate this quantity in the
semi-classical approximation where we focus on the dominant
contribution of only the lowest lying string states at levels
$N=0$ and $N=1$.

One can again classify the string paths according to the three
classes as in figure \ref{fig2}. The contribution of disconnected
paths will be just as in section \ref{opfquant}. In the TG phase
they vanish because the area of the string paths are infinite and
in the BH phase they yield the square of the one-point function
found in section \ref{opfquant}. In addition one also have to
consider the disconnected paths corrected by bulk-exchange
diagrams as in figure \ref{fig2}. We shall consider the
contribution of these latter diagrams in the end of this section.
First we focus on the connected string paths, see figure
\ref{fig3}.

%We gauge fix by $\sigma= X^1$ and $\tau=X^0$.
The connected  path integral of the string is given by summing
over all paths the string can travel between the space-time points
$(r,x_1,\vec{x}_\perp)= (0,0,\vec{0})$ and $(r,x_1,\vec{x}_\perp)= (0,x,\vec{0})$. Here $\vec{x}_\perp$ denote the
coordinates transverse to $r$ and $x$.
We denote these points as
$I_{in}$ and $I_{out}$ respectively. As displayed in figure
\ref{fig3} for the connected classical saddle, in the limit $L\gg
1$ these paths can naturally be divided into three parts, see figure \ref{fig3}:
\begin{enumerate}
\item The paths between the space-time points
$I_{in}=(0,0,\vec{0})$ and $I_i$.
\item  The paths between the  points
$I_i$ and $I_f$.
\item  The paths between the  points  $I_f$ and $I_{out} =  (0,x,\vec{0})$.
\end{enumerate}

This division of path integrals into separate regions in
space-time is a non-trivial operation that is described at length
in section \ref{opfquant}. Here we will not go through the same
derivation again but only highlight the computation.

Just as in section \ref{opfquant} we write formally divide the
full path integral by inserting complete set of states at $I_i$
and $I_f$ as, \be\lab{tpf1} \la P^*(x) P(0)\ra_{conn} \approx \int
dI_i dI_f \sum_{\chi\in {\cal H}_{\perp}} {\cal
F}(\chi,\Psi_i,I_i)  \Delta_{IR} (\chi,I_i,I_f) {\cal
F}^*(\chi,\Psi_f,I_f), \ee where $\Psi_i$ and $\Psi_f$ denote the
initial and  final string-wave functions and the sum is over the
physical string states in the Fock-space of the string. The
function ${\cal F}(\chi,\Psi_i,I_i)$ denotes the overlap of the
string state $\Psi_i$ at the point $I_{in}$ and the state $\chi$
at the point $I_i$. Similarly ${\cal F}(\chi,\Psi_f,I_f)$ denotes
the overlap of the string state $\Psi_f$ at the point $I_{out}$
and the state $\chi$ at the point $I_f$. The approximation in
(\ref{tpf1}) is due to the assumption that the propagation in the
intermediate paths between $I_i$ and $I_f$ are governed by the IR
CFT as in section \ref{opfquant}. The properties of this IR CFT
are specified in section \ref{IRCFT}.

\begin{figure}[h!]\begin{center}
\includegraphics[width=9cm]{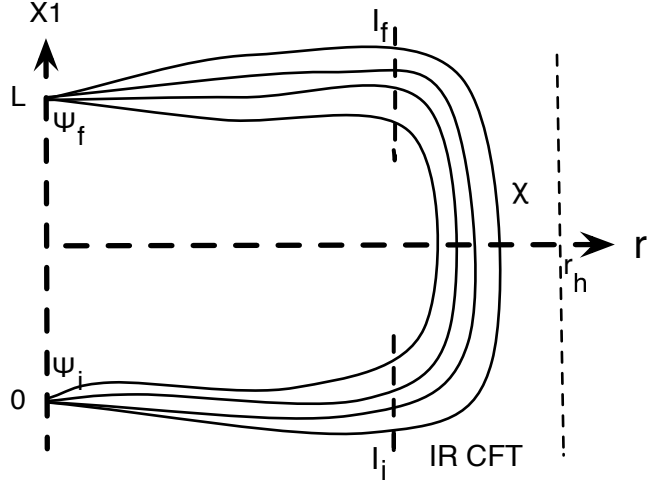}\end{center}
\caption{Quantum mechanical calculation of the spin-spin
correlator. The string paths are divided into three separate paths
by insertion of complete set of string states at $I_i$ and $I_f$.
For large $L$ and $r_h$ intermediate paths are governed by the IR
CFT.}\label{fig3}
\end{figure}

The propagation of the string in the paths 1st and 3rd class
depend on the full CFT, hence we cannot calculate. However, in the
limit $r_h\to\infty$ ($T\to T_c$) and for large values of $L$, the
dependence of the two-point function on $L$ is determined  by the
paths in the 2nd class where the propagation takes place in the
region where the IR CFT can safely be approximated by the
linear-dilaton CFT of section \ref{IRCFT}. Let us therefore
concentrate on the evaluation of this part of the path integral.

As discussed in section \ref{tpfclass}, one has to consider the
two parts of the correlator separately: \be\lab{tparts}
 \la P^*(x) P(0)\ra_{conn}  =  \la Re P(x) Re P(0)\ra_{conn}
  + \la Im P(x) Im P(0)\ra_{conn}.
\ee Only the $C{\cal T}^+$ ($C{\cal T}^-$) string  states
contribute in the real (imaginary) parts. First of all, we are
interested in determining the scaling of the longitudinal
correlation length $\xi(t)$ near $T_c$. For this purpose we focus
on the real part in what follows and return to the imaginary part
in the end.

\paragraph{Thermal gas phase:}

The propagation of closed string state $\chi$ from a point $x=0$
to\footnote{We drop the subscript in $x_1$ for notational
convenience.} $x=L$ in the linear-dilaton CFT is given by
\be\lab{tpf2}
 \Delta_{IR} (\chi,0,L)  \propto  \int dp_x~ e^{-i p_x(\chi) L}
\int_{|z|<1}
 \frac{d^2z}{|z|^2} z^{L_0(\chi)-1} \bar{z}^{\tilde{L}_0(\chi)-1},
\ee where $p_x(\chi)$ denotes the momentum of the state $\chi$ in
the $x$ direction. The exponential factor in (\ref{tpf2}) arise
from the center-of mass part of the vertex operator insertions at
$x=0$ and $x=L$, see section \ref{opfquant}.

 The integral over $z$ restricts to the
mass-shell states (\ref{mashel1}) and (\ref{mashel2}) on which the
momentum $p_x$ becomes,
 \be\lab{pxstar} p_x^* = -i \le( \frac{2}{\a'}\le(N+\tilde{N} - 2\ri)
 + p_\perp^2 +p_r^2 +2i m_0 p_r + (2\pi kT)^2
 +\le(\frac{w}{2\pi T \a'}\ri)^2 \ri)^\half,
\ee where $p_\perp$ now denotes the momenta transverse to the
x-direction in the spatial dimensions. As mentioned  before the
boundary condition for the closed string restricts only to the
states with winding number $w=1$. This cannot change along the
propagation  in the limit $g_s\to 0$, unless the string paths fall
onto the horizon, that does not happen for the connected string
diagram. We therefore focus on the case $w=1$. One should of
course satisfy the level-matching condition (\ref{mashel2}),
$N-\tilde{N} + k=0$.

 In order to read off the scaling of the
correlation length from the real part of the propagator then one
has to consider the contribution of all $C{\cal T}^+$ states that
has winding $w=1$. The propagation amplitude of a state $\chi$ in
this region is given by \be\lab{tpf22}
 \Delta_{IR} (\chi,0,L)  \propto  \int dp_r d^{d-2}p_{\perp}e^{-i p_x^*(\chi) L}.
\ee The momenta $p_r$ and $p_\perp$ in (\ref{pxstar}) are  to be
integrated over after substituting (\ref{tpf2}) in (\ref{tpf1}).
This can easily be seen to produce a factor of $L^{d-3}$. The
correlation length then can be obtained by state $\chi_{min}$ for
which the expression $-ip_x^*$ in (\ref{pxstar}) for
$p_r=p_\perp=0$  is minimum. \be\lab{corlenpx} \xi(t)^{-1} = i
p_x^*(\chi_{min})\bigg|_{p_\perp=p_r=0} \ee

%When it comes to the winding mode however, one has to be careful
%because the case of the thermal gas and the black-hole are
%qualitatively different in regard to winding. In the former case
%one has to specify to the winding sector $w=1$ because this is the
%winding number of the string on the boundary, and the winding
%number cannot change along the propagation because breaking or
%rejoining of a  string is an $\cO(g_s)$ process.

Consequently, in the thermal gas (disordered) phase the
correlation length is given by \be\lab{corTG1} \xi_{TG}^{-1} =
\le( \frac{2}{\a'}\le(N+\tilde{N} - 2\ri)
  +\le(\frac{1}{2\pi T \a'}\ri)^2 \ri)^\half\bigg|_{min}.\ee
The minimum $C{\cal T}^+$ state is clearly given by the tachyon
$N=\tilde{N}=0$. Level matching then sets $k=0$ and  one obtains
\be\lab{corTG2} \xi_{TG}^{-1}= \le( -\frac{4}{\a'}
  +\le(\frac{1}{2\pi T \a'}\ri)^2 \ri)^\half.\ee
For an arbitrary temperature this is a positive number. However we
observe that it vanishes precisely at the temperature when the
winding tachyon mode\footnote{This is a misnomer as the ``winding
tachyon" is actually massive for an arbitrary temperature above
Hagedorn.} becomes massless: \be\lab{Hagedorn} T_H^{b} =
\frac{1}{4\pi\ell_s}. \ee For critical strings in flat space-time,
this special radius was shown to correspond to the {\em Hagedorn
temperature} of the string. Furthermore \cite{AtickWitten} argued
that upon turning on an infinitesimal $g_s$ this point becomes a
{\em first order} phase transition in the string partition
function.

Here, similar arguments combined with our finding section
\ref{model}  that the linear-dilaton with mild subleading
corrections has a continuous transition would imply that, in the
case of a linear-dilaton background rather than the flat space,
the same point indicates a {\em continuous} transition into
formation of black-holes. In the limit of vanishing $g_s$, this
argument thus determines the phase transition temperature as,
\be\lab{trtemp} T_c\bigg|_{g_s\to 0} = T_H^{b} =
\frac{1}{4\pi\ell_s}. \ee From (\ref{corTG2}) it is also clear
that one obtains mean-field scaling for the correlation length,
\be\lab{mfsxi} \xi_{TG}(t) \to \frac{\ell_s}{2\sqrt{2}}
|t|^{-\half}, \qquad as \,\,\, t\to 0, \ee in the bosonic
linear-dilaton CFT. (We recall the definition $t= (T-T_c)/T_c$.)

In case of the fermionic CFT a similar calculation in the NS-NS
sector yields a limiting temperature  that we again propose to
coincide with the continuous Hawking-Page transition:
\be\lab{trtempf} T_H^{f} = T_c\bigg|_{g_s=0}  =
\frac{1}{2\sqrt{2}\pi\ell_s}, \ee with similar mean-field scaling
of the correlation function. On can consider variants of the CFT
by applying non-standard boundary conditions on the $r$-direction,
but the result does not change. As long as there exists a winding
tachyon one obtains mean-field scaling provided that one
identifies the transition temperature with the Hagedorn
temperature.

%We note that the non-winding tachyon does not cause any problem in the calculation because
%the boundary conditions for the string propagation---that corresponds to the Polyakov loop--- have
%winding number $w=1$ and in the limit $g_s\to 0$ winding cannot change along the propagation from
%boundary to boundary.
%This is another example showing the  big difference  between finite N and infinite N.
%In the former case

Our result for the real part of the connected contribution is then
summarized as, \be\lab{resTG} \la Re P(x) Re P(0)\ra_{TG} \propto
\frac{e^{-\frac{L}{\xi_{TG}(t)}}}{L^{d-3}}, \ee where the
correlation length near $T_c$ is given by (\ref{mfsxi}).

Now, we consider the imaginary part in (\ref{tparts}).  The only
difference is that now we have to sum over the intermediate states
with $C{\cal T}^-$ quantum numbers and with winding $w=1$. The
lowest lying such state is given by the NS-NS two-form
fluctuations in the $N=\tilde{N}=1$ level. For the $w=1$ state,
level matching again requires $k=0$. It becomes clear that this
contribution is subdominant with respect to (\ref{resTG}) in the
large $L$ limit: \be\lab{resTG1} \la Im P(x) Im P(0)\ra_{TG}
\propto \frac{e^{-\frac{L}{\xi_{0}(t)}}}{L^{d-3}}, \ee where
$\xi_0(t)$ asymptotes to a constant at the transition $\xi_0(t)\to
2\pi T_c\ell_s^2$. Using the value of $T_c$ obtained  in
(\ref{trtemp}) this constant is, \be\lab{xiNS} \xi_0(t)\to
\frac{\ell_s}{2}, \qquad as \,\,\, t\to 0. \ee Recalling that the
disconnected diagrams in the thermal gas phase vanish because they
correspond to paths with infinite area, our final result then is,
\be\lab{resTGfin} \la P^*(x) P(0)\ra_{TG} \propto
\frac{e^{-\frac{L}{\xi_{TG}(t)}}}{L^{d-3}} + \cdots, \ee with
correlation length near $T_c$ is given by (\ref{mfsxi}). The
ellipsis denote contributions from higher mass states in the
spectrum such as (\ref{resTG}).

As an aside we also observe that the classical computation  of the
previous section gave ``almost" the right answer except that it
missed the tachyonic contribution. One can check this by observing
that the contribution from the 1st level mass states is {\em
precisely} the same as   the classical result (\ref{sscon}),
(\ref{scon}). We already calculated this in (\ref{resTG1}) for the
B-field, and one has the same result for the other states at the
same level, namely the dilaton and the graviton fluctuations.

\paragraph{Black-hole phase:}

Let us now consider the Euclidean black-hole. Here as well, the
winding will be protected along the propagation of the closed
string as long as the string does not fall on the horizon. This is
indeed the case for the connected paths.
%The crucial difference between the one-point function section \ref{opfquant} is that there the string paths ended on the horizon, hence
%the winding was not expected to be protected.
The calculation of the connected paths is very similar to  the
thermal gas case above, with one important difference: here
$T>T_c$ and a naive application of (\ref{corTG1}) would give rise
to an imaginary correlation length. There is another important
difference though: the BH becomes linear-dilaton strictly in the
limit $r_h\to\infty$ and for any temperature less than $T_c$ we
expect additional contributions to the mass spectra.

From the space-time point of view there is a finite horizon for
finite $r_h$ and the computation of the mass spectrum---as
fluctuations in space-time follow from applying normalizable
boundary condition at the horizon. This generally gives rise to a
discrete spectrum and one expects a correction to the mass spectra
which is supposed to vanish only in the strict $r_h\to\infty$
limit, due to the presence of the horizon. The only invariant
quantity that would be a candidate for such a correction term then
is the {\em black-hole mass} $m_{BH}$. Consequently we expect that
the mass spectra be shifted {\em positively} by a term
proportional to $m_{BH}$ near $r_h\lesssim \infty$. On the other
hand it is easy to see that the ADM mass of the black-hole
\cite{GKMN2} is proportional to the string-frame Ricci scalar that
was computed in \cite{exotic1} and it was found that,
\be\lab{BHmass} m_{BH}^2 \propto R_s \propto e^{\kappa \f_h}
\ell_s^{-2} \propto t \ell_s^{-2}, \ee where the last relation
follows from (\ref{D17}). Consequently, the relation
(\ref{corTG2}) should be modified in the limit $r_h\to \infty$ in
Euclidean black-hole as, \be\lab{corBH1} \xi_{BH} \to \le(
-\frac{4}{\a'}
  +\le(\frac{1}{2\pi T \a'}\ri)^2 + c_{bh} t \ri)^\half,\qquad as\,\,\, t\to 0^+ \ee
where $c_{bh}>0$ is some constant that we cannot determine
unfortunately \footnote{One can try to obtain an effective action
for the winding tachyon on the Euclidean black-hole and determine
its spectrum by applying normalizability both in the UV and near
horizon. However, there will still be undetermined coefficients in
the effective action and the constant would not be determined. The
only way to determine it is to construct the small BH as a
marginal deformation of the linear-dilaton CFT and obtain the
exact spectrum.}. Thus we find, \be\lab{corBH2} \xi_{BH } \to
\sqrt{c_{bh}-8}~t^\half, \qquad as\,\,\, t\to 0^+. \ee It should
be checked that $c_{bh}>8$ for consistency of the  picture we
present here. However we do not have any means to check this at
present.

The result (\ref{corBH2}) summarizes the dominant contribution  in
the large L limit to the {\em connected} string paths. As the
tachyon is a
 $C{\cal T}^+$ state (\ref{corBH2}) yields the leading term in
 the real part of the connected two-point function:
 \be\lab{resBH1}
\la Re P(x) Re P(0)\ra_{conn} \propto
\frac{e^{-\frac{L}{\xi_{BH}(t)}}}{L^{d-3}}, \ee with
(\ref{corBH2}).

How about the imaginary part? Just as in the thermal gas case
above the leading connected  contribution to the imaginary part is
given by the $w=1$ NS-NS two-form with the same answer
(\ref{resTG}): \be\lab{resBH2} \la Im P(x) Im P(0)\ra_{conn}
\propto \frac{e^{-\frac{L}{\xi_{0}(t)}}}{L^{d-3}}, \ee with
$\xi_0$ behaving as (\ref{xiNS}) in the limit $t\to 0$.
%There are sub-dominant contributions  from the higher mass states, such as the graviton, the dilaton and the NS two-form  fluctuations.
%They end up giving a contribution of the form $\exp(-L\xi_0^{-1})/L^{d-3}$ where $\xi_0$ is given by (\ref{xiNS}).
%Consequently, the leading piece in the connected diagram is (\ref{})

Another crucial difference between the black-hole and the
thermal-gas is that, for a finite horizon (any $r_h$ other than
strict $r_h=\infty$) there are {\em disconnected} string paths and
similarly exchange diagrams of the sort we described in section
\ref{tpfclass}. The two disconnected string paths that fall into
the horizon as in figure \ref{fig2} give square of the one-point
function that we already calculated in section \ref{opfquant}:
\be\lab{disconBH} \la P(x)  P(0)\ra_{dis} \propto |\vec{M}|^2, \ee
where $|\vec{M}|$ is given by (\ref{magnacarta}) in the limit
$t\to 0$. This contribution is order $\cO(g_s^{-2})$ \cite{YK} and
it should be treated separately.

There are also {\em exchange diagram} contributions of the form
figure \ref{fig2}. These are of the same order in $g_s$ expansion
as the connected diagrams and their contribution is simply given
by the sum over all possible bulk modes  that can couple to the
disconnected string paths  between the boundary and the horizon,
see figure \ref{fig2}. As the disconnected string paths do fall on
the horizon, the string with a winding $w=1$ boundary condition at
$r=0$ can unwind at $r_h$ even when $g_s\to 0$. Therefore the
winding number is not conserved in the exchange diagrams in the
black-hole phase and we should include the contribution from the
non-winding $w=0$ states. Contribution of the exchange-diagram in
figure \ref{fig2} is then given by the propagator  of a bulk-mode
with a $d-1$ dimensional mass $m_\perp^2$:
\be\lab{bulkexc} \la
P(x)^* P(0)\ra_{exc, m_\perp} \propto L^{-(d-3)}e^{-\sqrt{(2\pi
kT)^2 + m_\perp^2}L}. \ee The dominant contribution is always
given by the $k=0$ modes, hence we consider the case $k=0$ in what
follows. Again we divide this into the real and imaginary parts
which receive contributions from the $C{\cal T}^+$ and $C{\cal
T}^-$ bulk states respectively.

We first focus on the imaginary part. The lowest mass  $C{\cal
T}^-$contribution is the fluctuations of the B-field that are
massless from the $d-1$ dimensional point of view:  $m_\perp=0$.
We refer to appendix \ref{Appflucs} for a derivation. Consequently
we obtain \be\lab{bulkexcBH} \la Im P(x) Im P(0)\ra_{exc} \propto
\frac{1}{L^{(d-3)}} + \cdots \ee This is the transverse part of
the spin-spin two point function in the superfluid (ordered) phase
and the fluctuation of the B-field in this phase is identified
with the Goldstone mode. There is no analogous term in the thermal
gas (disordered) phase because the disconnected string paths that
the exchange mode couple to, have infinite area and this
contribution vanishes.

Now we consider the real part.  The lowest mass  $C{\cal T}^+$
bulk mode is the tachyon and we have to compute its mass $m_\perp$
in the $d-1$ dimensional point of view. This is a dangerous mode
because contribution of a negative mass state to the two-point
function implies non-unitarity in the dual spin-system. The
analysis of the bulk spectrum can easily be turned into a
Strum-Liouville eigenvalue problem, see appendix \ref{Appflucs}.
In this appendix we show that, {\em in the interesting cases of
two and three spatial dimensions, the tachyonic mode can be
avoided only when $\kappa = 2$.} Quite conveniently the case
$\kappa=2$ corresponds to a {\em second order} transition---the
most interesting case! We conclude that consistency of the entire
analysis can only be established for second order phase
transitions in 2D or 3D (spatial).

Let us focus on this interesting case and denote the minimum value
of the $d-1$ dimensional tachyon spectrum as $m_g$. Then, the
dominant contribution to the real part of the (\ref{bulkexc}) is
\be\lab{bulkexcRe} \la  Re P(x) Re P(0)\ra_{exc} \propto
L^{-(d-3)}e^{- m_g L}. \ee Comparison of (\ref{resBH1}) and
(\ref{bulkexcRe}) shows that the former dominates in the large L
limit and near the transition $t\approx 0$. One then finds for the
final result, \be\lab{finBHRe} \la  Re P(x) Re P(0)\ra_{BH}
\propto L^{-(d-3)}e^{- \xi_{BH}^{-1} L} + \cdots \ee with
mean-field scaling of the correlation length (\ref{corBH2}). In
the imaginary part, the comparison of (\ref{resBH2}) and
(\ref{bulkexcBH}) shows that  the latter dominates for large $L$
and the final result is \be\lab{finBHIm} \la  Im P(x) Im
P(0)\ra_{BH} \propto \frac{1}{L^{(d-3)}} + \cdots \ee One should
add to these two, the disconnected contribution (\ref{disconBH}).
Comparison with the spin-model result (\ref{sscorg}), using the
identifications (\ref{idt21}) and (\ref{idt22}) shows perfect
agreement:{\em In the limit of weak gravitational interactions
$g_s\to 0$, we obtained exactly the same correlator with the XY
model, with mean-field scaling exponents} $\eta = 0$, $\beta = \nu
= 1/2$. How the mean-field scaling can be altered will be
discussed in section \ref{discuss}.

\subsection{D-strings and vortices}
\lab{hooft}

In the original picture of \cite{Malda2}\cite{Rey} and their
subsequent generalizations, the relevant field theory on the
boundary is $SU(N)$ gauge theory and the Wilson loop in question
traces the path of an ``electrically charged'' fundamental field
in the theory, i.e. the ``electric quarks". The motivation to
relate the Wilson loops with the open strings is obvious in the
D-brane picture, where the fundamental strings couple to
electrically charged fields on the D-brane. Similar considerations
also suggest that the Wilson loop that traces a ``magnetically
charged'' particle, i.e. the 't Hooft loop \cite{thooft}, should
be related to the D1 branes \cite{Witten}. Indeed, this picture is
very suggestive and the various computations in the context of
QCD-like holographic models confirm the field theoretic
expectations \cite{GKN}.

In the LGT-spin model equivalence, the Polyakov loops are mapped
onto the spin operators on the spin-model side. Similarly, it is
very suggestive to relate the vortices, with the 't Hooft loops
that are dual to the D1 branes on the gravity side. Thus we
propose that the D1 brane configurations are the right tools to
probe the vortex dynamics in spin-models.

In this section we perform some basic checks with the D-strings
and show that indeed one obtains  the expected qualitative
behavior of the correlation functions. We reviewed the expected
spin-model result in the case of two-dimensions at the end of
appendix \ref{statmech}. We denote the operator that creates a
vortex that is localized at point $x$ by $v(x)$ and similarly the
anti-vortex by $\bar{v}(x)$. The proposal is the following chain
of dualities:
\begin{equation}\label{vduality}
  \la v(x) \ra \lra \la tr~P~e^{-\int \tilde{A}_0}\ra \lra
  e^{-S_{D1}}
\end{equation}
where the second object is the 't Hooft-Polyakov loop in the dual
gauge theory and $S_{D1}$ denotes the on-shell value of the dual
D-string configuration.

\subsubsection{One-point function}

Both in this section and in the next we consider the classical calculation
of the D-strings. A semi-classical calculation in the sense of section \ref{opfquant} and
\ref{tpfquant} is saved for future work.

First of all, we note that the vortex charge in the spin-model
should be dual to the D-brane charge on the string theory side. As
mentioned at the end of section \ref{statmech}, the total charge
of vortices in a configuration in 2D should vanish. The equivalent
statement on the gravity side would be that the total number of
$D1$s and $\bar{D}1$s which wrap the sub-manifold spanned by
$r,\,\,t$ coordinates should be equal in $D=4$. This is indeed the
case in $D=4$ as the D1s are charged, and the gauge-field that
couple them in the flat transverse space has a log-divergence.
This means that when one considers an ensemble of $D1$ and
$\bar{D}1$s, the configurations with non-equal numbers of the two
species have vanishing Boltzman weights in the partition function.
{\em This provides the first basic check in favor of associating
the vortices in the spin-model by the D-strings.} Note that the
argument applies equally-well when the target space is TG or BH
geometry. Thus we obtain,
\begin{equation}\label{vD2}
  \la v(x) \ra_{TG} = \la \bar{v}(x) \ra_{TG} = \la v(x) \ra_{BH} = \la \bar{v}(x)
  \ra_{BH} = 0, \qquad for \,\,\, d-1=2.
\end{equation}
In higher dimensions the argument above does not apply and one can
have configurations with non-trivial charge.\footnote{The dual
analogous objects to higher dimensional vortices (vortex lines,
planes, monopoles etc) in higher dimensions are the $Dp$ branes
with $p = d-2$.} Therefore, in higher-dimensions the expectation
value is determined by evaluation of the on-shell D-string action.

The boundary condition for a single D-string is just as in the
case of F-strings, above: it ends on the boundary at point $x$ and
wraps around the time-circle. The action is given by,
\begin{equation}\label{Dact}
  S_{D1} = -T_1 \int d^2\sigma e^{-\f} \le( det[h_{ab} +
  b_{ab}]\ri)^\half,
\end{equation}
where $T_1$ is the D-string tension and we defined,
\begin{equation}\label{hbdef}
h_{ab}= g^s_{\m\n} \6_a X^{\m} \6_b X^{\n}, \qquad b_{ab}=
B_{\m\n} \6_a X^{\m} \6_b X^{\n}.
\end{equation}
Here $g^s_{\m\n}$ is the string-frame metric (\ref{sBH}) in the
case of the BH phase. In the TG phase the metric is given by the
replacement $f\to 1$ and $A\to A_0$ in (\ref{sBH}).\footnote{The
D-string also couples to the gauge field on it. In fact the only
gauge-invariant combination (under ``big" transformations $A_{\m}
\to A_{\m} + \a_{\m}$) is of the form $b_{ab} + f_{ab}$ where
$f_{ab}$ is the pull-back of the gauge-field strength on the
D-brane. One can make the gauge choice $A_{\m}=0$. This choice
does not affect the discussion below.}

It is straightforward to find the on-shell action that corresponds
to a single D-string hanging from the boundary to some point
$r_f$: from (\ref{Dact}), (\ref{hbdef}) and (\ref{sBH}):
\begin{equation}\label{DactD}
  S_{D1} = -\frac{2T_1}{T} \int_{\epsilon}^{r_f} dr e^{-\f}
  \sqrt{e^{4A_s} + b^2},
\end{equation}
where $b$ is the constant value of the B-field on the r-t subspace
\footnote{As mentioned before we take the $B$-field to be either pure gauge or constant. Here we entertain the second possibility.}:
\begin{equation}\label{bdef}
b\equiv  B_{r0} = const.
\end{equation}
We compute the on-shell action separately on the TG and the BH
geometries for $d-1>2$ in App. \ref{AppT}. In the BH case string
hangs down to the horizon $r_f= r_h$, whereas in the TG case
$r_f=\infty$. The result is that {\em the on-shell action is
finite both on the TG and one the BH geometry.}
\begin{equation}\label{vDh}
  \la v(x) \ra_{TG} \ne 0, \quad \la v(x) \ra_{BH} \ne 0, \qquad for \,\,\, d-1>2.
\end{equation}
This is unlike the F-string case which diverges on the TG
geometry, and yields $\la \vec{m} \ra =0$. The reason for
finiteness here is clear in (\ref{DactD}): The only potential
divergence\footnote{The standard UV divergence $\epsilon\to 0$ can
easily be cured by adding counter-terms as in App. \ref{AppPP}.}
would be in the TG case where the upper limit of the integration
is $r_f=\infty$. The factor $A_s\to 0$ in that limit, however the
action is still finite because exponential suppression provided by
the $\exp(-\f)$ term in the action (\ref{Dact}); note that $\f$
grows linearly near the singularity, (\ref{fs}).

{\em This provides a second non-trivial check on the proposal.} It
matches the dual statement in the XY model is that the vortices
play no role in determining the phase of the system for higher
than two-spatial dimensions. We will thus consider the case of
$d-1=2$ below.

\subsubsection{Two-point function}

As the one-point function vanishes, the first non-trivial object
is the two-point function $\la \bar{v}(x) v(0)\ra$ in $d-1=2$.
This is dual to a connected $D1-\bar{D}1$ configuration,
completely analogous to the F-string case described in section
\ref{ssCor}. The boundary conditions are exactly the same as in
that case. Here too, we confine our interest in the classical computation
in order to see whether the association of D-strings with vortices of the
spin-systems pass the basic qualitative tests.

The computation is non-trivial and it is presented  in
App. \ref{AppT}. The result is as follows:

\begin{enumerate}
\item{The thermal gas:} We denote the difference between the
end-points of the $D1-\bar{D}1$ configuration as $L=|x|$. As in
section \ref{ssCor}, one can consider three contributions:

1) {\bf The disconnected D-string contribution:} This is given by
a disconnected $D1$ brane and a $\bar{D}1$ brane hanging from the
boundary and extends up to the singularity at $r=\infty$. This
contribution is dual to $|\la v \ra|^2$ and vanishes in $d-1=2$,
for the reason described above. It is finite for $d-1>2$.

2) {\bf The connected D-string contribution:} In App. \ref{AppT},
we show that there is a maximum value $L_{max}$ {\em that is
independent of $T$} and above which, there exists no connected
D-string solution. Therefore this configuration is replaced by the
{\em exchange diagram} directly analogous to the diagram described
in section \ref{ssCor}.

3) {\bf The exchange diagram:} For $L>L_{max}$ this is the only
non-trivial contribution. It is given by a $D1$ and a $\bar{D}1$
connected by exchange of bulk modes that couple to the D1s. For
large $L$ it is proportional to the propagator of the lowest mass
bulk mode $\exp(-m_D(T) L)$. As in section \ref{ssCor}, this mass
is bounded from below, $m_D(T)>0$ for all $T<T_c$ as there are no
massless modes.

We learn that the total result on the TG geometry for large $L$
is,
\begin{equation}\label{vvD1}
  \la \bar{v}(x) v(0) \ra_{TG} \sim e^{- m_D L}\qquad L\gg 1,\,\,
  d-1=2,
\end{equation}
where $m_D$ should be determined by a study of fluctuations around
the TG geometry. On the spin-model side this means that there is a
finite correlation length between vortices and anti-vortices in
the high-T phase. {\em This is in accord with the expectation that
one obtains a plasma of vortices and anti-vortices in the high T
phase of the 2D XY-model.}

\item{The black-hole:} The computation on the BH geometry is
completely analogous to the TG case above: 1) The disconnected
configuration vanishes in $d-1=2$ and is non-zero for higher
dimensions. 2) There exists a $L_{max}(T)$ above which the
connected D-string configuration does not exist. This time,
however $L_{max}$ is a function of T. One can show (see App
\ref{AppT}) that $L_{max}$ is finite for any $T$. Thus if we are
interested in the qualitative result for large $L$, then it is
again determined by {\em the exchange diagram}. The exchange
diagram in the BH case differs than the TG case above, in that,
there exists a massless bulk excitation that couples to the
D-string. It is given by the zero mode of the $B$-field as in
section \ref{ssCor}. Thus one finds,
\begin{equation}\label{vvD2}
  \la \bar{v}(x) v(0) \ra_{BH} \sim e^{{\cal O}(\log L)} \qquad L\gg 1,\,\,
  d-1=2,
\end{equation}
which gives a power-law with a T-dependent power. In order to
determine the power, one should calculate the ${\cal O}(\log L)$
terms in the exchange diagram. We postpone this computation to
future work, and content ourselves with the qualitative result
(\ref{vvD2}).
%The result implies that the energy of a
%vortex-anti-vortex pair, which is given by
%\begin{equation}\label{Evv}
% E_{v\bar{v}} = - T \log\le[ \la \bar{v}(x) v(0) \ra_{BH}\ri] = {\cal O}(\log L),
% \qquad for \,\,
%  d-1=2,
%\end{equation}
%This is the same as the Coulomb gas in 2D and confirms
%expectations from the XY-model side.
\end{enumerate}

All in all, comparison of (\ref{vvD1}) and (\ref{vvD2}) with
(\ref{vv}) provides a non-trivial check on the proposal.

\subsection{Vanishing of the second sound}
\lab{grss}

As reviewed in the end of appendix \ref{statmech}, the  speed of
sound that is associated with the phase fluctuations $\psi$ should
vanish {\em linearly} in the mean-field approximation as $T\to
T_c$ in the ordered phase. The effective action for this Goldstone
mode  can directly be obtained from the gravity action
(\ref{action}) because it maps onto the B-field on the gravity
side. We recall that in the large N limit, the Landau action is
given by the on-shell gravity action (\ref{grlan3}). Therefore to
obtain the effective action of the Goldstone mode we should
consider quadratic fluctuations of the B-field around the on-shell
value. We recall that the in the BH phase the phase of the
mean-field acquires an expectation value given by
\be\lab{phasevev} \psi = \int_M B \ee where $M$ is the submanifold
of the blackhole spanned by the $(r,x_0)$ coordinates. Goldstone
mode corresponds to x-dependendent fluctuations $\psi \to \psi +
\delta \psi(x)$. It is crucial that this fluctuation {\em cannot
be gauged away} by a gauge transformation of the B-field of the
form \be\lab{gaugetrans} B_{r0} \to B_{r0} + \6_r \xi_0 - \6_0
\xi_r. \ee The shift in $\psi$ corresponds to a ``big" gauge
transformation, it is a topological symmetry in the problem
\cite{Witten}\footnote{I am grateful to Sean Hartnoll for a
discussion on the various issues in this section.}.

A technical but important point is to regulate the divergence of the on-shell
action near boundary by subtracting a counter-term action. One can obtain this
counter-term action by the standard methods of holographic renomalization \cite{SkenderisReview}
but an easier way---that is equivalent for our purposes here---is to
subtract the on-shell thermal gas action. Consequently the Landau functional in the large N limit will be given by
\be\lab{lanfunc}
F_L \propto  \Delta {\cal A} = {\cal A}_{BH} - {\cal A}_{TG}
\ee
This is valid only in the black-hole phase i.e. $T>T_c$ which corresponds to the low T superfluid phase of the spin system.
The energy of the Goldstone mode should be computed in reference to $F_L$ that corresponds to the ground state energy
of the spin-model:
\be\lab{energy}
F_L(\psi + \delta\psi) - F_L(\psi) \equiv \delta_\psi F(\psi) \propto \Delta {\cal A}(\psi+\delta\psi) - \Delta {\cal A}(\psi).
\ee
 Thus, we substitute \be\lab{delps} B_{\m\n} =
B_{r0} + \delta\psi(x), \qquad B_{r0} = {\rm pure\,\,\, gauge}. \ee in
(\ref{action}) and obtain, \be\lab{act2psi} \delta_\psi F_L(\psi)
\propto \Delta{\cal A}_0 + C_\psi(r_h) \int d^{d-1} x \6_i
\delta\psi\6_j \delta\psi, \ee where $g$ denotes the BH metric
(\ref{BH}) and $g_0$ denotes the TG metric (\ref{TG}) and $\Delta
{\cal A}_0$ is the difference between the parts of the on-shell
values of the action that do not depend on $B$. For a second order
transition this piece vanishes at criticality as \be\lab{vanishA0}
\Delta {\cal A}_0 \sim t^2. \ee The coefficient $C_\psi$ in
(\ref{act2psi}) on the other hand contains the desired information
on the sound speed (and the energy of the Goldstone mode). It is
given by \be\lab{coefcpsi} C_\psi(r_h) = \int_0^{r_h} \!\!dr
\sqrt{g} e^{-\frac{8}{d-1}\f}g^{ij} g^{rr} g^{00}
-\!\!\int_0^{\infty} \!\!dr
\sqrt{g_0}e^{-\frac{8}{d-1}\f_0}g_0^{ij} g_0^{rr} g_0^{00}. \ee On
the spin model side, this is proportional to the kinetic term
$c^2_\psi \int \nabla \delta\psi\cdot \nabla\delta\psi d^{d-1}x$,
hence the speed of sound for the Goldstone mode is given by the
coefficient $C_\psi(r_h)$. One finds, \be\lab{ssgr} c^2_\psi
\propto  C_\psi(r_h) = \int_0^{r_h} dr e^{-\frac{8}{d-1}\f
+(d-5)A} -\int_0^{\infty} dr e^{-\frac{8}{d-1}\f_0+(d-5)A_0} \ee
First of all,  we observe that the energy of the Goldstone mode
\be\lab{engold} E_\psi \propto C_\psi \int d^{d-1} x \6_i
\delta\psi\6_j \delta\psi, \ee is finite for any $r_h$. This is a
crucial requirement to be able to associate the fluctuations of
the B-field with the Goldstone mode.

Secondly, we find that
$c_\psi^2$  indeed vanishes  as $t\to 0$
($r_h\to\infty$) because the BH background $(A,\f)$ asymptotes to
the TG background  $(A_0,\f_0)$ in the limit $r_h\to\infty$ where
the BH mass vanishes. This is a nice check already because it also
confirms that $c_\psi^2$ vanishes {\em only} in a continuous
transition which requires that the saddle solutions coalesce in
the transition region.

A more non-trivial check however is to see whether the  scaling
exponent for the vanishing rate is indeed the one expected from
the mean-field scaling, i.e. whether $c_\psi^2 \sim t$ or not. We
recall that\footnote{See section \ref{what} for a discussion on
this issue.} mean-field scaling is expected whenever an operator
in the dual field theory is related to the fluctuations of the
bulk action on the gravity side, in the limit of weak
gravitational interactions $g_s\to 0$. Using the asymptotics of
the BH background functions in (\ref{fs}) and (\ref{As}) we find
that the contributions from $A$ and $\f$ in the exponent conspire
nicely to produce \be\lab{ssgr1} c^2_\psi \propto
\int_{r_h}^{\infty} dr e^{-\sqrt{V_\infty} r_h} \propto
e^{-\sqrt{V_\infty} r_h}, \qquad r_h\gg 1. \ee Finally, use of
(\ref{D18}) yields, as $t\to 0$,
%%%%%%%%%%%%%%%%%%%%%
\be\lab{ssgr2}
c^2_\psi  \propto  \left\{ \begin{array}{ll}
t^{\frac{2}{\kappa}}, & \mathrm{case\,\, i}, \\
e^{-2\le(\frac{t}{C}\ri)^{-\frac{1}{\a}}}, & \mathrm{case\,\, ii}.
\end{array}\right.
\ee
%%%%%%%%%%%%%%%%%%%%%
The case of a second order transition corresponds to  $\kappa=2$
in case i, see equation (\ref{kappaNC}) and we indeed find the
expected mean-field behavior: \be\lab{ssmf} c_\psi^2 \sim t,
\qquad as\,\,\, t\to 0. \ee We note that the precise form of the
kinetic term for the B-field in (\ref{action}) is crucial in
reproducing the desired behavior. This form stems from the
non-critical string action in $d+1$ dimensions after a Weyl
transformation to the Einstein frame. As the physical results
should be independent of the frame, one can of course produce the
same result directly in the string frame.

In fact, the calculation is more transparent in the  string frame.
The asymptotics as $r_h\to\infty$ are given  by the linear-dilaton
background, where the string-frame metric becomes
flat\cite{exotic1} \be\lab{stas} g_{s,\n\m} = e^{\frac{4}{d-1}\f}
g_{\m\n} \to \delta_{\m\n}. \ee On the other hand the gravity in
the string frame is given by \be\lab{acst} {\cal A}_s \propto \int
d^{d+1} x \sqrt{g} e^{-2\f} (dB)^2. \ee As the metric becomes
flat, the scaling of $c^2_\psi$ is entirely determined by the
factor $e^{-2\f}$ above: \be\lab{ssgr3} c^2_\psi \propto
e^{-2\f_h}, \qquad \f_h\gg 1, \ee where $\f_h$ is the value of the
dilaton on the horizon. Use of (\ref{D17}) now produces the same
result as in (\ref{ssgr2}).

It is also easier to investigate possible $\a'$  corrections in
the string frame. Higher derivative corrections to the B-field can
be schematically represented as \be\lab{acstcor} {\cal A}_s
\propto \sum_{k,l,m=0}^{\infty}  c_{klm} \int e^{-2\f} (dB)^{2k}
R^{(l)} (d\f)^{2m}, \ee where  $c_{klm}$ are some unknown
constants---that are supposed to be determined by the world-sheet
sigma model---and $R^{(l)}$ represents higher curvature
invariants, e.g. $R^{(0)} = 1$ and $R^{(1)}=R$, $R^{(2)} \sim R^2
+ R_{\a\beta\gamma\delta} R^{\a\beta\gamma\delta}$ etc. Sum over
non-trivial cross contractions between $dB$, $R$ and $d\f$ terms
are also implied to be included in this schematic expression.

The linear-dilaton solution in the asymptotic  region is $\a'$
exact, therefore the form of the background functions $\f$ and
$g_{s,\m,\n}$ are not subject to $\a'$ corrections in the far IR.
All of the curvature invariants in the string frame vanish as
shown in \cite{exotic1}. On the other hand only the term $k=1$
above can contribute to the quadratic term in $\delta \psi$
because the leading order piece in B is constant, see equation
(\ref{delps}). Therefore one finds, \be\lab{acstcor2}
\delta^2{\cal A}_s \to \sum_{m=0}^{\infty}  c_{m} \int e^{-2\f}
(d\delta\psi)^2 (d\f)^{2m}. \ee Finally, we note that all of the
dilaton invariants also go over to a constant in the
linear-dilaton background, hence one still obtains (\ref{ssgr})
with a renormalized overall coefficient. We conclude that we do
not expect the $\a'$ corrections change the linear scaling in
(\ref{ssmf}).

\section{A proposal for gravity-spin model correspondence in the
general case} \lab{sec63}

Here, we would like to return the discussion of section 2 and
promote the gravity-spin model duality that we advocated in the
case of $U(1)$ models to the general case.

We are interested in employing gravitational techniques to learn
about the dynamics of the spin model around the transition. We
want to map the spin model to a lattice gauge theory, which than
will be related to a gravitational background in the continuum
limit. Unlike the derivation of the spin-model from the LGT as
reviewed in
 Appendix \ref{sec2}, the opposite map from the spin model to the gauge
theory is non-trivial. There are two sources of complication:
\begin{enumerate}
\item {\bf Non-uniqueness:} The map may be non-unique.
Clearly, there may be many gauge theories that share the same
center symmetry. First of all, this may be due to the fact that
the center symmetry of different gauge groups may be the same. As
an extreme example, the centers of $SU(2)$, $Sp(N)$ with arbitrary
$N$, $SO(N)$ with odd $N$ and $E(7)$ are all isomorphic to $Z_2$.
Thus for example, the critical phenomena (if exists) in any of
these theories should be described by the universality class of
the Ising model in $d$ dimensions. Secondly, the deformation of
the pure gauge theory by addition of any {\em adjoint matter}
leaves the center symmetry intact. Thus, generally, the
equivalence maps a spin model with symmetry ${\cal C}$ to a set of
LGTs with various gauge groups $G$ and matter $M$:
%%%%%%%%%%%%
\be\lab{map1} SM_{\cal C} \longrightarrow \{ LGT[G,M] \}, \ee
%%%%%%%%%%%
where $Center[G] = {\cal C}$.
\item {\bf Non-existence:} Another source of complication has to do
 with non-existence of such a map {\em in the continuum limit}.
In fact, it is not easy to find continuous critical phenomena in
gauge theories. As an example, among the {\em pure} Yang-Mills
theories with gauge group $SU(N)$, only in the case of $N=2$, and
possibly the case of $N=\infty$ (see the discussion at the end of
(\ref{sec1})) exhibit continuous confinement-deconfinement
transition. All the rest is believed to have first order
transitions.

Therefore, given the map (\ref{map1}), the critical phenomena in
the continuum limit of RHS may be non-existent. Define the
subspace  $(G^*,M^*)$ of the gauge groups and adjoint matter
$(G,M)$ that appear in (\ref{map1}), such that in the continuum limit
$LGT[G^*,M^*]\to GT[G^*,M^*]$ the criticality prevails. Then, we
can extend the map (\ref{map1}) to the continuum limit:
%%%%%%%%%%%%
\be\lab{map2} SM_{\cal C} \longrightarrow \{ LGT[G,M] \}
\longrightarrow \{ GT[G^*,M^*] \}, \ee
%%%%%%%%%%%
where again $Center[G^*] = { \cal C}$.
\end{enumerate}

We conclude that, the map (\ref{map2}) may or may not exists, and
even if exists, it may not be unique. We observe, however that the
non-uniqueness is a positive fact, in the sense that, it provides
us with a greater space of gauge theories to scan in search for
continuous critical phenomena. Indeed, it may be possible to find
critical phenomena either by changing the gauge group $G$ (while
keeping the center $C$ intact) or by changing the (adjoint) matter
content $M$.

Now, the last step of the procedure is to employ the gauge-gravity
correspondence to map the RHS of (\ref{map2}) onto a gravitational
background $GR[G,M]$. Suppose that the gauge-gravity
correspondence holds for the {\em subspace} $(G^{**},M^{**})$ of
the pairs $(G^*,M^*)$ that appear in (\ref{map2}). Then, we can
extend the map as,
%%%%%%%%%%%%
\be\lab{map3} SM_{\cal C} \longrightarrow \{ LGT[G,M] \}
\longrightarrow \{GT[G^*,M^*]\} \longrightarrow
\{GR[G^{**},M^{**}]\} , \ee
%%%%%%%%%%%
where again $Center[G^{**}] = {\cal C}$. The last map is the
highly non-trivial gauge-gravity correspondence that is assumed to
hold for arbitrary N, (not necessarily in the large N limit). By
the standard lore of the gauge-gravity correspondence, the center
symmetry ${\cal C}$  should correspond to a {\em bulk gauge
symmetry} ${\cal C}$ on the gravity side. For continuous ${\cal
C}$,  this can be a continuous isometry of the gravitational
background. In the case ${\cal C}$ is discrete it may be a
continuous isometry broken down to a discrete subgroup ${\cal C}$
by stringy effects.

Some comments are in order:
\begin{enumerate}
\item {\bf Top-bottom approach}: From the above procedure it is clear that one arrives at
 an operational definition of the gravity-spin model correspondence. Take a spin-model $SM_{\cal C}$ that exhibits continuous criticality. Then one
 should scan through all of the LGTs  $\{LGT[G,M]\}$ with various adjoint matter $M$ and gauge group $G$, such that  the critical phenomena
 persists in the continuum limit. This step, {\em in principle} can be done with Monte-Carlo simulation techniques. The outcome
 of this step would be the space of gauge theories $\{GT[G^*,M^*]\}$ in (\ref{map3}). The next step is to construct D-brane configurations
 that correspond to these gauge theories. In general this would only be possible for a subspace of theories $(G^{**}, M^{**})$.
 The next step then, is to take the decoupling limit of the D-brane  configurations to find the gravitational backgrounds $\{GR[G^{**}, M^{**}]\}$
that appear at the end of (\ref{map3}). The final step is to look
for the black-hole solutions and study the Hawking-Page transition
at finite temperature. One can then compute observables of the
spin model around criticality, such as the scaling of the
correlation functions of order parameters, critical exponents, the
transition temperature $T_c$ etc. by holographic techniques. This
operational definition of the duality corresponds to the so-called
top-bottom approach in the gauge-gravity duality, that is
unfortunately unpractical.

\item {\bf Bottom-up approach}: Instead, one may adopt a ``phenomenological'' approach and search for continuous critical phenomena
{\em directly on the gravity side}. This is the approach that we
take in this paper. The symmetries, the (bulk) matter content and
various dynamical phenomena (such as spontaneous symmetry
breaking) on the gravity side should then hint at what kind of
spin model that the gravity theory describes. Of course, one
should check by computations on the gravity side that the theory
indeed fulfill the basic expectations of the spin model. As a last comment
the bottom-up approach is not necessarily doomed by lack of predictive power.
As we argued in section 4, there exists a notion of universality exactly around the transition region
for models that are based on Einstein-dilaton gravity.

\item {\bf The ``large-N'' limit}: The meaning of the ``number of colors" and the large-N limit becomes
clear in this approach. On the gravity side, it corresponds to the
small $G_N$ limit where one arrives at a classical string theory
in which the interactions between the bulk fields can be ignored.
%In the weak curvature regime, this corresponds to a saddle-point
%approximation where one retains on the dominant saddle in the
%gravity action.
On the spin-model side in the case ${\cal C} = Z_N$ for $N>4$ the
corresponding gauge group can only be $SU(N)$. Then the limit $N
\to\infty$ corresponds to the limit where the $Z_N$ invariant
spin-model becomes $U(1)$-invariant. In this case---{\em ``number
of colors" correspond
 to the number of spin states that the spin vector $\vec{s}$ on a
  lattice site can attain}.
 %In other words, it is the large q limit of a q-state Potts model.
Of course the large-color limit exists only in theories with gauge
groups $SU(N)$, $Sp(N)$, $SO(N)$ (with possible additions of
$U(1)$ factors). In the other cases, one cannot study the
spin-model in a large-N approximation.

 \item {\bf Discrete ${\cal C}$:} One may then think that it is never possible to study a spin model with discrete symmetry, by a gravitational
 theory in the small $G_N$ limit. This is not necessarily the case. As an illustration, consider the Ising model in d-dimensions. The symmetry
 group is $Z_2$ and one of the gauge groups that has this, as a center symmetry is $Sp(N)$ with arbitrary $N$. Thus the center
 remains $Z_2$ also in the large-N limit! Therefore one can make the following proposal for a gravity dual of the Ising model:
 Consider a D-brane set up that is dual to YM theory with $Sp(N)$ gauge group with additional adjoint matter content $M$ chosen
 such that the theory exhibits continuous critical phenomena. Then consider the background that is dual to this configuration. Then
 the black-hole solution near the Hawking-Page transition should fall into the same universality class as the Ising model. By this procedure,
 it may then be possible to analytically calculate the critical exponents of an Ising model in any dimension d.

\end{enumerate}

\section{Discussion}
\lab{discuss}

\subsection{Summary}

This paper has two related purposes. The first one is to advocate
a particular approach to holography in condensed matter systems.
We proposed to establish the link between certain spin systems and
gravity through the better understood case of gauge-gravity
correspondence and the IR equivalence between gauge theories and
spin-models. The latter is expected to hold only around
criticality (in the continuum limit). Therefore a gravity-spin
model duality is expected to hold only near the phase transition
region. In particular one should not rely on the gravitational
description in the UV of the spin-system.

On the other hand, precisely around the critical region, where the
spin-system is strongly correlated, the dual gravity description
is expected to simplify as the higher-derivative corrections
become smaller. We showed that this expectation indeed holds in a
specific gravity model based on non-critical string theory.

This example also hints at a kind of  universality in the dual
gravity theory which only arises in the transition region: we
found that regardless of the details of the gravity theory, the
physics around criticality is governed by a linear-dilaton CFT.
Moreover focusing on the lowest states in the CFT at levels $N=0$
and $N=1$ imply mean-field scaling---in the semi-classical
approximation where only the lowest lying states are kept in
string propagation---regardless of the matter content  of the CFT.

We emphasize that the approach formulated in the previous section
in principle allows for a top-bottom constructions in AdS/CMT.
What more can be learned from gravity around the critical region
is  described in section \ref{what}.

A second purpose of the paper was to construct a model of
holographic superfluidity based on continuous  Hawking-Page
transitions in gravity. A duality between gravity and spin-models
of the type described above provides motivation for this model but
it could have been constructed with no reference to such
arguments. Indeed all one needs  from the phenomenological
perspective is a gravitational model 1) with some mechanism of
spontaneous breaking of global $U(1)$ symmetry in a continuous
transition and 2) a bulk field that is charged under this $U(1)$
which would serve as a dual of the order parameter. In the model
that we studied here the $U(1)$ is the topological shift symmetry
of the NS-NS two-form field that breaks down at a continuous
Hawking-Page transition and the fields that are charged under this
$U(1)$ are the winding modes of the string around the time-circle.
Viewed from this perspective, one wonders if a gravity model can
be obtained in a more direct fashion by truncating the string down
to the bulk dynamics of  gravity, dilaton an Abelian gauge field
and the winding modes. In this approach one expects to study an
effective action of the sort, \be\lab{efac} S \sim \int e^{-2\f}
\le(R + \half (\6\f)^2  + V(\f) + \half |DT|^2 - \half m_T^2
|T|^2\ri) \ee where $T$ ($T^*)$ is the winding tachyon with $w=+1$
($w=-1$) and mass $-m^2_T$ and it is minimally coupled to a $U(1)$
gauge field  through $D = \6 -i A$. The gauge field may arise
either from reduction of the $B$-field on the time-circle or
gauging the aforementioned topological shift symmetry of the
B-field. In this effective theory, the Goldstone mode in the
superfluid phase is given by the phase of the $T$ field. The
system would be in the ordered phase when (the particular mode
that corresponds to the order parameter in the fluctuations of)
$T$ becomes normalizable above a certain $T_c$.

Such models have the same flavor as the ones in \cite{Gubser, Sean} and more recently in \cite{Elias}. One immediate
future direction is to understand holographic implications of a model such as (\ref{efac}).

Another immediate  future work concerns going beyond the mean-field scaling at criticality. We showed that the lowest mass sector of the linear-dilaton CFT gives rise to mean-field scaling. We named such a restriction to the lowest lying modes in the tree-level string  path integrals as the ``semi-classical"
approximation. Then, the question is what happens beyond the semi-classical approximation? Can one produce exponents beyond mean-field scaling
in this manner? Can one obtain universal exponents of the 3D XY-model by summing up contributions of all string states?

Further future directions are listed below.

\subsection{Outlook}

\begin{itemize}

\item{Embedding in string theory}

Clearly, it is of great interest to look for examples of the
proposed correspondence in a consistent truncation of {\em critical} string
theory. The very recent papers \cite{Minwalla}\cite{Minwalla1} may
be relevant for this enterprize\footnote{We thank Yaron Oz for mentioning
possible relevance of these works.}. Another relevant work is \cite{Skenderis1,Skenderis2}
where it was shown that  linear-dilaton type geometry universally arises from non-conformal branes.

 Explicitly put, one should find a consistent truncation of string theory
which possess small slack-hole solutions that exhibit continuous
Hawking-Page transitions. We observe that the asymptotic form  of
the scalar potential in (\ref{Vs}) and (\ref{case1}) is sum of
exponentials that quite generically appears in consistent
truncations of IIB and IIA critical string theory. We shall leave
this investigation for future work.

\item{An explicit D-brane set-up?}

Even if one finds examples of continuous Hawking-Page transitions
in string theory, this would not necessarily give control over the
microscopic condensed matter system that we want to describe. On
the other hand,  the prescription proposed in section 6 in
principle goes beyond this and allows for a top-bottom approach.

Therefore, one should search for examples of gauge theories with
gauge group $G$ and adjoint matter such that the theory exhibits
criticality at some finite $T_c$. There are indeed examples of
this. In \cite{Minwalla2}, the authors studied $SU(N)$ with
adjoint matter on $S^3$, in the large N limit and showed that for
certain choices of the matter, the theory exhibits a {\em second
order} deconfinement transition at finite temperature, at weak
coupling. This happens when the coefficient of the quartic term in
the effective action for the Polyakov loop is negative. Whether
this transition prevails in the limit when the radius of the
sphere becomes large (the case relevant here), or whether it is
continuously connected to a transition at strong coupling is
unclear, but it is probable.

\item{Discrete center}

We note that, the proposal advocated in this paper can also be
applied to spin-models with discrete symmetry groups in principle.
In most of the paper we focused on the $SU(N)$ LGT in the large N
limit. Going beyond the large N limit seems to be a difficult
enterprize at the moment, however one may consider other gauge
groups such as $SO(N)$ and $Sp(N)$ with adjoint matter, in the
large N limit. The latter is particularly interesting, because it
has the center $Z_2$ for arbitrary N, hence also in the large N
limit. It is very tempting to employ the ideas developed in this
paper to this particular case to arrive at a gravitational
description of the 3D Ising model.

\item{Other critical exponents}

One can also study critical behavior in other quantities. One such quantity is
the susceptibility:
%%%%%%%%%%%%%
\be\lab{suscep} \chi =\lim_{h\to 0} \frac{d |\vec{M}|_h}{dh} \sim
t^{-\gamma}, \qquad as\,\,\, t\to 0. \ee
%%%%%%%%%%%%%
Here $h$ denotes and external magnetic field and the expectation
value $\vec{M} = \la \vec{s} \ra$ is taken in the ensemble with an
external magnetic field present, i.e. the Hamiltonian of the spin
model replaced by $H \to H + \vec{h}\cdot \sum_i \vec{s}_i$.

How to generate such an external magnetic field in the gravity
picture? One hint is that the external magnetic field should break
the $U(1)_B$ invariance explicitly by analogy with the dual
spin-model. This may happen through a Chern-Simons type coupling
in the gravity action $\int d^{d+1}x B_2 \wedge H_{d-1}$ where
$B_2$ is the NS-NS two form and $H$ is an appropriate RR form. The
role of the magnetic field would be played by a constant $H$-form
on the d-1 dimensional space part. It would be very interesting to
investigate this issue in the future and eventually compute the
critical exponent $\gamma$ in (\ref{suscep}) by gravitational
methods.

%Susskind, in \cite{Susskind} showed that the XY model with a
%magnetic field can be obtained on the LGT side by turning on a
%matter field $\Phi_a$ in the adjoint representation on each
%lattice site\footnote{In \cite{Susskind} only the Abelian case is
%discussed but we believe that it can easily be generalized to
%non-Abelian lattice gauge theories.}. The contribution of this
%field to the LGT Hamiltonian is of the form $m Q_a Q_a$ where
%$Q_a$ is the conjugate momentum to $\Phi_a$ and $m$ is some
%constant. Then, if one goes through the same analysis of section
%\ref{sec2}, one finds that the dual spin-model is an XY model with
%an external field $h\sim m^{-1}$. In an hypothetical D-brane
%picture the field $\Phi_a$ should correspond to one of the extra
%dimensions transverse to the brane.

\item{The UV geometry}

We observed that most of the interesting scaling behavior in the
observables of the spin-model depend on the IR geometry on the
gravity side. We did not have to specify the UV geometry so far,
but we tacitly assumed that it becomes asymptotically AdS, for
consistency in holographic applications. From a practical point of
view, the UV geometry will be important if one desires to obtain
the full form of the n-point functions, not just the scaling with
$t$. In \cite{exotic1} we indeed constructed analytic kink
solutions that fulfill this promise. These solutions interpolate
between an asymptotically AdS geometry (with constant dilaton) in
the UV towards an asymptotically linear-dilaton geometry in the
IR. It will be very interesting to study correlation functions
holographically obtained from these backgrounds.

The specification  of the IR geometry follows from physical
requirements of the spin-model near the transition region $T_ c$.
On the other hand, the black-hole with temperature $T$ is argued
to correspond to the super-fluid phase with temperature $1/T$.
Then, one can ask whether we can also produce the expected
behavior of the super-fluids at very low temperatures, by
specifying the high T regime of the black-hole that corresponds
to {\em the UV geometry}: One basic feature of the two-fluid model
for super-fluidity is that the (normal) speed of sound, that is
associated with fluctuations in the magnetization vanish as $T\to
0$. This is certainly {\em not} a behavior expected from an
asymptotically AdS geometry in the UV which would correspond to a
conformal fluid with $c_s^2 = 1/3$. We conclude that the kink
solution that flows from AdS to linear-dilaton \cite{exotic1}
would not do the job here. One possible way  to proceed may be to
consider the non-conformal brane solutions
\cite{Skenderis1,Skenderis2} which on one hand allow for a
holographic computation of observables, and on the other, there is
a chance to find backgrounds with $c_s^2\to 0$ in the UV.

Do we really expect to find a background as a  solution to
two-derivative Einstein-dilaton theory, that would produce the
desired behavior  in the entire range $T\in (0,T_c)$ of the
super-fluid? The answer is most probably negative. Let us suppose
for a moment that such a background exists as a solution to the
full $d+1$ dimensional non-critical string theory. As we showed in
\cite{exotic1}, the curvature invariants in the IR vanish in the
string frame, therefore a two-derivative approximation is expected
to work in near the transition $T \approx T_c$ On the other hand,
the invariants away from the transition region are determined by
the intrinsic string scale $\ell_s$. This means that in a
two-derivative approximation one deals with a background with
$\ell/\ell_s\sim 1$. To conclude: we indeed expect non-trivial
$\a'$ corrections in the UV region and the two-derivative
approximation presented here is expected to give reliable results
only near the transition region.

\item{The two fluid model of super-fluidity:}

We only performed the computation of the speed  of sound for the
Goldstone mode. It would be also very interesting to look at the
dissipation coefficients associated with these fluctuations. In
the two-fluid model of super-fluidity, one deals with a coupled
system of pressure and entropy waves of the two-component
superfluid, cf. \cite{SonHydro} for a recent review. The pressure
waves are dual to metric and dilaton fluctuations, whereas the
entropy waves are associated with fluctuations of the B-field. It
would be very interesting to work out this coupled system of
fluctuations in order to determine the associated dissipative
fluid dynamics\footnote{It seems that one needs to turn on a
Chern-Simons term of the form $\int B \wedge H_{RR}$ where
$H_{RR}$ is a d-1-form turned on the spatial directions in order
to achieve such a mixing.}.

\item{Spin models with non-Abelian symmetry groups}

One fundamental restriction of the approach in section 6 is that the spin
symmetry cannot be non-Abelian as it follows from the center
symmetry of the corresponding lattice gauge theory. On the other
hand a very important model for superconductors involves the
$O(3)$ model\footnote{See \cite{Liu2} for a recent proposal for a
holographic description of this model.}. Whether one can overcome this
restriction in our set-up is an interesting question. In phenomenological models
such as \cite{GubserPwave}, one can achieve this simply by considering black-holes
with non-abelian charges.

In our perspective, one idea is to consider the enhanced symmetries of string theory at special radii \cite{PolchinskiBook1}.
When the sting is compactified on the time-circle one obtains  $U(1)_G\times U(1)_B$ symmetry at an
arbitrary radius. The second one is spontaneously broken in the BH phase. At special a radius $T = T_s = (2\pi\ell_s)^{-1}$
 (in bosonic NCST) however one obtains  an enhanced symmetry $SU(2)_L\times SU(2)_R$.
 If this radius corresponds to the transition temperature
$T_c$ then one may be able to obtain a model with the desired behaviour within our set-up.

\end{itemize}

\addcontentsline{toc}{section}{Acknowledgments}
\acknowledgments

\noindent It is a pleasure to thank  Alex Buchel, Carlos Nunez,
Yaron Oz, Marco Panero, Erik Plauschinn, Giuseppe Policastro, Sean
Hartnoll, Elias Kiritsis, Stefan Vandoren and especially to Henk
Stoof for valuable discussions. The author is supported by the
VIDI grant 016.069.313 from the Dutch Organization for Scientific
Research (NWO).

\newpage

\appendix
\renewcommand{\theequation}{\thesection.\arabic{equation}}
\addcontentsline{toc}{section}{Appendices}
\section*{APPENDIX}

\section{Simplest example of the LGT-spin equivalence}
\lab{sec2}

Let us review how the LGT-spin equivalence works in the simplest
case of the $U(1)$ lattice gauge theory in $d$ dimensions. Through
this example we will illustrate that the temperature of the spin
model is inversely related to the temperature of the (lattice)
gauge theory which also holds in the most general case.

In the Hamiltonian formalism, the lattice theory is defined by:
\begin{equation}\label{u1h}
  H = \frac{g^2}{2a} \sum_{(r,\hat{n})} E^2(r,\hat{n}) -
 \frac{1}{2ag^2}  \sum_{\Gamma} \le(V[\Gamma] + V^{\dagger}[\Gamma]\ri).
\end{equation}
Here, $g$ is the coupling constant, $a$ is the lattice spacing.
The first sum is over the links $(r,\hat{n})$ on a d-dimensional
square lattice ($r$ denotes  the lattice site, $\hat{n}$ denotes
the direction of the link that originates from this site) and $E$
denote the electric fields residing on these links. The first sum
above yields the electric energy. The second one is over the
elementary plaquets $\Gamma$. The $V$s denote the Wilson lines on
these plaquets. This gives the magnetic energy. The partition
function of gauge invariant states at temperature $T_l$ is given
by,
\begin{equation}\label{Parfunc}
  Z_{lat}(T_l) = {\rm Tr}' e^{-H/T_l},
\end{equation}
where the prime reminds us that we have to impose the Gauss' law
on the states in the ensemble. Consequently, the sum above is over
the gauge invariant states $|\psi\ra$ which should satisfy,
\be\lab{gis} H |\psi\ra = E |\psi\ra, \qquad \Gamma_r |\psi\ra=0.
\ee Here, the  second equation imposes the Gauss' condition on the
states; the operator $\Gamma_r$ is the lattice analog of
$\nabla\cdot E$ on each lattice site r : \be\lab{Gamma} \Gamma_r =
\sum_{\hat{n}} E(r,\hat{n}). \ee In the strong coupling limit, one
can drop the magnetic energy term in (\ref{u1h}).

Now, the sum is only over the electric link variables and the
prime can be removed by a suitable Lagrange multiplier $\a$: (at
strong-coupling):\cite{Polyakov, Susskind}
%%%%%%%%%%%%%%%%%%%%%%%
 \be\lab{u1}
Z_{lat}(T_l)  = \int_{-\pi}^{\pi} \prod_r d\a(r) \prod_{links}
\sum_E \exp\le( -\frac{g^2}{2aT_l}E(r,\hat{n})^2 +
i[\a(r)-\a(r+\hat{n})]E(r,\hat{n}) \ri). \ee
%%%%%%%%%%%%%%%%%%%%%%%
The integral over $\a$ imposes the Gauss' law. Using the Poisson
summation formula,
%%%%%%%%%%%%%
\be\lab{Poisson} \sum_E e^{cE^2+i\a E} \propto \sum_m
e^{-\frac{1}{4c}(\a+2\pi m)^2}, \ee
%%%%%%%%%%%%%
the sum over the $E$ can be performed:
%%%%%%%%%%%%%
\be\lab{u2} Z_{lat}(T_m)  = \int_{-\pi}^{\pi} \prod_r d\a(r)
\prod_{links} \sum_{m_r} \exp\le( -\frac{a
T_l}{g^2}[\a(r)-\a(r+\hat{n}) + 2\pi m_r]^2\ri). \ee
%%%%%%%%%%%%%
This is the Villain approximation to the {\em Heisenberg model}
for ferromagnetism in d-dimensions. It is in the same universality
class with the Heisenberg model\cite{Susskind}. In particular, for
$d=2$ this becomes the famous XY model in 2 dimensions, where the
BKT scaling was first observed \cite{BKT}.

The salient feature of the Heisenberg model is that it exhibits
order in the low T phase (that corresponds to the, high T
deconfined phase of the Abelian LGT),
%%%%%%%%%%%%%
\be\lab{order} \la e^{i\a(R)} e^{i\a(0} \ra \propto 1, \qquad
as,\,\,\,\,\,\, R\to\infty, \ee
%%%%%%%%%%%%%
and disorder  in the high T phase (that corresponds to the low T,
confined phase of the LGT),
%%%%%%%%%%%%%
 \be\lab{DISorder}
 \la e^{i\a(R)} e^{i\a(0} \ra \propto E^{-R/\xi}, \qquad as \,\,\,\,\,\, R\to\infty,
 \ee
%%%%%%%%%%%%%
where $\xi$ defines the correlation length. Some comments are in
order:
\begin{itemize}
\item The computation  can be generalized to the non-Abelian
case \cite{Polyakov, Susskind}.

\item The computation is performed in the strong coupling limit where one can ignore the
magnetic energy in (\ref{u1h}). This constraint can easily be
loosened  and the equivalence prevails also if one considers the
magnetic piece \cite{Polyakov, Susskind}.

\item{One can generalize to add adjoint matter, as the center
symmetry remains intact under addition of adjoints.}
\end{itemize}

\section{Relation between non-critical strings and the
linear-dilaton theory} \lab{lindilNC}

It is long known that the two theories are intimately connected
\cite{Myers}. The connection is made precise in a beautiful work
by Chamseddine \cite{Chemseddine} which we would like to review
here.

For simplicity we  consider bosonic matter. Then the non-crirtical
string theory in $d-1$ spatial dimensions with flat Euclidean
target-space metric can be defined by the world-sheet action
\be\lab{wsaction} {\cal A}_{ws}  = \frac{1}{\pi \a'} \int_M
d^2\sigma \sqrt{h} \phi (R+\Lambda)  + \lambda \chi(M) + \mu
\int_M d^2\sigma \sqrt{h}  + \frac{1}{4\pi\a'} \int_M d^2\sigma
\sqrt{h} h^{ab} \6_a X^i \6_b X^i, \ee where we ignored coupling
to $B$-field for simplicity. Here $\phi$ is a scalar field
introduced in \cite{Chemseddine} in order to ameliorate evaluation
of higher genera diagrams. Its presence was also motivated on
physical grounds \cite{Chemseddine}. The matter index runs from
$i=1$ to $i=d-1$. The manifold M can be with arbitrary genus
$\chi(M)= 2(1-g)$, and the constant $\lambda$ is determined by the
asymptotic value of the dilaton. The constant $\Lambda$ is a free
parameter and $\mu$ and is subject to renormalization. It was
argued in \cite{Chemseddine} that the $\phi$ coupling in the
action apparently overcomes the difficulties, ``the c=1 problem''
encountered in the study of non-critical strings. The path
integral over the world-sheet metric can be performed  in the
conformal gauge $h_{ab} = e^{\sigma_L} \hat{h}_{ab}$ and results
in the Liouville action for the field $\sigma_L$ producing
additional world-sheet terms $\mu e^{\sigma_L}$ and  $\Lambda
e^{\sigma_L}$. Then the effective renormalized action after
gauge-fixing involves the matter part as in (\ref{wsaction}), the
renormalized gravity action, the ghost part that arise from the
reparametrization fixing and the induced Liouville action:
\bea\lab{wsaction2} {\cal A}_{ws}  &=& \frac{1}{2\pi} \int_M
d^2\sigma \sqrt{h} \phi (\hat{R} + \Delta_{\hat{h}} \s_L + \Lambda
e^{2\s_L}) + \frac{1}{4\pi\a'}
\int_M d^2\sigma \sqrt{h} h^{ab} \6_a X^i \6_b X^i  +\lambda \chi(M)  \nn\\
{}&+& \frac{1}{2\pi} \int_M d^2\sigma \sqrt{h} b_{ab} \nabla^a c^b
+ \frac{1}{\pi} \int_M d^2\sigma \sqrt{h} \le[a(\half \hat{h}^{ab}
\6_a \s_L\6_b \s_L + \s_L\hat{R}) +\mu e^{2\s_L}\ri]. \eea Here
$a$ is a constant that vanishes only in the critical case (e.g.
$a=(25-d)/12$ for the bosonic case). The world-sheet terms $\mu
e^{\sigma_L}$ and $\Lambda e^{\sigma_L}$ can be shown to
correspond to conformal primaries, hence one can treat them as
marginal deformations of the theory with $\mu=\Lambda=0$. The full
theory can be shown to be free of conformal anomaly and has a
well-defined OPE among the fields $X^i(z)$, $\phi(z)$ and
$\sigma_L(z)$.

The valuable  observation of \cite{Chemseddine} is to interpret
(linear combinations of) $\f$ and the Liouville-field $\sigma_L$
as two new additional dimensions of the target-space. The
resulting theory is described by the new action (for the case
$\mu=\Lambda=0$), \be\lab{newaction} {\cal A}_{ws}  =
\frac{1}{\pi} \int_M d^2\sigma \sqrt{\hat{h}}~v_\m X^\m \hat{R} +
\frac{1}{4\pi\a'} \int_M d^2\sigma \sqrt{\hat{h}} \hat{h}^{ab}
\6_a X^\mu \6_b X^{\nu}\eta_{\m\n}+ \frac{1}{2\pi} \int_M
d^2\sigma \sqrt{h} b_{ab} \nabla^a c^b, \ee where $\m$ runs from 0
to $d$ and $\eta_{\m\n}$ is the flat Minkowski space metric. The
additional dimensions $X_0$ and $X^{d}$ is given in terms of
$\sigma_L$ and $\phi$ of the non-critical string as,
\be\lab{fredef} X^0 =  \sqrt{\frac{6\a'}{25-d}} \phi, \qquad X^d =
2\sqrt{\frac{25-d}{6\a'}}\sigma_L + \sqrt{\frac{6\a'}{25-d}} \phi,
\ee and the coefficient $v_{\m}$ in (\ref{newaction}) satisfies
the condition \be\lab{condv} v^\m v_\m =
\frac{25-d}{6\a'}\,\,\,\,\,\, bosonic; \qquad v^\m v_\m =
\frac{9-d}{4\a'}\,\,\,\,\,\, fermionic, \ee where we also show the
condition in the fermionic case for reference.This is nothing else
but the linear-dilaton theory that arises in the IR limit  of our
geometry. Therefore the non-critical string theory in $d-1$
spatial dimensions is equivalent to a $d+1$ dimensional
linear-dilaton theory.

The main advantage of mapping the linear-dilaton theory to the
non-critical string theory is  that the latter provides a
well-defined CFT. The spectrum  as well as the arbitrary genus
path integrals can be evaluated \cite{Chemseddine}. Another great
advantage is that one can generalize this construction to include
fermions on the world-sheet and $N=1$ world-sheet super-symmetry.
This opens the Pandora's box and a rich variety of linear-dilaton
theories can be constructed with various possible GSO projections,
twisted or shifted boundaries for the bosons with various
combinations of NS or R fermions including the heterotic case.

\section{Some background in statistical mechanics}
\lab{statmech}

In this section we review some standard background in statistical
mechanics that we need in the following section.
%The reader may
%wish to skip this and continue with the next section where we
%derive the results reviewed here on the gravity side.

We take the XY model as our example although the approach can be
very general. Consider the spin-model that is described by the
Hamiltonian,
\begin{equation}\label{SpinH}
  H = -J \sum_{\la ij \ra} \vec{s}_i\cdot \vec{s}_j + \cdots
\end{equation}
where $J$ is the interaction strength which is positive for a
ferromagnet and negative for an anti-ferromagnet, $\vec{s}$ is the
spin vector that rotates on a plane and $\la ij \ra$ denotes
nearest neighbor pairs on the lattice. The dimensionality of the
system only shows up in the number of nearest neighbors of a
lattice cite.

\subsection{Landau action}

The Landau action can be defined from the partition function in a
formal way, with no resort to any approximation, see for example
\cite{Henk}. One first introduces a continuous spin density,
\begin{equation}\label{SpinC}
  \vec{s}(x) = \sum_i \delta(x - x_i) s_i.
\end{equation}
Than one defines the {\em local magnetization} as in the following
formal identity
\begin{equation}\label{SpinM}
  1 = \int {\cal D} \vec{m} \delta[\vec{m}(x) - \vec{s}(x) ],
\end{equation}
inserts the RHS of this identity in the partition function
\begin{equation}\label{SpinZ}
  Z =Tr~\le[ \int {\cal D} \vec{m}\delta[\vec{m}(x) - \vec{s}(x)]
  e^{-\beta H}
\ri]
\end{equation}
and finally performs the sum over the spin degrees of freedom
$\vec{s}_i$. This yields,
\begin{equation}\label{SpinL}
  Z_L = \int {\cal D} \vec{m} e^{-\beta~F_L[\vec{m}]}
\end{equation}
that defines the Landau action $F_L$. This is formal but exact. Of
course, the last step of summing over the spin degrees of freedom
is practically impossible in most of the spin systems.

\subsection{Landau approximation}

This corresponds to keeping the most dominant contribution $m_L$
in (\ref{SpinL}):
\begin{equation}\label{Landau}
  Z_L \approx e^{-\beta F_L(m_L)}.
\end{equation}
The most dominant contribution $m_L$ is determined by minimizing
$F_L$. Furthermore, near an order-disorder phase transition the
total magnetization
\begin{equation}\label{Mag}
  \vec{M} = \frac{1}{V_{d-1}} \int d^{d-1}x~\vec{m}(x),
\end{equation}
should go to zero. Then the $O(2)$ symmetry of the XY model
dictates the following general form of the Landau action:
\begin{equation}\label{LanAct}
  F_L = \int d^{d-1}x \le( \a_0(T) |\6 \vec{m}(x)|^2 + \a_1(T) |\vec{m}(x)|^2 + \half \a_2(T) |\vec{m}(x)|^4 +\cdots  \ri)
\end{equation}
where the ellipsis stand for higher (even) powers in $m$.

Basic quantities to be computed, which determine the phase diagram
of the spin system involve the functions $\a_0(T)$, $\a_1(T), \dots$\footnote{In a more complicated system, for
example with a chemical potential, these functions depend on
additional variables.} For example the point $\a_1(T_c)=0,
\a_2(T_c)\ne 0$ corresponds to a second order transition. The
point $\a_1(T_c)= \a_2(T_c) = 0$ corresponds to a tri-critical
point, etc.  A further simplification occurs in the case of
positive $J$ (ferromagnet), when the ground state of the system
should avoid fluctuations in local magnetization $\vec{m}(x)$
because they increase the energy of the system, hence one can
ignore the kinetic term in (\ref{LanAct}) and one can set
$\la\vec{m}(x)\ra=\vec{M}$. This means that the ground state has
isotropic magnetization.

\subsection{Mean-field approximation}

Mean-field approximation  is a standard method to compute the
Landau action (\ref{LanAct}). One expands around a mean-field
$\vec{s}_i = \vec{M} + \delta \vec{s}_i$ where $\vec{s}_i$ denotes
fluctuations around the mean value $\vec{M}$. One substitutes this
in the Hamiltonian (\ref{SpinH}). Ignoring the terms of second and
higher order in $\delta s$ corresponds to the mean-field
approximation. One can clearly compute the partition function,
hence the Landau action $F_L$ analytically within this
approximation. One immediately obtains,
%%%%%%%%%%%%%%%%%%
\be\lab{Zmf} Z = Tr~e^{-\frac{H}{T}} = e^{-\frac{F_L}{T}} =
e^{-\frac{NzJ}{T}|\vec{M}|^2}\prod_i\int_{-\pi}^{\pi} d\theta_i
e^{\frac{2zJ}{T}|\vec{M}| \cos(\theta_i)} \ee
%%%%%%%%%%%%%%%%%%
where $N$ is the total number of sites on the lattice and $z$ is
the number of nearest neighbors. Evaluating the integrals one
obtains the following Landau action:
%%%%%%%%%%%%%%%%%%
\be\lab{Lanmf} F_L^{MF} = -NT\log(2\pi) + NzJ|\vec{M}|^2 -
N\log\le[J_0\le(\frac{2zJ|\vec{M}|}{T}\ri)\ri]. \ee
%%%%%%%%%%%%%%%%%%
The first term corresponds to the entropy and can be ignored for
our purposes here, as it is identical on both phases. The rest
corresponds to the energy of the system. Expanding energy near
$T_c$ where $\vec{M}$ is small, one obtains,
%%%%%%%%%%%%%%%%%%
\be\lab{LanmfEXP} F_L^{MF} = |\vec{M}|^2 N J \le( z -
\frac{z^2J}{T}\ri) +\cdots \ee
%%%%%%%%%%%%%%%%%%
At the second-order transition the mass term should vanish. Thus
one obtains the mean-field value of the transition temperature:
%%%%%%%%%
\be\lab{TcMF} T_c^{MF} = z~J.\ee
%%%%%%%%%
For a square lattice in $d-1$ spatial dimensions $z= 2(d-1)$.

%Expanding $F_L$, obtained by this way, near $T_c$, one computes
%observables like $T_c$ etc.
%
In the context of Landau approximation\footnote{What we mean by
the Landau approximation is summarized by eqs. (\ref{Landau}) and
(\ref{LanAct}) without further specification of the Landau
coefficients $\a_1(T)$, $\a_2(T)$ etc.}, the further {\em mean-field
approximation} means that the Landau coefficients admit a Taylor
series expansion near $T_c$. For example
%%%%%%%%%%%%%%
\be\lab{MFa} \a_{1,MF}(T) = \a_c (T-T_c) + \cdots \ee
%%%%%%%%%%%%%%
and the basic data to determine involve the quantities $T_c$ and
$\a_c$ in this case. This linear dependence is explicit in
(\ref{LanmfEXP}).

Clearly, the mean-field approximation is crude and one can compute
the aforementioned observables to a greater accuracy by the
renormalization group methods or Monte-Carlo simulations. We shall
see below, how the gravitational techniques can go beyond the
mean-field approximation.

\subsection{Gaussian fluctuations}
Fluctuations  around the mean value $\vec{M}$ yield crucial
information on the spin-model, in particular the spin-spin
correlation function and the associated critical exponents near
$T_c$. To compute them, one substitutes\footnote{We drop the
$\delta$ in front of $\vec{s}$ from here on, for notational
convenience.} \be\lab{flus} \vec{m}(x) = \vec{M} + \vec{s}(x), \ee
in (\ref{LanAct}) and expand to second order, ignoring higher
order terms in the mean-field approximation.

One introduces the {\em correlation length} as the natural length
scale,
\begin{equation}\label{corlen}
  \xi(T) = \sqrt{\frac{\a_0}{\a_1(T)}},
\end{equation}
in (\ref{LanAct})\footnote{Strictly speaking this corresponds to
the correlation length for $T<T_c$. In the low-T regime it differs
by a factor of $1\/Sqrt{2}$. However, we are mainly interested in
the scaling of $\xi$ near $T_c$ and the scaling is the same in the
mean-field approximation, from below and above.}. The calculation
of the spin-spin correlation function in the ordered phase
($T<T_c$) within this approximation is standard, see for example
\cite{Henk}.

The only crucial point is that, in the case of {\em spontaneous
symmetry breaking}, as in the XY-model, one has to decompose the
correlation function in parts longitudinal and transverse to the
direction of magnetization $\vec{M}$. Let us denote the components
of $\vec{m}$ by $m_i$. Introducing the unit vector along the
direction of magnetization in the system,
\begin{equation}\label{magun}
  v_i = \frac{M_i}{|\vec{M}|}.
\end{equation}
Then one finds,
\begin{equation}\label{sscor}
  \la m_i(x)~m_j(0)\ra =  |\vec{M}|^2 v_iv_j +
  \frac{T}{4\pi\gamma} \frac{e^{-L/\xi(T)}}{L^{d-3}}
  v_iv_j  + \frac{T}{4\pi\gamma} \frac{1}{L^{d-3}} (\delta_{ij} -
  v_i v_j),
\end{equation}
where the result only depends on the radial distance $L =
|\vec{x}|$ by rotational symmetry. Here the first term comes from
the disconnected part of the correlator, and it is present only in
the ordered phase. The second and third terms are the pieces
longitudinal and transverse to the magnetization respectively. The
longitudinal piece arise from massive fluctuations in the Mexican
hat potential where the typical mass of the fluctuations is given
by $m_l = \xi^{-1}$. This attenuation term is missing in the
transverse correlator because the fluctuations correspond to the
massless Goldstone mode.

The result (\ref{sscor}) is valid in the {\em mean-field
approximation} where we only treat Gaussian fluctuations. This
approximation will break down if the system is strongly
correlated. Generally in condensed matter systems, strong
correlations arise around a phase transition.
Therefore we expect that the gravity dual becomes a
good description in the transition region. The notion of strong
fluctuations is quantified by the Ginzburg criterion:
%%%%%%%%%%
\be\lab{Ginzburg} \xi^{5-d} \ll \frac{4\pi\gamma^2}{\a_2 T_c}, \ee
%%%%%%%%%%
where $\a_2$ and $\a_0$ are the Landau coefficients in
(\ref{LanAct}). For fluctuations of $\vec{s}$ we see that the
mean-field approximation unavoidably breaks down near the
transition where $\xi$ becomes very large\footnote{One can be more
careful by considering the amplitude and phase fluctuations of
$\vec{s}$ separately. In the former case the coefficients $\a_0$
and $\a_2$ stay constant at $T_c$ and from (\ref{Ginzburg}) one
finds that strong correlations are indeed unavoidable in the
transition region where $\xi$ diverges. In the latter case also
the constant $\a_0$ vanishes near the transition, see below. In
the mean-field scaling $\a_0 \sim t$ and $\xi \sim t^{-\half}$.
Therefore one finds that, only for uninteresting dimensions $d>9$
the mean-field approximation is expected to be good for the phase
fluctuations.}.

Beyond the mean-field approximation, one has to take into account
non-trivial self-energy corrections to the correlator that
generically result in the anomalous exponent $\eta$. Therefore the
generic form of the correlator is similar to (\ref{sscor}) but
with the additional anomalous dimension, in addition to the
engineering dimension in the correlator:
\begin{equation}\label{sscorg}
  \la m_i(x)~m_j(0)\ra =  |\vec{M}|^2 v_iv_j +
  \frac{T}{4\pi\gamma} \frac{e^{-L/\xi(T)}}{L^{d-3+\eta}}
  v_iv_j  + \frac{T}{4\pi\gamma} \frac{1}{L^{d-3+\eta}} (\delta_{ij} -
  v_i v_j),
\end{equation}
The mean-field approximation corresponds to $\eta=0$.

Equation (\ref{sscorg}) gives the correlator in the ordered phase
that is dual to the black-hole solution in gravity.  Above the
transition $\la\vec{M}\ra$  vanishes and there is no Goldstone
mode. Hence, the correlator is given by the second piece of
(\ref{sscorg}). This is dual to the thermal graviton gas phase of
gravity.
%However, we shall only be interested in the scaling of observables in the ordered phase below.

In a second order phase transition, the correlation length
diverges as $T\to T_c$, as \be\lab{xisca} \xi(t) \sim t^{-\nu},
\ee where $\nu$ defines a critical exponent. In the mean-field
approximation $\nu = 1/2$.
%The fact $\xi$ diverges near the phase
%transition can be seen in the gravity side as follows. At the
%transition point values of the  on-shell actions on the thermal
%gas and on the black-hole geometries become the same. Futhermore,
%in a continuous transition, the mass of the longitudinal
%deformation that deforms the black-hole into the thermal gas
%solution also vanishes. As $\xi \propto m_l^{-1}$, we see that
%gravity also predicts divergence of $\xi$ at the transition. We
%will calculate the critical exponent $\nu$ from the gravity dual
%below.

Another important point concerns the scaling of $\vec{M}$ near the
transition. It vanishes in a continuous transition, as it is the
order parameter. As explained above one can ignore the kinetic
term in (\ref{LanAct}) in  the ground state of a ferromagnetic
system. Vanishing of $\vec{M}$ near $T_c$ is characterized by the
critical exponent $\beta$\footnote{Not to be confused with the
perimeter of the time-circle $\beta = 1/T$. We  use the same
notation for these quantities and which one is meant should be
clear from the context.}:
\begin{equation}\label{beta}
  \vec{M} \sim t^{\b}.
\end{equation}
In the Landau theory the expectation value $\vec{M}$ is determined
from (\ref{LanAct}) as, \be\lab{MLan} \la  \vec{m}(x) \ra =
\vec{M} = \sqrt{\frac{|\a_1(T)|}{\a_2(T)}}. \ee As $\a_2$ and $\a_0$
stays constant at $T_c$, we see from (\ref{corlen})  that, {\em in
the mean-field approximation} the scaling of $\vec{M}$ and $\xi$
are inversely related, \be\lab{Msca} \vec{M} \propto \xi^{-1}
\propto  t^{\nu}. \ee Therefore, it suffices to determine how
$\vec{M}$ scales  in order to obtain the scaling of $\xi$ {\em in
the mean-field approximation}.

Futhermore, in the mean-field theory the coefficient $\a(T)$ in
(\ref{LanAct}) is assumed to be linear in $t$ near $T_c$,
(\ref{MFa}).
%%%%%%%%%%%%%
%\be\lab{asca} \a(T) = \a_c (T-T_c) + \cdots, \qquad as \,\,\, T\to
%T_c. \ee
%%%%%%%%%%%%%
Comparison with (\ref{MLan}) and  (\ref{Msca}) then shows that in
this approximation:
%%%%%%%%%%%%%
\be\lab{MFexp} \b_{MF} = \nu_{MF} = \half. \ee
%%%%%%%%%%%%%

\subsection{Vanishing of the second sound}

We note that, the vanishing of $\vec{M}$ as $T\to T_c$ from below
implies vanishing of the sound velocity associated with the phase
fluctuations. To see this we represent the
fluctuations $\vec{s}(x)$ in (\ref{flus}) as, \be\lab{fluscomplex}
\vec{m}(x) \to \le(|\vec{M}| + \rho(x)\ri) e^{i \psi +
i\delta\psi(x)}. \ee Substituting this in (\ref{LanAct}), one
obtains the kinetic term for the phase fluctuations,
\be\lab{flucLan}
  \delta F_L \sim \int d^{d-1}x \le( \gamma(T) |\vec{M}|^2 (\delta\psi(x))^2 + \cdots  \ri)
\end{equation}
Therefore, the speed of sound associated with the phase
fluctuations vanish near $T_c$, \be\lab{ssound} c_{\psi} \sim
t^{2\b}, \qquad t\to 0, \ee and the rate it vanishes is determined
by the critical exponent associated with the magnetization
(\ref{beta}).

In the derivation above, we used the Landau approximation and only
kept the leading terms in fluctuations. Therefore, within this
picture the magnetization critical exponent should be the
mean-field one, (\ref{MFexp}). This means that in this picture one
obtains, \be\lab{ssound2} c_{\psi} \sim t, \qquad t\to 0. \ee An
important check for the proposed gravity-spin model correspondence
here is to derive the same scaling law on the gravity side. This
is done in section \ref{grss}.

\subsection{BKT theory}
Finally, we consider the XY-model in two spatial dimensions. As
well-known, in less than three dimensions, long-range order is
destroyed by the IR divergences in fluctuations of the order
parameter \cite{Coleman}, i.e. there are no Goldstone bosons.
However Berezinskii, Kosterlitz and Thouless \cite{BKT} observed
that the 2D XY-model still serves as a good model for
superfluidity. The main observation is that, although there is no
long-range order in the standard sense, there exists a {\em
topological order} below a certain $T_c$, where the vortex- anti
vortex pairs condense. Above $T_c$ the system is the
``deconfined'' phase where the vortex anti-vortex pairs are
liberated and one has a plasma of vortices. All of this is of
course very similar to what happens in QCD, with the replacement
of quarks with ``magnetic" quarks.

Vortices are charged objects. One assigns vortex charge $Q_v = \pm
1$ for the vortices and anti-vortices respectively. The total
vortex charge in a configuration should vanish in two-dimensions
because the gauge field has an IR divergence and the energy of an
unbalanced configuration would diverge, hence its Boltzman factor
in the ensemble vanishes.

What do we expect from the behavior of vortex correlation
functions above and below $T_c$? Let us denote the operator that
creates a vortex, localized at point $x$ by $v(x)$ and the
operator that creates an anti-vortex by $\bar{v}(x)$. For the
reason described above, one cannot have any non-trivial
expectation value neither below nor above the transition, as it
would break the vortex charge\footnote{The TG (BH) phase of
gravity is dual to the high (low) T phase of the XY-model, whence
we denote the vortex correlators accordingly.}:
\begin{equation}\label{v}
  \la v(x) \ra_{TG} = \la \bar{v}(x) \ra_{TG} = \la v(x) \ra_{BH} = \la \bar{v}(x)
  \ra_{BH} = 0.
\end{equation}
The phase of the system can be probed by the two-point function of
the vortex-anti-vortex pair, however. One finds that the two-point
function is exponential in the high T phase, hence there exists a
correlation length, whereas it has power-law in the low T phase:
\begin{equation}\label{vv}
  \la \bar{v}(x) v(0) \ra_{TG} \sim e^{-L/\xi(T)}; \qquad
  \la \bar{v}(x) v(0) \ra_{BH} \sim L^{p(T)}.
\end{equation}
where $L= |x|\gg 1$ and p is some power.

In systems with more than two spatial dimensions, one can still
consider vortex configurations, however they would not have as
significant effects on the phase of the system as in 2D. The
relative objects in higher dimensions would be the vortex-lines,
planes etc, that are analogous to monopole configurations in gauge
theories.

\section{Fundamental string action}\lab{AppD}

Here we fill in the details of the computations in section
\ref{sec8}. The F-string action involves two terms\footnote{The
role of a non-trivial B-field is already discussed in the test and
we shall ignore it here.}:
%%%%%%%%%%%%%
\be\lab{D0}
S_{NG}  = S_G + S_\f
\ee
%%%%%%%%%%%%%
where
\bea\label{D1}
  S_G  &=& \frac{1}{2\pi\ell_s^2}\int_{\s_0}^{\s_f} d\s d\tau~\sqrt{h} h^{ab}\6_a X^{\m} \6_b X^{\n} G_{\m\n},\\
S_\f &=& \frac{1}{2\pi}\int_{\s_0}^{\s_f} d\s d\tau~\sqrt{det\,\,
h_{ab}}~R^{(2)}~\f(X(\s)),\lab{D2} \eea
%%%%%%%%%%%%%
where $\ell_s$ is the string length, $r_f$ is some turning point
of the string embedding  that will be specified in the following,
$R^{(2)}$ is the Ricci scalar that corresponds to the world-sheet
metric $h_{ab}$,
 and $G_{\m\n}$ is the BH metric in the string frame:
%%%%%%%%%%%%%
\begin{equation}\label{D3}
  ds^2_s = e^{2A_s(r)}\le( f^{-1}(r)dr^2 + dx_{d-1}^2 + dx_0^2 f(r) \ri), \qquad  A_s(r) = A(r) + \frac{2}{d-1}\f(r),
\end{equation}
%%%%%%%%%%%%%
The on-shell value of the action depends on the boundary
conditions of the string. In this paper we consider three separate
cases: \bn \item The Polyakov loop, \item The Polyakov correlator
(The Wilson loop),
\item The 't Hooft loop. \en

\subsection{The Polyakov loop}\lab{AppP}

First of all, we fix the world-sheet diffeo-moprhism invariance as
%%%%%%%%%%%%%%%
\be\lab{D4} \s= x_0, \qquad  \tau=r, \ee
%%%%%%%%%%%%%%%
where $x_0$ is the Euclidean time.

We consider  two separate geometries  that the string is embedded: a) the thermal gas solution with
topology $S^1\times B_d$ where $B_d$ is a d-dimensional ball and
b) the black-hole with topology $D^2\times B_{d-1}$ where $D^2$
is a 2 dimensional disk.

The boundary of $B_d$ is $S^{d-1}$ but we are interested in the
flat limit where $S^{d-1}\to R^{d-1}$. Therefore, in both cases
the boundary of space-time becomes $S^1\times R^{d-1}$. The string ends
on a curve $C$ on the boundary where $C = S^1\times P$, $P$ being
a point x on $R^{d-1}$ that we can take as the origin with no loss
of generality. In case a, the only string solution with $C$ as
the boundary is the semi-infinite cylinder $S^1\times R$ where R
is isomorphic to the radial coordinate r and $P$ is the point that
corresponds to the endpoint of the line $R$ at $r=0$. In case b,
the only string solution that ends on $C$ is isomorphic to $D^2$,
hence it wraps the entire $D^2$ part of the bulk geometry.

Clearly, the action (\ref{D1}) diverges in case a because $\sigma_f= r_f = \infty$. Therefore
(\ref{D0}) will diverge unless there is some cancellation between
(\ref{D1}) and (\ref{D2}). In the following we show that
(\ref{D2}) is finite in all of the cases under consideration. Therefore the result in case a is
that the Polyakov-loop vanishes.

Now, consider the case b, i.e. the black-hole geometry. As
explained above, the string wraps a $D^2$.  Then, the radius of
$D^2$ is given by $r_h$. Clearly, both (\ref{D1}) and (\ref{D2})
are finite hence contribute to the energy for an arbitrary but
finite $r_h$. However, we are interested in the limit $T\to T_c$
i.e. $r_h\to\infty$ and we ask how do (\ref{D1}) and (\ref{D2})
scale with $r_h$.

Let us first consider the scaling of $S_\f$ with $r_h$ in the
limit $r_h\to\infty$. The world-sheet metric $h$ in the gauge
(\ref{D4}) is given by,
%%%%%%%
\be\lab{D5} ds_{ws}^2 =  e^{2A_s} \le( dx_0^2 f + dr^2/f \ri) \ee
%%%%%%%
One finds,
%%%%%%
\be\lab{D6}
\sqrt{h} R^{(2)} = - 2A_s'~f' - 2~f~A_s''  - f'',
\ee
%%%%%%
where prime denotes $d/dr$.  Using eqs. (\ref{fs}) and
(\ref{delAs}) one finds that,
 %%%%%%%%%%%%%%%%%%%%%
\be\lab{delAs}
A_s(r) \propto \left\{ \begin{array}{ll}
e^{-\kappa' r}, & \mathrm{case\,\, i}, \\
r^{-\a}, & \mathrm{case\,\, ii},
\end{array}\right.
\ee
%%%%%%%%%%%%%%%%%%%%%
where $\kappa' = \kappa V_\infty/2$ and $\a$ are positive
constants. Thus we find that, even though $\f$ diverges linearly
as $r\to\infty$, the integrand in $S_\f$ vanishes in this limit.
Therefore the contribution of $S_\f$ is finite  in the case b,
also in the  limit $r_h\to\infty$.

Now we consider the scaling of $S_G$ with $r_h$.  The metric in
the string frame is given by (\ref{D3}):
%%%%%%%%%%%%%%
%\begin{equation}\label{D10}
%  ds^2_s = e^{2A_s(r)}\le( f^{-1}(r)dr^2 + dx_{d-1}^2 + dx_0^2 f(r) \ri), \qquad  A_s(r) = A(r) + \frac{2}{d-1}\f(r),
%\end{equation}
%%%%%%%%%%%%%%
From (\ref{D1}) one immediately finds that,
%%%%%%%%%%
\be\lab{D11} S_G = \frac{T^{-1}(r_h)}{2\pi\ell_s^2}
\int_{\eps}^{r_h} e^{2A_s(r)}dr. \ee
%%%%%%%%%%
where $\eps$ is the UV cut-off that should be removed by a proper
counter-term action, but it will be irrelevant for the scaling of
the $S_G$ with $r_h$.

In passing we note the trivial result on the TG solution. This corresponds to (\ref{D11})
where $r_h$  replaced with $\infty$ and $A_s$ replaced with  $A_{s,0}$. As $A_{s,0}$
goes to zero in the limit $r\to\infty$ (\ref{D11}) is divergent and the magnetization that
corresponds to exponential of $-S$ vanishes.

In the limit $r_h\to\infty$ in case of the BH, one has $S_G\propto
\lim_{r_h\to\infty} e^{2A_s(r_h)} r_h$, whence it diverges
linearly, whereas $S_\f$ remains finite as  we showed above. Thus,
using the fact that $A(r_h)\to 0$ (\ref{delAs}), one finds,
%%%%%%%%%%
\be\lab{D12} P[C] \propto e^{-S_{NG}}\propto  \exp\le(-
\frac{T^{-1}(r_h)}{2\pi\ell_s^2}~r_h\ri ),  \qquad r_h\to\infty
\ee
%%%%%%%%%%
The next task is to express $r_h$ in terms of the normalized
temperature $t$ (\ref{t}). We know how $\f_h$ can be expressed in
t from (\ref{ast}). Thus if suffices to find $r_h$ in $\f_h$ in
large $\f_h$ limit. This is given by \ref{fs}:
%%%%%%%%%%
\be\lab{D16}
\lim_{\f_h\to\infty} \f_h =\frac{\sqrt{V_\infty}}{2}~r_h.\ee
%%%%%%%%%%
Now, (\ref{ast}) yields,
%%%%%%%%%%%%%%%%%%%%%
\be\lab{D17}
\f_h  = \left\{ \begin{array}{ll}
-\frac{1}{\kappa} \log(t/C), & \mathrm{case\,\, i}, \\
\le(\frac{t}{C}\ri)^{-\frac{1}{\a}}, & \mathrm{case\,\, ii},
\end{array}\right.
\ee
%%%%%%%%%%%%%%%%%%%%%
Therefore (\ref{D16}) gives,
%%%%%%%%%%%%%%%%%%%%%
\be\lab{D18}
r_h  = \left\{ \begin{array}{ll}
-\frac{2}{\sqrt{V_\infty}~\kappa} \log(t/C), & \mathrm{case\,\, i}, \\
\frac{2}{\sqrt{V_\infty}}\le(\frac{t}{C}\ri)^{-\frac{1}{\a}}, & \mathrm{case\,\, ii}.
\end{array}\right.
\ee
%%%%%%%%%%%%%%%%%%%%%
Substitution of (\ref{D18}) in (\ref{D12}), use of the fact that
$T\to T_c$ in the limit $r_h\to\infty$ finally yields
%%%%%%%%%%%%%%%%%%%%%%%%%%%%%%%%%%%%%%%%%%%%%%%%%%%%%%%%%%%%%%%%%
\bea
  \mathrm{Case\,\, i:}\qquad e^{-S_{NG}} &\propto& t^{\frac{4}{\kappa~V_s}}\qquad
  t\to 0 \lab{ezcase1}\\
  \mathrm{Case\,\, ii:}\qquad e^{-S_{NG}} &\propto& e^{\frac{4}{V_s} \le(\frac{t}{C}\ri)^{-\frac{1}{\a}}}, \qquad
  t\to 0 \lab{ezcase2},
\eea
%%%%%%%%%%%%%%%%%%%%%%%%%%%%%%%%%%%%%%%%%%%%%%%%%%%%%%%%%%%%%%%%%
%where the constants $\a,\k$ and $C$ are defined in (\ref{case1}) and (\ref{case2}).
In the arguments above, we ignored the issue of renormalizing the
UV divergence in the action (\ref{D11}). The renormalization can
be done by subtracting the self-energy of the single quark that
corresponds to a single disconnected string solution that hangs
from $\epsilon$ to $r_f$. This renormalization is considered in
detail at the end of App \ref{AppPP} below. The counter-term
action is  the same (up to a factor of 2) as there. The same
conclusion reached there---that ignoring the renormalization does
not affect the leading term in $L$ in the large $L$ limit---is
also valid here.

\subsection{The Polyakov loop correlator}
\lab{AppPP}

We compute the on-shell string action that corresponds to the Polyakov-loop correlator here. We will consider the BH geometry, and the same problem
on the TG geometry can be obtained from our result below, see below eq. (\ref{P9}).

The fundamental string action is \be\lab{P1} S_{F1} =
\frac{1}{2\pi\ell_s^2} \int d\tau d\sigma \sqrt{h} \le[\le(h^{ab}
g^s_{\m\n}  + i \e^{ab}B_{\m\n}\ri) \6_aX^{\m} \6_bX^{\n} +
\ell_s^2 \f(X) R^{(2)}\ri], \ee where $h$ is the induced metric,
$g^s_{\m\n}$ is the target-space metric of the BH geometry in the
string-frame
\begin{equation}\label{P1s}
  ds^2_s = e^{2A_s(r)}\le( f^{-1}(r)dr^2 + dx_{d-1}^2 + dx_0^2 f(r) \ri), \qquad  A_s(r) = A(r) + \frac{2}{d-1}\f(r),
\end{equation}
and $R^{(2)}$ is the world-sheet Ricci scalar. The string that
corresponds to the to Polyakov-loop correlator $\la P^*(x)
P(0)\ra$ is a connected string solution with end-points $x$ and
$0$ on the boundary. With no loss of generality, we take these
points to lie on the same axis that we call $x_1$. The string that
connects these points extends  towards the deep-interior of the
$d+1$ dimensional target-space in the r-direction. Thus, a good
choice of the gauge-fixing is given by $\sigma = x_0$, $\tau=x_1$
where $t$ is the Euclidean time coordinate with perimeter $1/T$.
The string should also wind-around the time-circle. As we look for
a solution that only depends on $r$, the $\tau$-integral factors
out and yields a multiplicative factor of $1/T$.

First of all, the $B$-coupling cannot arise here because it yields
an imaginary contribution, whereas the Polyakov-loop correlator is
manifestly real. Thus we can drop the second term in (\ref{P1}).
Secondly, one can show that the contribution of the
dilaton-coupling is sub-dominant with respect to the first term in
(\ref{P1}). One can see this as follows. Let us assume that indeed
the dilaton-coupling is sub-dominant. Then, the induced metric is
solely determined by the first term in (\ref{P1}) \be\lab{P2}
h_{ab} =  \6_aX^{\m} \6_bX^{\n} g^s_{\m\n}. \ee Given this one can
compute the Ricci scalar: \bea\lab{P3}
\sqrt{h} R^{(2)} & = & \le( f+Z^2\ri)^{-\frac32} \bigg[ -Zf^{'2} + 4f^2(A_s'Z' + Z A_s'') + 2Z^3 (2A_s'f'+f'')\nn\\
{}& &+ 2f(Z'f'+2Z^3A_s'' +Z(A_s'f'+f'')\bigg] \eea where we
defined, \be\lab{P4} Z(r) = \le(\frac{dx_1}{dr}\ri)^{-1} =
\sqrt{f(r)} \sqrt{\frac{e^{4A_s(r)}f(r)}{e^{4A_s(r_f)}f(r_f)}-1}.
\ee In this paper, we are interested in how the Polyakov-loop
correlator scales near $T_c$ and for large $L$. The latter is
given by, \be\lab{P5} L = 2\int_{\eps}^{r_f} dx_1 =
2\int_{\eps}^{r_f} Z^{-1}(r) dr, \ee where $\e$ is some cut-off
near the boundary\footnote{We will comment on how to remove the
cut-off by appropriate renormalization in the end of this
Appendix. However, as we are interested in the limit
$r_f\to\infty$ in this paper, the $\e$ dependence can be kept, it
will not contribute to the results.}. The limit $T\to T_c$ and $L$
large corresponds to \be\lab{P6} r_h\gg r_f\gg 1. \ee In the limit
$r_h\gg 1$ one can show that the blackness function can be
replaced by $f\approx 1$ everywhere except $r=r_h$ where it
vanishes (see \cite{GKMN2} for a derivation, that immediately
carries over here). At the same time for $r_f\gg 1$ the
scale-factor $A_s(r_f)\to 0$. Then, using the sub-leading terms in
$A(r)$ (\ref{delAs}), one finds, \be\lab{P7} Z \sim
e^{-\kappa\sqrt{V_{\infty}}/4~r} \quad\quad ({\rm Case\,\, i});
\qquad Z \sim   r^{-\a/2} \quad\quad ({\rm Case\,\, ii}), \ee The
constants $\kappa$ and $\a$ are defined in  (\ref{case1}) and
(\ref{case2}).

In the limit $r_f\to\infty$ the dilaton $\f(r_f)$  diverges as in
(\ref{fs}), therefore one may worry that the last term in
(\ref{P1}) contributes significantly. On the other hand, we see
from (\ref{delAs}) that the $A_s'$ factors (and similarly
$A_s''$)factors in (\ref{P3}) are suppressed exponentially (case
i) and with power-law (case ii) in the region $r\gg 1$:
%%%%%%%%%%%%%%%%%%%%%
\be\lab{P71}
A_s'(r) \sim  \left\{ \begin{array}{ll}
%\frac{\tilde{C}\kappa}{\kappa+\tilde{\g}}e^{-\kappa\f} \le(\gamma_E + \psi^{(0)}(-\frac{\kappa}{\tilde{\g}})\ri)+\cdots, & \mathrm{case\,\, i}, \\
e^{-\kappa\sqrt{V_\infty}~r/2}, & \mathrm{case\,\, i}, \\
r^{-\a}, & \mathrm{case\,\, ii},
\end{array}\right.
\ee
%%%%%%%%%%%%%%%%%%%%%
and the $Z$ factors in (\ref{P3}) are suppressed as (\ref{P7}).
Using the latter in (\ref{P5}) we find, in the region $L\gg 1$:
\be\lab{P72} L \sim   e^{\kappa \sqrt{V_\infty}/4~r_f} \quad\quad
({\rm Case\,\, i}); \qquad    L\sim   r_f^{-\a/2+1} \quad\quad
({\rm Case\,\, ii}), \ee Thus, we conclude that, in the regime
$r_h\to\infty$, and $r_f\gg 1$, the dilaton contribution to the
string action (the last term in (\ref{P1})) scales as
%%%%%%%%%%%%%%%%%%%%%
\be\lab{P73}
S_{\f} \sim  \left\{ \begin{array}{ll}
L^{-1}, & \mathrm{case\,\, i}, \\
L^{\frac{2-\a/2}{1+\a/2}}, & \mathrm{case\,\, ii},
\end{array}\right.
\ee
%%%%%%%%%%%%%%%%%%%%%
in the region $L \gg 1$.

Below, we show that---with the assumption that one can drop the dilaton contribution in (\ref{P1})---the string action scales linearly in $L$:
\be\lab{P74}
S_{G} \propto L, \qquad L\gg 1,
\ee
where by $S_{G}$ we denote the on-shell contribution of the first term in (\ref{P1}) with the assumption that  the last term can be dropped.
{\em Thus, we can safely conclude that, our assumption in the  beginning of this discussion, namely that the dilaton-coupling in (\ref{P3}) does not contribute to the string-solution in the limit (\ref{P6}) is valid in case i, and also in case ii, unless $\a \leq 1$.}\footnote{The case of BKT scaling corresponds to $\a=2$ and it is in the safe region.} In the case of $\a=1$  our assumption above is violated as both $S_G$ and $S_\f$ and one should solve the full action in (\ref{P1}). In the case $\a<1$ the metric term $S_G$ is sub-dominant to $S_\f$ and one can turn the aforementioned argument in favor of $S_{\f}$, i.e. one can assume that $S_{\f}$ is the leading contribution in the limit $L\gg 1$. These cases provide interesting examples that the $\f$-coupling becomes crucial in determining the behavior of the Polyakov loop correlator (similarly the quark-antiquark potential in holography), however they are not of direct interest to us in this work.

Thus, we can drop the last two terms in (\ref{P1}) and the
solution for $h_{ab}$ is given by (\ref{P2}). The on-shell action
then becomes a Nambu-Goto action. Given the target-space geometry
(\ref{P1s}) finding the on-shell NG action is a standard exercise
(see for example  \cite{Kinar}): \be\lab{P8} S_{F1} =
\frac{1}{2\pi\ell_s^2~T} \int_{\epsilon}^{r_f}  dr
\frac{e^{2A_s(r)}}{\sqrt{1-
\le(\frac{e^{4A_s(r_f)}f(r_f)}{e^{4A_s(r)}f(r)}\ri)}}. \ee Then
the action as a function of $L$ is given by the parametric
solution of (\ref{P5}) and (\ref{P8}).

We are interested in the limit (\ref{P6}) where we can replace
$f\approx 1$ throughout the entire range of $r$, up to $r_f$ and
up to the value of $r_f$ slightly smaller than $r_h$ (in the limit
$r_h\to\infty$, $T\to T_c$.   Comparing   (\ref{P5}) and
(\ref{P8}) and using the fact that $A_s\to 0$ in the numerator of
(\ref{P8}) we conclude, (assuming $\a>1$ in case ii) \be\lab{P9}
S_{F1} \to  \frac{1}{2\pi\ell_s^2~T_c}~L, \qquad as\,\,\, L\to
\infty. \ee The on-shell action of the same type of connected
string solution in the TG geometry is given by replacing the
metric functions above by $f=1$, $A_s\to A_s^0$ and $\f\to \f_0$.
The same arguments above then directly carry over to this case.

Finally, let us discuss the renormalization of the action
(\ref{P8}): We regulated the action by inserting a cut-off at
$r=\epsilon$ close to the boundary. As one removes the regulator,
$\epsilon\to 0$, the on-shell action diverges, due to infinite
area of the space-time metric near the boundary\footnote{No where
in the paper, we explicitly specified the form of the metric near
the boundary, however we assume that it is asymptotically AdS.}.
One should remove the regulator by adding a counter-term action
designed to cancel the divergences. In the case of Polyakov-loop
correlator in AdS, this is done by subtracting the self-energie
that corresponds to two disconnected strings hanging from
$\epsilon$ to $r_h$ \cite{PP1}, \cite{PP2}: \be\lab{P10} S_{ren} =
\frac{1}{2\pi\ell_s^2~T} \int_{\epsilon}^{r_h} e^{2A_s(r)} dr. \ee
Clearly, subtraction of $S_{ren}$ from $S_{F1}$ in (\ref{P8})
removes the divergence at $\epsilon=0$. However, as criticized in
\cite{YK}, this makes the counter-term action temperature
dependent. Moreover, (\ref{P10}) is divergent in the limit
$r_h\to\infty$, that is the limit that we are interested in.
Instead, here we propose to regulate the action by subtracting the
self-energy of two quarks of the Polyakov-loop correlator as in
\cite{Kinar}: \be\lab{P11} S_{ren} = \frac{1}{2\pi\ell_s^2~T}
\int_{\epsilon}^{r_f} e^{2A_s(r)} dr. \ee Clearly this removes the
UV divergence in (\ref{P8}) in the limit $\epsilon\to 0$ and
yields a finite result for finite $L$. It is also apparent from
(\ref{P11}) that it diverges in the limit $L\to\infty$ but this
divergence is physical. The only point that we have to worry about
is that, it does not  diverge faster than (\ref{P9}). This is
required for the consistency of our discussion above, where we
ignored explicitly regulating the action.

Let us now determine precisely how it scales with $L$ in this
limit. The region $L\gg 1$ corresponds to $r_f\gg 1$. As $A_s\to
0$ in the integrand in this region, (\ref{P11}) scales linearly in
$r_f$. One can convert $r_f$ to $L$ by using (\ref{P72}). We
consider case i first. In this case one finds, \be\lab{P12}
S_{ren}^i \to \frac{1}{2\pi\ell_s^2~T} \frac{4}{\kappa
\sqrt{V_\infty}}~\log~L, \qquad L\gg 1. \ee From (\ref{Tc}) one
finally obtains, \be\lab{P121} S_{ren}^i \to
\frac{8}{V_s\kappa}~\log~L, \qquad L\gg 1. \ee As it scales like
$\log(L)$ our arguments above by neglecting the counter-term
action in the large $L$ region is thus justified. However, we note
that this term  {\em does} contribute the final result in the
spin-spin correlator as it affects the sub-dominant terms that are
denoted by ellipsis in (\ref{sscongr}). For example, in the
mean-field approximation (\ref{VsMF}) for a second-order phase
transition $\kappa=2$ one finds that the coefficient in
(\ref{P121}) is precisely 1. In order to determine the exact power
of the sub-leading terms in (\ref{sscongr}), one should take this
contribution into account in addition to possible other $\log~L$
terms that may arise from the expansion of the leading piece
(\ref{P8}) and possible quantum fluctuations of the string.

 In the case ii, a similar calculation shows that
 \be\lab{P13}
 S^{ii}_{ren} \sim L^{\frac{1}{1-\a/2}}.
 \ee
 As this is always sub-dominant to the linear behavior in (\ref{P9}) for $\a>0$,
 our discussion above by neglecting the counter-term action above is again justified.

 The same calculation should be done for dilaton-coupling in (\ref{P1}).  Of course the counter-term action has the same physical origin. It comes from
 the dilaton-coupling  in the string solution that corresponds to two quarks in the Polyakov-loop correlator. This is given by the dilaton-contribution
 for two disconnected strings hanging from $\epsilon$ to $r_f$:
\be\lab{P14} S_{ren, \f} = -\frac{1}{2\pi T} \int_{\epsilon}^{r_f}
(2A_s'(r)f'(r)+2f(r) A_s''(r) +f''(r))~\f(r) dr. \ee
 This is obtained by calculating the world-sheet Ricci scalar $R^{(2)}$ in (\ref{P1}) for the disconnected string solution.
 We ask whether this cancels out the UV divergence in the dilaton-term in (\ref{P2}) in the limit  $\epsilon\to 0$. The integrand of the latter is
 given by $\f(r)$  times (\ref{P3}). Near $r\approx 0$ the function $Z$ in (\ref{P4}) diverges, thus only the terms proportional to $Z^3$ in the square brackets in (\ref{P3}) survive. This is precisely the form that one has in (\ref{P14}), thus one indeed see that the counter-term given in (\ref{P14}) does the job. As it scales exactly the same way in $L$ as (\ref{P73}), we see that, also the {\em renormalized} dilaton-coupling is subdominant to the area term
 $S_G$, hence our arguments  by neglecting the renormalization above are still justified.

 \subsection{The `t Hooft loop}
\lab{AppT}

Here we detail the computation of the 't Hooft-Polyakov loop
correlator which we propose to correspond to the vortex-anti vortex pair $\la \bar{v}(x) v(0)\ra$. This is represented by a $D1-\bar{D}1$ brane pair
ending  on the boundary at  points $x$ and 0, and wrapping the time circle. Thus we want to compute the RHS of
\be\lab{T1}
\la \bar{v}(x) v(0)\ra \propto e^{-S_{D1}},
\ee
with
\begin{equation}\label{T2}
  S_{D1} = -T_1 \int d\s d\tau e^{-\f} \le( det[h_{ab} +
  b_{ab}]\ri)^\half,
\end{equation}
where $T_1$ is the D-string tension and we defined,
\begin{equation}\label{T3}
h_{ab}= g^s_{\m\n} \6_a X^{\m} \6_b X^{\n}, \qquad b_{ab}=
B_{\m\n} \6_a X^{\m} \6_b X^{\n}.
\end{equation}
Here $g^s_{\m\n}$ is the string-frame metric
\begin{equation}\label{T4}
  ds^2_s = e^{2A_s(r)}\le( f^{-1}(r)dr^2 + dx_{d-1}^2 + dx_0^2 f(r) \ri), \qquad  A_s(r) = A(r) + \frac{2}{d-1}\f(r),
\end{equation}
in the BH phase. In the TG phase the metric is given by the
replacement $f\to 1$ and $A\to A_0$ in (\ref{sBH}).

We choose the gauge, $\sigma = x_0$ and $\tau=x_1$. Here $x_1$ is
one of the spatial directions of the spin-model on which the
points $x$ and 0 lie. From (\ref{T3}) and (\ref{T4}) one finds,
\bea
h_{00} &=& e^{2A_s}f; \,\,\, h_{11} = e^{2A_s}\le( 1 +  \le(\frac{dr}{dx_1}\ri)^2 \frac{1}{f} \ri); \,\,\, b_{10} = - b_{01} = b  \le(\frac{dr}{dx_1}\ri)\lab{T5}\\
h_{10} &=& h_{01} = b_{00} = b_{11} = 0,\lab{T6} \eea where $b$ is
a constant given by the $(r,x_0)$ component of the B-field $b=
B_{r0}$. From (\ref{T2}) we find the action for the $D1-\bar{D}1$
pair,
\begin{equation}\label{T7}
  S_{D1} = -\frac{2}{T_1 T} \int_{\epsilon}^{r_f} dr e^{-\f} e^{2A_s} \sqrt{f +\le(\frac{dr}{dx_1}\ri)^2( 1+ b^2e^{-4A_s})}.
\end{equation}
The action only contains derivatives of $r$ explicitly, thus the
corresponding Hamiltonian should be a constant of motion. Let us
define the following functions for notational simplicity:
\be\lab{T8} \tilde{f}^2 = e^{4A_s-2\f}f,\qquad  \tilde{g}^2 =
e^{4A_s-2\f}(1+ b^2e^{-4A_s}). \ee Then the Lagrangian in
(\ref{T7}) is \be\lab{T9} \cL = \sqrt{\ft^2 + \dot{r}^2 \gt^2}.
\ee The canonical momentum that corresponds to $r$ and the
Hamiltonian is given by, \be\lab{T10} p_r = \frac{d\cL}{d\dot{r}}
= \frac{\gt^2\dot{r}}{\sqrt{\ft^2+\dot{r}^2\gt^2}}, \,\,\, \cH  =
p_r \dot{r} - \cL = -\frac{\ft^2}{\cL}. \ee We are interested in
the {\em connected} $D1$ - $\bar{D}1$ pair. This is given by a
curve on the $(x_1,r)$ plane that ends on the
points\footnote{$\epsilon$ is the boundary cut-off. We shall keep
it explicit here, and it can be removed by renormalizing the
D-string action by adding counterterms   as in section
{\ref{AppPP}}.}$(x,\epsilon)$ and $(0,\epsilon)$  and has a
turning-point at $r_f$, at which $\dot{r}=0$. We assume that the
curve is symmetric around the turning point which corresponds to
the point $(x/2,r_f)$ on the $(x_1,r)$ plane. As the Hamiltonian
is conserved and independent of $x_1$ it can be computed at $r_f$.
Changing variable to $r$ one has, $\cH = - \ft(r_f)$. Then the
Lagrangian is given by $\cL  = \ft^2(r)/\ft(r_f)$ and one obtains
from this, the first-order e.o.m that determines the shape of the
curve: \be\lab{T11} \le(\frac{dx_1}{dr}\ri)^{-1} = \pm
\frac{\ft(r)}{\gt(r)} \frac{1}{\ft(r_f)} \sqrt{\ft^2(r) -
\ft^2(r_f)}. \ee One chooses the plus sign for $x_1\in (0,x/2)$
and the minus sign for $x_1\in (x/2,x)$.  Substituting this in the
Lagrangian (\ref{T9}) one obtains the on-shell action,
\be\lab{T12} S_{D1} = -\frac{2}{T_1T}\int_{\epsilon}^{r_f} dr
\frac{\gt(r)}{\sqrt{1-\frac{\ft^2(r_f)}{\ft^2(r)}}}. \ee The
distance $L= |x|$ between the end-points on the boundary is given
by, \be\lab{T13} L =\int_0^x dx_1 = 2 \int_{\epsilon}^{r_f} dr
\frac{\gt(r)\ft(r_f)}{\ft^2(r)}\frac{1}{\sqrt{1-\frac{\ft^2(r_f)}{\ft^2(r)}}},
\ee where we used (\ref{T10}). The on-shell D-string action, hence
the vortex-anti-vortex correlator in (\ref{T1})  is given by the
parametric solution of (\ref{T12}) and (\ref{T13}) in favor of
$S_{D1}(L)$.

In fact, this solution will only be valid for particular values of
$L$ less than some $L_{max}$ {\em both for the BH and the TG
background}. The reason is that, the connected D-string solution
ceases to exist beyond this value. To see this we note, first of
all, that the integrand in (\ref{T13}) is positive definite, hence
$L$ increases with increasing $r_f$. Then, in case of the BH, when
$r_f=r_h$ at which the $D1$ and the $\bar{D}1$ falls into the
horizon. This corresponds to the maximum value of $L$ for the
connected D-string solution. At the technical level, one can see
this by observing that the integrand in (\ref{T13}) vanishes for
$r_f=r_h$, hence $L$ is bounded by the value $L_{max} = L(r_h)$.
Beyond this point, the RHS of (\ref{T1}) is dominated by {\em the
exchange diagram} as explained in section \ref{hooft}.

Let's now consider the TG solution. The distance $L$ is again
given by (\ref{T13}) but  this time the metric functions are given
by $A= A_0$ and $f=1$. One also has, $\f= \f_0$: \be\lab{T14} L =
2 \int_{\epsilon}^{r_f} dr
\frac{\gt_0(r)\ft_0(r_f)}{\ft_0^2(r)}\frac{1}{\sqrt{1-\frac{\ft_0^2(r_f)}{\ft_0^2(r)}}},
\qquad \tilde{f}_0^2 = e^{4A_{s,0}-2\f},\,\,\, \tilde{g}_0^2 =
e^{4A_{s,0}-2\f}(1+ b^2e^{-4A_{s,0}}). \ee In order to see that
(\ref{T14}) is bounded from above in the entire range
$r_f=\epsilon$ to $r_f=\infty$, one divides the range into two
parts $(\epsilon,r_1)$ and $(r_1,r_f)$\footnote{This argument is
first given in \cite{GKN}. We warn the reader that there are typo
errors in that reference. Here we prefer to present the argument
independently.}, where  $r_1$  is large enough so that we can
assume that in the second range the background functions are
approximately given by their asymptotic forms: \be\lab{T15} \f(r)
\approx \f_\infty~r, \qquad A(r) \approx -A_\infty~r. \ee Let us
denote the two contributions in (\ref{T14}) from the ranges
$(\epsilon,r_1)$ and $(r_1,r_f)$ as $L_1$ and $L_2$ respectively.
One first shows that $L_1$ is bounded from above: \bea\lab{T16}
L_1 &=&  2 \int_{\epsilon}^{r_1} dr
\frac{\gt_0(r)}{\ft_0(r)}\frac{1}{\sqrt{\frac{\ft_0^2(r)}{\ft_0^2(r_f)}-1}}
= 2 \frac{\ft_0(r_f)}{\ft_0(r_1)}\int_{\epsilon}^{r_1} dr
\frac{\gt_0(r)}{\ft_0(r)}\frac{1}{\sqrt{\frac{\ft_0^2(r)}{\ft_0^2(r_1)}-
\frac{\ft_0^2(r_f)}{\ft_0^2(r_1)}}} \nn\\ {}&<& 2
\frac{\ft_0(r_f)}{\ft_0(r_1)}\int_{\epsilon}^{r_1} dr
\frac{\gt_0(r)}{\ft_0(r)}\frac{1}{\sqrt{\frac{\ft_0^2(r)}{\ft_0^2(r_1)}-
1}} = 2 \frac{\ft_0(r_f)}{\ft_0(r_1)} L(r_1). \eea In the second
line we used the fact that $\ft_0(r)$ is a monotonically
decreasing function\footnote{This is clear from the definition
(\ref{T14}). The exponent is $4A_0(r) + 2\f_0(r)/3$ where $A_0$ is
the Einstein frame scale factor. By assumption, $A_0$ is a
monotonically decreasing and $\f_0$ is a monotonically increasing
function. We also know that the combination $4A_{s,0}(r) = 4A_0(r)
+ 8\f_0(r)/3$ is monotonically decreasing and having an asymptotic
minimum at $r=\infty$. Thus $4A_0(r) + 2\f_0(r)/3$ should be
monotonically decreasing, with no minimum.}. Therefore the only
possible divergence in $L$ can come from $L_2$: \be\lab{T17} L_2 =
2 \int_{r_1}^{r_f} dr
\frac{\gt_0(r)}{\ft_0(r)}\frac{1}{\sqrt{\frac{\ft_0^2(r)}{\ft_0^2(r_f)}-1}}
\approx  \frac{\sqrt{1+b^2}}{\f_\infty}
\int_0^{2\f_\infty(r_f-r_1)}\frac{dy}{e^y-1}, \ee where we used
that, by assumption the background functions are given by the
asymptotic forms (\ref{T15}) in this range of $r$. We see that
this is bounded from above. In the limit $r_f\to\infty$ one finds,
\be\lab{T18} \lim_{r_f\to\infty} L_2(r_f) \approx
\frac{\pi\sqrt{1+b^2}}{\f_\infty}. \ee In fact, with little more
effort, one can show that the RHS of (\ref{T18}) is the upper
bound on $L_2$. Thus $L$ is bounded from above and there is a
maximum value $L_{max}$ that is reached at some point
$r_f=r_{max}$. It is clear from the calculation above that this
point is independent of temperature in the TG phase. Beyond this
point, the connected D-string solution ceases to exist and the
vortex correlator is determined by the exchange diagram, cf.
section \ref{hooft}.

As a last comment, we observe that the calculation above  could be applied directly to the BH case, just by replacing the functions $\ft_0$ and
$\gt_0$ by $\ft$ and $\gt$. The crucial point about the monotonicity of the function $\ft$ is guaranteed just like in the footnote below, given that $f$ is also monotonically decreasing. Thus, also in the BH geometry, one has a point $L'_{max}(r_h)$ above which the connected diagram does not exist.
It is an interesting question, whether this point is before or beyond the horizon. Namely, one can ask the question whether $L'_{max}(r_h) < L(r_h)$ or not. If so, then the connected D-string solution would cease to exist even before it falls into the horizon! The answer will be determined by the
precise background functions of the holographic model, however, this point does not change our arguments in section \ref{hooft} that only depends of existence of {\em some} $L_{max}$.

\section{Spectrum of bulk fluctuations}
\lab{Appflucs}

\subsection{Graviton and dilaton}
We first consider the  spectrum of the dilaton and the graviton fluctuations.
We shall carry out the calculation with Minkowski signature for convenience. The
results can easily be translated in Euclidean time by analytic continuation.
There are two independent modes with spin-0 and spin-2. The first
one is given by a mixture of the isotropic fluctuations of the
metric components $h_{11}= h_{22} = \cdots = h_{d-1,d-1}$ and the
dilaton fluctuations $\delta \f$. We denote this mode as $h_0$.
The second one is is just the transverse traceless shear
fluctuations that will be denoted by $h_2$.
%\footnote{These are the
%usual bosonic physical states of the string at the level
%$N=\tilde{N}=1$ decomposed into a scalar and transverse-traceless
%spin 2. There is also the anti-symmetric component, i.e. the
%fluctuations of the B-field, at this level. These are always
%massless $m_r=0$ and they correspond to the Goldstone mode of
%superfluid phase in this setting, as described in section
%\ref{sec3}. As emphasized in that section, they only contribute in
%the phase factor in the string path integral and thus irrelevant
%for computing the magnetization $|\vec{M}|$.}
The equations of motion for these modes on a generic BH background
(\ref{BH}), are obtained by decomposing  $h_{0,2}(r,x) = e^{-i p_i
x_i + i \o x^0} h_{0,2}(r)$. They can be found for example in
\cite{exotic1},\cite{GKMiN}. In the Einstein frame one obtains
\footnote{Strictly speaking these equations are correct only when
either of $|\vec{p}|^2 = p^ip^i$ or $\o^2$ vanish. Otherwise there
may be some more complicated mixing terms. We will keep this
combination and in the end of the computation we will be
interested either of these two cases.} \bea\lab{fluc1} h''_0 +h'_0
\le((d-1)A' + \frac{f'}{f} + 2 \frac{X'}{X}\ri) +
 h_0\le( \frac{\o^2-|\vec{p}|^2}{f^2}-  \frac{f'}{f}\frac{X'}{X} \ri)&=&0,\\
h''_2 +h'_2 \le((d-1)A'  + \frac{f'}{f}\ri) + h_2\le(
\frac{\o^2-|\vec{p}|^2}{f^2}\ri)&=&0, \lab{fluc2} \eea where the
function $X$ is given by $X(r) = 2/((d-1)\sqrt{d})~\f'/A'$ and it
asymptotes to a constant $X \to -1/\sqrt{d}$ in the IR region
$r\to\infty$. The analogous fluctuation equations on the thermal
gas background us given by setting $f=1$ in these equations.

From (\ref{fluc1}) and (\ref{fluc2}) we see that the spin-0 and
spin-2 modes become degenerate in this far IR region, hence it
suffices to consider only the latter.   As argued before and shown
in \cite{GKMN2} in the limit $r_h\to\infty$ the function $f$
approaches to 1 and $A$ approaches to the scale factor of the TG
solution (\ref{TG}) $A\to A_0$. Therefore, in this regime of
interest we want to solve, \be\lab{fluc3} h''_2 +h'_2 (d-1)A'_0  +
h_2~(\o^2-|\vec{p}|^2)=0. \ee One can easily transform this
equation to a Schrodinger form  by $h_2 = \tilde{h}_2
\exp(-(d-1)A/2)$: \be\lab{fluc4} -\tilde{h}_2'' + V_S(r)
\tilde{h}_2 = (\o^2-|\vec{p}|^2) \tilde{h}_2, \qquad V_S =
\frac{d-1}{2}A_0'' + \frac{(d-1)^2}{4} {A_0'}^2. \ee The
asymptotics of the function $A$ (\ref{As}) imply that the
Schrodinger potential asymptotes to a constant in the far IR:
\be\lab{schros} V_S \to m_0^2 +
\cO(e^{-\kappa\frac{\sqrt{V_\infty}}{2}r}),\qquad m_0^2 =
\frac{V_\infty}{4} \ee and for the form of the subleading
corrections we refer to \cite{exotic1}.

On the black-hole background the fluctuation equation
(\ref{fluc2}) is solved by imposing normalizability near the
boundary and incoming boundary condition (or normalizibility in
the Euclidean signature) at the horizon. Consider first the case
$|\vec{p}|=0$. Then one gets a discrete spectrum of $\o^2$ on the
BH. As $r_h$ is taken to $\infty$, i.e. near the phase transition
region $T\to T_c$ the spectrum becomes nearly continuous with the
lowest lying state determined by the asymptotic constant in
(\ref{schros}). Consequently, at finite but large $r_h$ the
spectrum of states are of the form, \be\lab{spec} \o_0 = m_0=
\frac{\sqrt{V_\infty}}{2}, \qquad \o_1 = m_0 +
\cO(e^{-\kappa\frac{\sqrt{V_\infty}}{2}r_h}), \,\,\, etc. \ee The
constant $m_0$ is the same as the one that appears in the
derivative of the dilaton (\ref{ldcftus}). This is not a
coincidence but required for the consistency of the theory. What
we learned is that, the gravitational fluctuations on the
linear-dilaton background is always gapped with a gap factor
$m_0$, i.e. \be\lab{gap} \o^2 > m_0^2, \qquad for\,\,
N=\tilde{N}=1. \ee

In the actual calculation of the two-point function in section
\ref{tpfclass} and \ref{tpfquant} we need the Euclidean spectrum
with compact time. One finds that the exchange diagram is of the
form $\exp(- m L)$ where $m$ is always bounded as in (\ref{gap})
from below.

\subsection{B-field}

We consider the fluctuations around $B_{r0}= const$ that we denote
as $\delta \psi$.  The spectrum of fluctuations are obtained from
the  equation of motion for the B-field $d*dB=0$. We decompose
$\delta\psi(r,x_0) = e^{-i p_i x_i + i\o x^0} \delta\Psi(r)$. On
simply obtains  $\o^2 = 0$, hence the spectrum of fluctuations
from the point of view of $d-1$ dimensions are also massless. This
means that in the exchange diagrams of section (\ref{tpfclass})
and (\ref{tpfquant}), the contribution from the lowest ${\cal
CT}^-$ modes is massless, in accord with existence of the
Goldstone mode in the superfluid phase.

\subsection{Tachyon}

The tachyon action is \cite{PolchinskiBook1} (in the string frame),
\be\lab{tachact}
{\cal A}_T \sim \int \sqrt{g_s} e^{-2\f} \le(g_s^{\m\n} \6_\m T \6_\n T - \frac{4}{\ell_s^2} T^2\ri),
\ee
where the metric and the dilaton reads,
\be\lab{bhapp}
  ds^2 = e^{2A_s(r)}\le( f^{-1}(r)dr^2 + dx_{d-1}^2 + dx_0^2 f(r) \ri), \qquad \f=
  m_0 r,
\end{equation}
and $m_0$ is defined in (\ref{ldcftus}). We shall carry out the calculation on the black-hole, and the thermal gas result will
be obtained simply by setting $f=1$.

We fluctuate $T = \la T\ra + T(r) e^{-i p^i x^i + \o x_0}$  in the
action. We note that these fluctuations do not mix with the
dilaton fluctuations for $\la T \ra = 0$ which is indeed what we
assume throughout the paper: the only non-trivial profiles in the
background are the metric and the dilaton. It is straightforward
to obtain the fluctuation equation from (\ref{tachact}) and
(\ref{bhapp}). As we are interested in the spectrum of $\o$ we set
$p^i=0$:
 \be\lab{tachfluc2} T'' + \le[ (d-1)A'_s - 2m_0+
\frac{f'}{f} \ri] T'  + \le( \frac{4}{\ell_s^2} \frac{e^{2A_s}}{f}
+ \frac{\o^2}{f^2} \ri) T= 0. \ee This can be
transformed into a Schodinger form by the change of variable $T =
\exp(-(d-1)A/2 + m_0 r -\half \log f) \tilde{T}$ with the result,
 \be\lab{tachfluc3}
-\tilde{T}'' + V_T \tilde{T} = \frac{\o^2}{f^2}
\tilde{T}, \qquad V_T = \frac{d-1}{2}A_s'' + (\frac{d-1}{2} A_s' -
m_0 + \frac{f'}{2f})^2 + \half(\frac{f''}{f} - \frac{f'^2}{f^2})-
\frac{4}{\ell_s^2} \frac{e^{2A_s}}{f}.
 \ee
 Let us first consider the simpler case of the thermal gas
   background that is obtained from (\ref{tachfluc3}) by setting $f=1$:
 \be\lab{tachfluc4}
V_T\bigg|_{TG} = \frac{d-1}{2}A_s'' + (\frac{d-1}{2} A_s' - m_0)^2
- \frac{4}{\ell_s^2} e^{2A_s}.
 \ee
In the TG phase in the IR $A_s$ vanishes as (\ref{delAs})
\be\lab{AsIR} A_s(r) \to a_1 e^{-\kappa m_0 r}\ee with some
positive coefficient $a_1$. The Schrodinger potential then
becomes, \be\lab{tachfluc5} V_T\bigg|_{TG} = m_0^2 -
\frac{4}{\ell_s^2} + a_1~e^{-\kappa m_0 r} \le(
\half(d-1)(\kappa+2)\kappa m_0^2 -\frac{8}{\ell_s^2}\ri) +
\cO(e^{-2m_0r}).
 \ee
For a moment let us consider the pure linear-dilaton geometry. In
this case the exponential correction term in (\ref{tachfluc5}) is
absent $a_1=0$, and one obtains the exact answer as,
\be\lab{tachfluc6} V_T\bigg|_{LD} = m_0^2 - \frac{4}{\ell_s^2} =
\left\{ \begin{array}{ll}
\frac{1-d}{6\ell_s^2} & \textrm{bosonic}\\
\frac{1-d}{4\ell_s^2} & \textrm{fermionic}
\end{array} \right.
 \ee
where we used the no-anomaly condition (\ref{m0sq}). The tachyon
in the fermionic case comes from the ground state of the NS
fermions and has the mass $m_T^2 = -2/\ell_s^2$. We re-derived the
well-known result that the ``tachyon" in dimensions 2 or less is
actually a stable mode (recalling that our total number of
dimensions is $d+1$). This is of course a consistency check.

Coming back the issue of the spectrum, the result
(\ref{tachfluc5}) indicates that the fluctuations in the deep
interior of the thermal gas geometry, in the vicinity of the phase
transition $T\to T_c$\footnote{This is the only regime the
world-sheet CFT becomes linear dilaton} are tachyonic. Luckily we
do not need this lowest mode in the calculation of the two-point
function in section \ref{tpfquant} because the entire propagation
is governed by  modes with winding mode $w=1$ which are
non-tachyonic.

However, we needed this mode in the calculation of the two-point
function in the {\em black-hole} phase, cf. section
\ref{tpfquant}. Now let us inspect the spectrum of tachyon
fluctuations on the black-hole. This is determined by the equation
(\ref{tachfluc2}). The blackness function near the horizon behaves
as \be\lab{black} f \to 4\pi T (r_h-r), \qquad r\to r_h. \ee Then
the fluctuation equation becomes the standard form, \be\lab{stbh}
T'' - T'(r_h-r)^{-1} + \frac{\tilde{\o}^2}{(r_h-r)^2}T \approx 0,
\qquad r\to r_h, \ee where $\tilde{\o} = \o/4\pi T$. This can be
solved by changing  the variable as $r_h-r = \exp(-u)$ and the
solution near the horizon becomes, \be\lab{solhor} T \to T_+ e^{i
\tilde{\o} u} + T_- e^{-i \tilde{\o} u}. \ee The incoming one
corresponds to $T_+=0$. To inspect the issue of the tachyon, we
can change to the Euclidean metric by $\o \to -i \o$ and the
incoming solution of the Minkowskian BH corresponds to the
normalizable solution of the Euclidean one. This means that the
Euclidean spectrum is always discrete and bounded from below.
However we still have to see whether there is a negative mode in
the limit $r_h\to \infty$ ($T\to T_c$). We  recall that in this
limit the BH geometry asymptotes to the TG geometry. In particular
\be\lab{asf} f(r)\to 1, \qquad for\,\,\, all\,\,\, r<r_h. \ee
Then, for any $r<r_h$ the corresponding Schodinger  potential is
given by (\ref{tachfluc5}) which becomes negative in the far $r$
region for $r<r_h$. Near $r_h$ it becomes positive again and
finally it diverges as $(r_h-r)^{-2}$ as $r\to r_h$. Then
existence of a negative discrete mode crucially depends on whether
the approach to the negative minimum that is given by
(\ref{tachfluc6}) is from above or below. This is determined by
the sign of the coefficient of the exponential term in
(\ref{tachfluc5}). Recalling that $a_1>0$, we observe that the
sign  is always positive for the interesting case of $\kappa=2$
which corresponds to a {\em second order transition} both for
$d-1=2$ and $d-1=3$. For a third or higher order transition it is
negative both for $d-1=2,3$. We conclude that in the cases of
interest, although there is a negative minimum in the tachyon
potential, the approach to this minimum is from above and the
potential can always be arranged (by choosing the form of the
next-to-subleading terms in the dilaton potential (\ref{Vs})) so
that there is no negative discrete mode in the spectrum. The same
cannot be said for fluctuations on the thermal gas, as explained
above.

It is a reasonable question to ask whether the tachyon of the
linear-dilaton CFT (on the thermal gas) can be  extrapolated to
the UV theory.   To answer this question one has to study the
tachyon potential in the UV. This can only be done in an heuristic
way. The reason is that, in the UV we do not have an exact CFT
description unlike in the IR and the $\a'$ corrections would
renormalize the following discussion. Nevertheless, let us pretend
that there are no $\a'$ corrections in order to see what possible
behavior can arise. In this paper we have not specified the UV
geometry, but in fact we always tacitly assume that the UV
geometry is AdS. In \cite{exotic1} we found analytic kink
solutions that flow from the UV in the AdS and linear-dilaton in
the IR. In the case of AdS the metric scale factor is $A \to
-\log{r/\ell} +\cdots$. Then we obtain, \be\lab{SchroUV} V_T
\approx (\frac{d-1}{2}+\frac{(d-1)^2}{4} -
\frac{8\ell^2}{\ell_s^2}) \frac{1}{r^2}, \qquad r\to 0. \ee This
will be bounded only when the term in the bracket is positive. In
the case of $d-1=3$ this gives the condition $\ell^2/\ell_s^2 <
15/32$. As mentioned above, this result is supposed to be
corrected by $\a'$ corrections. However it is reasonable to expect
that there will always be an upper bound on $\ell/\ell_s$    by
demanding that there is no $d-1$ dimensional tachyon in the
spectrum in the UV. On the other hand, the simplest way to achieve
this is to demand that there is no tachyon to start with i.e. the
spectrum in the $d+1$ dimensional theory is non-tachyonic and the
tachyon of the linear-dilaton theory only arises in the IR
effective theory.

%\section{The scaling of the the fundamental string near $T_c$}

\end{document}